\documentclass[11pt,draftcls,onecolumn]{IEEEtran}%
\usepackage{amsmath,amstext,amsfonts,amssymb,amsthm,bbold}
\usepackage{graphicx,subfigure,color}
\usepackage[english]{babel}
\usepackage[latin1]{inputenc}
\usepackage[T1]{fontenc}
\usepackage{amsmath}
\usepackage{amsfonts}
\usepackage{amssymb}
\usepackage{graphicx}%
\def \a {\mathbf{a}}
\def \B {\mathbf{B}}

\def \I {\mathbf{I}}
\def \i {\mathbf{i}}

\def \j {\mathbf{j}}

\def \k {\mathbf{k}}

\def \l {\mathbf{l}}

\def \Q {\mathbf{Q}}

\def \R {\mathbf{R}}

\def \T {\mathbf{T}}

\def \U {\mathbf{U}}

\def \W {\mathbf{W}}

\def \Tr {\mathrm{Tr}\, }
\renewcommand{\Im}{\mathrm{Im}}

\newcommand{\bs}{\boldsymbol}

\newtheorem{theorem}{Theorem}
\newtheorem{definition}{Definition}
\newtheorem{lemma}{Lemma}
\newtheorem{remark}{Remark}

\newtheorem{proposition}{Proposition}

\renewenvironment{proof}
{\noindent\textbf{Proof}:\small}{\hfill$\square$\bigbreak}
\definecolor{gray}{gray}{0.8}
\newcounter{countassum}
\newenvironment{assumption}
{
        \refstepcounter{countassum}
        \begin{flushleft}
        \noindent\textbf{(As \thecountassum)}:
        \it
}
{
        \end{flushleft}

}
\begin{document}

\title{Improved subspace estimation for multivariate observations of high dimension:
the deterministic signals case.}
\author{\IEEEauthorblockN
{ Pascal Vallet $^{1}$ \thanks{$^{1}$ Université de Paris-Est IGM LabInfo,
UMR-CNRS 8049, 5 Boulevard Descartes, 77454 Marne la Vallee Cedex 2, France,
E-mail: pascal.vallet@univ-mlv.fr} }
\and \IEEEauthorblockN
{ Philippe Loubaton $^{2}$ \thanks{$^{2}$ Université de Paris-Est IGM LabInfo,
UMR-CNRS 8049, 5 Boulevard Descartes, 77454 Marne la Vallee Cedex 2, France,
E-mail: philippe.loubaton@univ-mlv.fr} }
\and \IEEEauthorblockN
{ Xavier Mestre $^{3}$ \thanks{$^{3}$ Centre Tecnològic de Telecomunicacions
de Catalunya (CTTC), Av. Carl Friedrich Gauss 7, 08860 Castelldefels
(Barcelona), Spain, E-mail: xavier.mestre@cttc.cat} } \authorblockA{} }

\maketitle

\begin{abstract}
We consider the problem of subspace estimation in situations where the number
of available snapshots and the observation dimension are comparable in
magnitude. In this context, traditional subspace methods tend to fail because
the eigenvectors of the sample correlation matrix are heavily biased with
respect to the true ones. It has recently been suggested that this situation
(where the sample size is small compared to the observation dimension) can be
very\ accurately modeled by considering the asymptotic regime where the
observation dimension $M$ and the number of snapshots $N$ converge to
$+\infty$ at the same rate. Using large random matrix theory results, it can
be shown that traditional subspace estimates are not consistent in this
asymptotic regime. Furthermore, new consistent subspace estimate can be
proposed, which outperform the standard subspace methods for realistic values
of $M$ and $N$. The work carried out so far in this area has always been based
on the assumption that the observations are random, independent and
identically distributed in the time domain. The goal of this paper is to
propose new consistent subspace estimators for the case where the source
signals are modelled as unknown deterministic signals.   In practice, this 
allows to use the proposed approach regardless of the statistical properties
of the source signals.
In order to construct the proposed estimators, new technical results concerning the
almost sure location of the eigenvalues of sample covariance matrices of
Information plus Noise complex Gaussian models are established. These results are believed to 
be of independent interest.

\end{abstract}






\textbf{Notation: }Matrix (resp. vectors) quantities\ are denoted by boldfaced
capital (resp. lower case) letters. The $N\times N$ identity matrix is denoted
as $\mathbf{I}_{N}$. Trace and spectral norm will be denoted $\mathrm{Tr}%
\left[  \text{\textperiodcentered}\right]  $ and $\left\Vert
\text{\textperiodcentered}\right\Vert$ 
respectively, and $\left[  \text{\textperiodcentered}\right]^{T}$ and
$\left[  \text{\textperiodcentered}\right]^{H}$ represent the transpose and the
conjugate transpose. For a set
$\mathcal{U}$, we denote by $\mathrm{Int}(\mathcal{U})$ and $\partial
\mathcal{U}$ its interior and boundary respectively. Given a complex number
$z$, $\mathrm{Re}\left(  z\right)  $\ and $\mathrm{Im}\left(  z\right)  $
denote its real and imaginary parts respectively, $($\textperiodcentered
$)^{\ast}$ stands for complex conjugation and $\mathrm{i}$ denotes the
imaginary unit. The upper complex half plane is denoted by $\mathbb{C}_{+}$,
i.e $\mathbb{C}_{+}=\{z\in\mathbb{C}:\mathrm{Im}(z)>0\}$, and equivalently
$\mathbb{C}_{-}$ will denote the lower complex\ half plane. Similarly, 
$\mathbb{R}_{+}$ and $\mathbb{R}_{-}$ represent the set of all positive real numbers
and the set of all negative real numbers respectively. We will also write
$\mathbb{R}^{\ast}\equiv\mathbb{R}\backslash\left\{  0\right\}  $ and
$\mathbb{C}^{\ast}\equiv\mathbb{C}\backslash\left\{  0\right\}  $. For a given
contour $\mathcal{C}$ on the complex plane, $\mathrm{Ind}_{\mathcal{C}}(\xi)$
will denote the index of the contour with respect to a point $\xi\in
\mathbb{C}$. The support of a particular function $\phi$ will be denoted as
$\mathrm{supp}\left(  \phi\right) $, and $\mathcal{C}_{c}^{\infty}%
(\mathbb{R},\mathbb{R})$ will represent the set of compactly supported
real-valued smooth functions defined on $\mathbb{R}$.

\section{Introduction}

\label{sec:intro}Subspace estimation methods have been widely proposed in
multiple applications of communications and signal processing, such as
direction of arrival (DoA)\ estimation \cite{schmidt79}, beamforming
\cite{hung83}, channel identification \cite{MOUL971}, waveform estimation
\cite{LIU96A}, and many other general parametric estimation problems based on
multivariate observations \cite{Ottersten92}. In general terms, these
algorithms are applicable to the situation where a number of parameters needs
to be extracted from a set of multivariate observations, which are composed of
a noise part, with full-rank empirical correlation matrix, plus a signal
contribution that has low-rank empirical correlation matrix. By exploiting the
inherent orthogonality between the signal subspace (i.e. the subspace spanned
by the columns of the signal empirical correlation matrix) and the noise
subspace, one can try to extract the original parameters from the set of noisy
observations. In general terms, the resulting estimators are computationally
much more affordable and hence are generally preferred over other estimators
such as those based in the Maximum Likelihood (ML) principle, which generally
perform better but unfortunately involve an exhaustive search in a
multi-dimensional parametric space.

In order to formulate a generic subspace estimator, one must first infer the
eigenvectors of the correlation matrix of the observation. This is generally
difficult, because the correlation matrix of the multivariate observation is
generally unknown. In consequence, classical subspace estimation methods make
use of the empirical correlation matrix, and approximate the eigenvectors of
the true correlation matrix as the eigenvectors of the sample estimate. This
procedure is clearly optimal when the number of observations (denoted by $N$)
tends to infinity while the observation dimension (denoted by $M$) remains
constant. Indeed, under certain ergodicity assumptions, when $N\rightarrow
\infty$ for a fixed $M$, the sample correlation matrix of the observation
converges almost surely to the true one, and consequently when $N>>M$ the
sample eigenvectors (i.e. the eigenvectors of the sample correlation matrix)
tend to be very good representations of the true ones. In practical
applications, however, the number of available observations ($N$) and the
observation dimension ($M$) are comparable in magnitude, which leads to strong
discrepancies between the sample eigenvectors and the true ones. This
originates what is usually referred to as the breakdown effect of
subspace-based techniques \cite{tufts91}.

The fact that sample eigenvectors are not the best estimators of the true ones
has been known for decades, although the study of valid alternatives to the
classical estimators has been limited by the fact that investigations
basically concentrated on the regime where $N>>M$. However, it has been
recently suggested \cite{mestre2008modified} that finite sample size
situations (whereby $N$ and $M$ are comparable in magnitude) can be better
examined by investigating the asymptotic regime in which $M$ and $N$ converge
to $+\infty$ at the same rate, i.e. $M,N\rightarrow+\infty$, whereas
$c_{N}\equiv\frac{M}{N}$ converges towards a strictly positive constant. Using
Large Random Matrix Theory (LRMT) results, it was shown in
\cite{mestre2008modified} that traditional subspace estimators are
asymptotically biased in this asymptotic regime. Furthermore, consistent
estimators for this regime can be found, which outperform the traditional ones
for realistical values of $M$ and $N$. In this context, LRMT can be very
useful (1) to characterize how the sample eigenvectors differ from the true
ones in a scenario where $M$ and $N$ are comparable in magnitude and (2) to
derive alternative estimators of the eigenvectors that converge, not only when
$N\rightarrow+\infty$ for a fixed $M$, but also when $M,N\rightarrow+\infty$
at the same rate. This was more extensively demonstrated in
\cite{mestreeigsp08}\ and \cite{mestre2008improved}, which respectively
considered the characterization of the sample eigenvectors when
$M,N\rightarrow+\infty$ at the same rate, and proposed alternative consistent
estimators for these quantities in the new asymptotic regime.

Unfortunately, the work in \cite{mestreeigsp08}\ and \cite{mestre2008improved}
cannot be applied to the signal plus noise model considered here, unless the observations are
random multivariate quantities that are Gaussian, independent and
identically distributed in the time domain. In practice, however, there are multiple applications in which the
observation does not present this structure, and is better modelled as a
deterministic component (corresponding to the signal part) plus some additive
noise, that is generally Gaussian distributed. This model is usually referred
to as the \textquotedblleft information plus noise model\textquotedblright\ in
the LRMT literature \cite{dozier2007empirical}, as opposed to the more
classical \textquotedblleft sample covariance matrix model\textquotedblright%
\ \cite{silverstein95}, which was the one used in
\cite{mestre2008modified,mestreeigsp08,mestre2008improved}. The main objective
of this paper is to propose improved subspace estimators for the information
plus noise model, which will represent the case where the source signals are
as non-observable deterministic sequences. In order to obtain these estimators,
new mathematical results related to the almost sure location of the eigenvalues 
of the empirical covariance matrix of a Gaussian information plus noise model are derived. 
These results are believed to be of independent interest.

The rest of the paper is organized as follows.\ Section \ref{section:problem}
introduces the information plus noise model associated with the 
specific application addressed here:  the determination of multiple directions of arrival (DoA)
using an array of antennas. The main objectives of the paper in mathematical
terms are also formulated. Section \ref{section:background} provides some
general facts related to the convergence of the eigenvalues
of the empirical correlation matrix for the information plus noise model. 
It is further explained in Section \ref{section:support} that
the eigenvalues of the sample correlation matrix tend to concentrate around
some clusters when both $M,N\rightarrow+\infty$ at the same rate. 
A very simple description of the position of these
asymptotic\ eigenvalue clusters is also provided. It is in particular 
shown that each cluster is associated with a set of consecutive eigenvalues of true covariance matrix
of the observation. Section \ref{section:exactseparation}
presents an intermediate result that has its own interest. In brief, it is
shown that, for sufficiently large $M$, $N$, with probability one no
eigenvalues of the sample correlation matrix will be located outside the
asymptotic eigenvalue clusters. Furthermore, the number of 
sample eigenvalues that are located in each of these clusters is directly related
to the dimensionality of the corresponding eigenspace of the true covariance matrix. 
In order to focus on the applicative context of the paper, this claim is proved for the cluster 
associated with the noise subspace, but it can be extended easily to the other clusters.  
This fact generalizes the results derived in \cite{bai1998no}
and \cite{bai1999exact} in the context of source signals independent identically distributed in 
the time domain. In contrast with \cite{bai1998no} and \cite{bai1999exact}, 
the results presented in this paper, inspired by the approach developed in \cite{capitaine2009largest}, are 
only valid in the complex Gaussian case. 
The above mentioned results are then used in Section \ref{section:estimationeigenvalues} in order to
derive an estimator of the localization function of the subspace estimate that is 
consistent not only when $N\rightarrow+\infty$
for fixed $M$, but also when $M,N\rightarrow+\infty$ at the same rate. Section
\ref{section:simu} provides some numerical examples that illustrate the
effectiveness of the proposed estimators. Finally Section
\ref{section:conclusions} concludes the paper. Most of the technical
derivations have been relegated to the appendices. \\

The results of this paper have been partly presented in the short conference paper \cite{vallet2009improved}. 


\section{Problem statement}

\label{section:problem} In order to motivate and illustrate the signal model
that is used in this paper, we consider the following DoA estimation problem.
Assume that $K$ narrow band deterministic source signals $(s_{k}%
)_{k=1,\ldots,K}$ are received by an antenna array of $M$ elements, $K<M$.
The corresponding $M$ dimensional observation signal $\mathbf{y}_{n}$ (at
discrete time $n$) can be mathematically described as
\[
\mathbf{y}_{n}=\mathbf{A}\mathbf{s}_{n}+\mathbf{v}_{n}%
\]
where $\mathbf{A}=\left[  \mathbf{a}(\theta_{1}),\ldots,\mathbf{a}(\theta
_{K})\right]  $ is an $M\times K$ matrix that contains the steering vectors of
the $K$ sources, $\mathbf{s}_{n}$ is a $K\times1$ column vector containing the
transmitted signals from the $K$ sources at time instant $n$, and where
$\mathbf{v}_{n}$ is an additive Gaussian white noise component with zero mean
and covariance matrix $\mathbb{E}\left[  \mathbf{v}_{n}\mathbf{v}_{n}%
^{H}\right]  =\sigma^{2}\mathbf{I}_{M}$. We assume that $\mathbf{y}_{n}$ is available from $n=1$ to $n=N$, and that
$M<N$, or equivalently that $c_{N}=\frac{M}{N}$ is strictly less than $1$. It
is possible to generalize our results to the situation where $c_{N}>1$,
although the presentation of the corresponding results would however
complicate the developments of the present paper.

We denote by $\mathbf{Y}_{N}=\left[  \mathbf{y}_{1},\ldots,\mathbf{y}%
_{N}\right]  $ the $M\times N$ observation matrix, which can be readily
written as
\begin{equation}
\mathbf{Y}_{N}=\mathbf{A}\mathbf{S}_{N}+\mathbf{V}_{N} \label{eq:modele-bis}%
\end{equation}
where $\mathbf{S}_{N}=\left[  \mathbf{s}_{1},\ldots,\mathbf{s}_{N}\right]  $
and $\mathbf{V}_{N}=\left[  \mathbf{v}_{1},\ldots,\mathbf{v}_{N}\right]  $.
From this matrix, we can define the empirical spatial correlation matrix of
the observation as $\hat{\mathbf{R}}_N \equiv\frac{1}{N}\mathbf{Y}_{N}%
\mathbf{Y}_{N}^{H}$, whereas the empirical spatial correlation matrix
associated with the noiseless observation will take the form $\frac{1}%
{N}\mathbf{A}\mathbf{S}_{N}\mathbf{S}_{N}^{H}\mathbf{A}^{H}$. It is worth
pointing out here that, since the number of signals is assumed to be lower
than the number of antennas ($K<M\,$), the steering matrix $\mathbf{A}$
will\ always be a tall matrix and therefore the empirical spatial correlation
matrix of the noiseless observation will never be full rank. In other words,
the minimum eigenvalue of the matrix $\frac{1}{N}\mathbf{A}\mathbf{S}%
_{N}\mathbf{S}_{N}^{H}\mathbf{A}^{H}$ will always be zero and will have
multiplicity equal to $M-K$.

In order to simplify the notation in the subsequent exposition, we define the
matrices $\boldsymbol{\Sigma}_{N}$, $\mathbf{B}_{N}$, $\mathbf{W}_{N}$ as
\begin{equation}
\boldsymbol{\Sigma}_{N}=\frac{\mathbf{Y}_{N}}{\sqrt{N}},\;\mathbf{B}_{N}%
=\frac{\mathbf{A}\mathbf{S}_{N}}{\sqrt{N}},\;\mathbf{W}_{N}=\frac
{\mathbf{V}_{N}}{\sqrt{N}} \label{eq:def-normalisation}%
\end{equation}
so that (\ref{eq:modele-bis}) can be equivalently formulated as
\begin{equation}
\boldsymbol{\Sigma}_{N}=\mathbf{B}_{N}+\mathbf{W}_{N} \label{eq:modele-ter}%
\end{equation}
where $\boldsymbol{\Sigma}_{N}$ is the (normalized) matrix of observations,
$\mathbf{B}_{N}$ is a deterministic matrix containing the signals contribution
and $\mathbf{W}_{N}$ is a complex Gaussian white noise matrix with i.i.d. entries that have
zero mean and variance $\sigma^{2}/N$. We denote by $\boldsymbol{\Pi}_{N}$ the
orthogonal projection matrix on the \textquotedblleft noise
subspace\textquotedblright, which in our context is defined as the orthogonal
complement of the column space of matrix $\mathbf{A}$. In the following, we
assume that the empirical correlation matrix of $\mathbf{S}_{N}$ defined by
$\frac{1}{N}\mathbf{S}_{N}\mathbf{S}_{N}^{H}$ is full rank. Therefore, the
noise subspace coincides with the kernel of the empirical correlation matrix
of the noiseless signal, namely $\mathbf{B}_{N}\mathbf{B}_{N}^{H}.$

Let $\left\{  \gamma_{k}^{(N)}\right\}  _{k=1,\ldots,M}$ denote the
eigenvalues of the empirical correlation matrix of the signal component,
namely $\mathbf{B}_{N}\mathbf{B}_{N}^{H}$, arranged in increasing order and
let $\left\{  \mathbf{e}_{k}^{(N)}\right\}  _{k=1,\ldots,M}$ denote the
corresponding unit norm eigenvectors. We note in particular that $\gamma
_{1}^{(N)}=\ldots=\gamma_{M-K}^{(N)}=0$ while the remaining eigenvalues are
strictly positive and that $\boldsymbol{\Pi}_{N}=\sum_{k=1}^{M-K}%
\mathbf{e}_{k}^{(N)}\left(  \mathbf{e}_{k}^{(N)}\right)  ^{H}$. The subspace
method for the determination of the $K$ directions of arrival (commonly
referred to as MUSIC algorithm) is based on the observation that the angles
$\left\{  \theta_{k}\right\}  _{k=1,\ldots,K}$ coincide with the $K$ solutions
of the equation $\mathbf{a}(\theta)^{H}\boldsymbol{\Pi}_{N}\mathbf{a}%
(\theta)=0$. In order to be able to use this last observation, it is in
practice necessary to estimate the function $\mathbf{a}(\theta)^{H}%
\boldsymbol{\Pi}_{N}\mathbf{a}(\theta)$ (usually referred to as the
\textquotedblleft localization function\textquotedblright) for each $\theta
\in\lbrack-\pi,\pi]$, or more generically to estimate the quantity
\[
\eta_{N}({\bf b})=\mathbf{b}^{H}\boldsymbol{\Pi}_{N}\mathbf{b}
\]
for each deterministic $M$-dimensional vector $\mathbf{b}$.

If $N\rightarrow+\infty$ while $M$ is fixed, the empirical correlation matrix
of the observations $\mathbf{\hat{R}}_{N}=\boldsymbol{\Sigma}_{N}%
\boldsymbol{\Sigma}_{N}^{H}$ of $\mathbf{Y}_{N}$ converges towards the matrix
$\mathbf{R}_{N}=\mathbf{B}_{N}\mathbf{B}_{N}^{H}+\sigma^{2}\mathbf{I}_{M}$ in
the sense that
\begin{equation}
\Vert\mathbf{\hat{R}}_{N}-(\mathbf{B}_{N}\mathbf{B}_{N}^{H}+\sigma
^{2}\mathbf{I}_{M})\Vert \rightarrow0\quad\mathrm{a.s.}
\label{eq:convergence-c=0}%
\end{equation}
where  a.s. represents the almost sure
convergence. We will denote by $\left\{  \hat{\lambda}_{k}^{(N)}\right\}
_{k=1,\ldots,M}$ the eigenvalues of $\mathbf{\hat{R}}_{N}$ arranged in
increasing order and by $\left\{  \hat{\mathbf{e}}_{k}^{(N)}\right\}
_{k=1,\ldots,M}$ the corresponding eigenvectors. The convergence result in
\eqref{eq:convergence-c=0} implies that for each $\theta$, $\hat{\eta}_{N}^{trad}({\bf a}(\theta)) -\eta
_{N}({\bf a}(\theta)) \rightarrow0\ \ \mathrm{a.s.}$ where $\hat{\eta}_{N}^{trad}({\bf a}(\theta))$ is the
traditional estimator of the localization function defined as
\begin{equation}
\hat{\eta}_{N}^{trad}({\bf a}(\theta))=\sum_{k=1}^{M-K}\mathbf{a}^{H}(\theta)\hat{\mathbf{e}%
}_{k}^{(N)}\left(  \hat{\mathbf{e}}_{k}^{(N)}\right)  ^{H}\mathbf{a}%
(\theta)\mathbf{.} \label{eq:def-eta_music}%
\end{equation}
In practice, predictions provided by the asymptotic regime corresponding to
letting $N\rightarrow+\infty$ for fixed $M$ are reliable only if $N$ is much
larger than $M$. However, this assumption may be quite restrictive in a number
of important application contexts. If $M$ and $N$ are comparable in magnitude,
then the asymptotic regime described by letting $M,N\rightarrow+\infty$ in
such a way that $c_{N}=\frac{M}{N}$ converges towards a non zero constant
appears to be more relevant. In this regime, the behavior of various classical
estimates are more complicated, and have to be studied carefully. In
particular, it can be shown that $\hat{\eta}_{N}^{trad}({\bf b}) -\eta_{N}({\bf b})$ does not
converge to $0$ when $M,N\rightarrow+\infty$ , which implies that the standard
MUSIC estimates are not consistent under this new asymptotic regime. The
purpose of this paper is to introduce an improved subspace estimate 
$\hat{\eta}_{N}^{new}({\bf b})$ of $\eta_N({\bf b})$ for each 
deterministic vector ${\bf b}$. The main feature of $\hat{\eta}_{N}^{new}({\bf b})$ 
is to be consistent if $M,N\rightarrow+\infty$ in such a way that
$c_{N}=\frac{M}{N}$ converges towards a non zero constant value. In order to
achieve this, we will heavily rely on results related to the asymptotic
behavior of the eigenvalue distribution of the empirical correlation matrix
$\mathbf{\hat{R}}_{N}$. It is however useful to mention that it is not established that
\begin{equation}
\label{eq:letop}
\sup_{\theta \in [-\pi, \pi]} |\hat{\eta}_{N}^{new}({\bf a}(\theta)) - \eta_N({\bf a}(\theta))| \rightarrow 0
\end{equation}
almost surely, a useful, but stronger property. We feel that the proof of (\ref{eq:letop}) would need mathematical technics different from 
those which are used in the present paper.


\section{Properties of the asymptotic eigenvalue distribution of matrix
$\hat{\mathbf{R}}_{N}$}

\label{section:background}

In this section, we will review some of the important properties related to
the asymptotic behavior of the eigenvalue distribution of the empirical
correlation matrix $\hat{\mathbf{R}}_{N}$ when $M,N\rightarrow+\infty$ in such
a way that $c_{N}=\frac{M}{N}$ converges towards a non zero constant, which
will be denoted as $c_{\ast}$. This implies that the observation dimension $M$
in principle depends on $N$, and should be denoted $M(N)$. We will however
drop this dependence on $N$ in order to simplify the exposition. Whenever it
is clear from the context, we will also drop the dependence on the number of
snapshots $N$ in matrices $\boldsymbol{\Sigma}_{N}$, $\B_{N}$, $\hat{\R}_{N}$,
eigenvalues $\hat{\lambda}_{1}^{(N)}$,\ldots, $\hat{\lambda}_{M}^{(N)}$ and
$\gamma_{1}^{(N)}$,\ldots,$\gamma_{M}^{(N)}$, as well as eigenvectors.

\begin{remark}
From now on, $N\rightarrow\infty$ will implicitly denote the limit as both
$M,N\rightarrow+\infty$ such that $\frac{M}{N}$ converges towards a non zero
constant $c_{\ast}$, where it is assumed that $0<c_{\ast}<1$. \newline
\end{remark}

\begin{remark}
All results that are presented in this paper are equally valid regardless of
the behavior of the number of sources $K$ when $N$ increases. In other words,
$K$ may scale up with $N$, or it may stay constant regardless of $N$.
\end{remark}

From now on, we assume that the spectral norms of matrices $({\bf B}_N)_{N \geq 1}$ 
remain bounded when $N \rightarrow \infty$, i.e. it exists $b_{max} > 0$ such that
\begin{equation}
\label{eq:Bbornee}
\sup_{N \geq 1} \| {\bf B}_N \| < b_{max} < \infty
\end{equation}

The eigenvalue distribution of $\hat{\mathbf{R}}_{N}$ is characterized by the
empirical distribution function of its eigenvalues, namely
\[
\hat{F}_{N}(\lambda)=\frac{1}{M}\mathrm{card}\{\hat{\lambda}_{k}^{(N)}%
:\hat{\lambda}_{k}^{(N)}\leq\lambda,\ k=1,\ldots,M\}
\]
where $\mathrm{card}$ denotes the cardinality of a set. For each $\lambda
\in\mathbb{R}$, the function $\hat{F}_{N}(\lambda)$ gives the proportion of
the eigenvalues of $\hat{\mathbf{R}}_{N}$ which are lower than or equal to
$\lambda$. Its associated probability measure, denoted $\hat{\mu}_{N}$, is
given by $\mathrm{d} \hat{\mu}_{N}(\lambda)=\frac{1}{M}\sum_{k=1}^{M}%
\delta(\lambda-\hat{\lambda}_{k}^{(N)})$ and is carried by $\mathbb{R}_{+}.$
In order to characterize the asymptotic behavior of $\hat{\mu}_{N}$, it is in
practice quite common to characterize the asymptotic behavior of its Stieltjès
transform. If $\mu$ is a positive finite measure (i.e. $\mu(\mathbb{R}) < \infty$), 
the Stieltjès transform of $\mu$
is the function $\Psi_{\mu}$ of complex variable defined as%
\begin{equation}
\Psi_{\mu}(z)=\int_{\mathbb{R}}\frac{\mathrm{d} \mu(\lambda)}{\lambda-z}
\label{eq:def_stieltjes}%
\end{equation}
We recall the following well-known properties of the Stieltjès transform,
which will be useful in the mathematical developments throughout the paper.

\begin{lemma}
\label{property:stieltjes} Let $\Psi_{\mu}$ be the Stieltjès transform of some
positive finite measure $\mu$ (i.e. $\mu(\mathbb{R}) < \infty$), and let us denote as $\mathcal{S}_{\mu}$ its
support. Then,

\begin{enumerate}
\item \label{enu:analytic} $\Psi_{\mu}$ is holomorphic on $\mathbb{C}%
\backslash\mathcal{S}_{\mu}$.

\item $\lim_{y \rightarrow +\infty} -i y \Psi_{\mu}(iy) = \mu(\mathbb{R})$

\item 
\label{enu:majoration}
$\forall z\in\mathbb{C}\backslash\mathbb{R}$,
\[
\left\vert \Psi_{\mu}(z)\right\vert \leq\frac{\mu(\mathbb{R})}{\left\vert \mathrm{Im}%
(z)\right\vert }%
\]
where $\mathrm{Im}(z)$ denotes the imaginary part of $z$. Moreover, 
$\forall z\in\mathbb{C}\backslash {\cal S}_{\mu}$
it holds that
\begin{equation}
\label{eq:inegalite-amelioree}
\left\vert \Psi_{\mu}(z)\right\vert \leq\frac{\mu(\mathbb{R})}{\mathrm{dist}(z,{\cal S}_{\mu})}
\end{equation}

\item \label{enu:C_+} $\Psi_{\mu}\in\mathbb{C}_{+}$ if $z\in\mathbb{C}_{+}$,
where $\mathbb{C}_{+}$ is the upper complex half plane.

\item 
\label{enu:Imzpsi}
If $\mu$ is carried by $\mathbb{R}_{+}$, then $z\Psi_{\mu}(z)\in
\mathbb{C}_{+}$ if $z\in\mathbb{C}_{+}.$

\item \label{enu:converse-stieljes} Conversely, if $\Psi$ is a function analytic in $\mathbb{C}_{+}$ satisfying
\begin{itemize}
\item $\Psi(z)$ and $z \Psi(z)$ belong to $\mathbb{C}_{+}$ if $z \in \mathbb{C}_{+}$
\item $\sup_{y >1} |iy \Psi(iy)| < +\infty$
\end{itemize}
then, $\Psi$ is the Stieljès transform of a positive finite measure carried by $\mathbb{R}_{+}$.

\item \label{enu:inversion_phi} $\forall\varphi\in\mathcal{C}_{c}^{\infty
}(\mathbb{R},\mathbb{R})$, (the set of compactly supported real-valued smooth
functions defined on $\mathbb{R}$), we have
\[
\int_{\mathbb{R}}\varphi(\lambda)\mathrm{d}\mu(\lambda)=\frac{1}{\pi}%
\lim_{y\downarrow0}\mathrm{Im}\left\{  \int_{\mathbb{R}}\varphi(x)\Psi_{\mu
}(x+\mathrm{i}y)\mathrm{d}x\right\}
\]

\end{enumerate}
\end{lemma}

Having recalled these basic properties of the Stieltjès transform of a
positive finite measure, let us now go back to the asymptotic characterization of
the empirical measure $\hat{\mu}_{N}$ or, quite equivalently, its Stieltjès
transform, which is defined for $z\in\mathbb{C}-\mathbb{R}_{+}$\ as
\begin{equation}
\hat{m}_{N}(z)=\int_{\mathbb{R}_{+}}\frac{\mathrm{d}\hat{\mu}_{N}(\lambda
)}{\lambda-z}=\frac{1}{M}\sum_{m=1}^{M}\frac{1}{\hat{\lambda}_{m}-z}.
\label{eq:def-stieltjesrandom}%
\end{equation}
It is worth pointing out that $\hat{m}_{N}(z)$ can be expressed as the
normalized trace of the resolvent matrix, which is a matrix-valued function
defined as
\begin{equation}
\mathbf{Q}_{N}(z)=\left(  \hat{\mathbf{R}}_{N}-z\mathbf{I}_{M}\right)
^{-1}=\left(  {\boldsymbol{\Sigma}}_{N}{\boldsymbol{\Sigma}}_{N}%
^{H}-z\mathbf{I}_{M}\right)  ^{-1} \label{eq:def-Q}%
\end{equation}
namely $\hat{m}_{N}(z)=\frac{1}{M}\mathrm{Tr}\left[  \mathbf{Q}_{N}(z)\right]
$. Except (\ref{eq:convergence-Qij}), the following results can be more or less immediately derived from
\cite{dozier2007empirical} (see also \cite{hachem2007deterministic})

\begin{theorem}
\label{theo:convergence-hatmu}There exists a deterministic probability
distribution $\mu_{N}$ carried by $\mathbb{R}_{+}$ such that $\hat{\mu}%
_{N}-\mu_{N}$ converges in distribution almost surely towards $0$ when
$N\rightarrow\infty$. The measure $\mu_{N}$, referred to in what follows as
the asymptotic eigenvalue distribution of matrix $\hat{\mathbf{R}}_{N}$, is
characterized by its Stieltjès transform $m_{N}(z)$ as
\begin{equation}
m_{N}(z)=\int_{\mathbb{R}_{+}}\frac{\mathrm{d} \mu_{N}(\lambda)}{\lambda-z}%
\label{eq:def-stieljes}%
\end{equation}
which is a solution of the equation
\begin{equation}
m_{N}(z)=\frac{1}{M}\mathrm{Tr}\left[  -z(1+\sigma^{2}c_{N}m_{N}%
(z))\mathbf{I}_{M}+\sigma^{2}(1-c_{N})\mathbf{I}_{M}+\frac{\mathbf{B}%
_{N}\mathbf{B}_{N}^{H}}{1+\sigma^{2}c_{N}m_{N}(z)}\right]  ^{-1}%
\label{eq:canonique-1}%
\end{equation}
for each $z\in\mathbb{C}-\mathbb{R}_{+}$. Let $\mathbf{T}_{N}(z)$ be the
$M\times M$ matrix valued function defined on $\mathbb{C}-\mathbb{R}_{+}$ by
\begin{equation}
\mathbf{T}_{N}(z)=\left[  -z(1+\sigma^{2}c_{N}m_{N}(z))\mathbf{I}_{M}%
+\sigma^{2}(1-c_{N})\mathbf{I}_{M}+\frac{\mathbf{B}_{N}\mathbf{B}_{N}^{H}%
}{1+\sigma^{2}c_{N}m_{N}(z)}\right]  ^{-1}.\label{eq:def-T}%
\end{equation}
Then, $\mathbf{T}_{N}(z)$ is holomorphic on $\mathbb{C}-\mathbb{R}_{+}$. Moreover, 
almost surely,
\begin{equation}
\lim_{N\rightarrow\infty} \left( \hat{m}_{N}(z)-m_{N}(z)\right)  =0\label{eq:convergence-traceQ}%
\end{equation}
for each $z\in\mathbb{C}-\mathbb{R}_{+}$. Finally, for each
$M$--dimensional deterministic vectors $\mathbf{u}_{N},\mathbf{v}_{N}$
such that $\sup_{N} \| {\bf u}_N  \| < \infty$ and  $\sup_{N} \| {\bf v}_N  \| < \infty$, it
holds that almost surely
\begin{equation}
\lim_{N\rightarrow\infty}\mathbf{u}_{N}^{H}\left(  \mathbf{Q}_{N}%
(z)-\mathbf{T}_{N}(z)\right)  \mathbf{v}_{N}=0\label{eq:convergence-Qij}%
\end{equation}
for each $z\in\mathbb{C}-\mathbb{R}_{+}$.
\end{theorem}

\begin{IEEEproof}
Convergence of $\hat{\mu}_{N}-\mu_{N}$ towards $0$ as well as the fact that
$m_{N}(z)$ is a solution to (\ref{eq:canonique-1}) is due to
\cite{dozier2007empirical}. As for the result in (\ref{eq:convergence-traceQ}%
), it is a well known consequence of the convergence of $\hat{\mu}_{N}-\mu
_{N}$ towards $0$. (\ref{eq:convergence-Qij}) is proved in the Appendix \ref{sec:individual-entries}.  
\end{IEEEproof}

Theorem \ref{theo:convergence-hatmu} is pointing out that the entries of the
resolvent $\mathbf{Q}_{N}(z)$ are almost surely asymptotically close to the
entries of the deterministic matrix function $\mathbf{T}_{N}(z)$ (this
statement follows from (\ref{eq:convergence-Qij}) by selecting $\mathbf{u}_{N}$ and 
$\mathbf{v}_{N}$ as two columns of $\mathbf{I}_{M}$); and that its
normalized trace, $\hat{m}_{N}(z)$ as defined in (\ref{eq:def-stieltjesrandom}%
), is almost surely asymptotically close to $m_{N}(z)$, one of the solutions
to the polynomial equation in (\ref{eq:canonique-1}). Furthermore, the random
measure $\hat{\mu}_{N}$ is also almost surely equivalent (in distribution) to
the deterministic measure $\mu_{N}$ in this asymptotic regime.

We denote by $\mathcal{S}_{N}$ the support of this measure $\mu_{N}$, which
will play a very important role in the following. The characterization of
$\mathcal{S}_{N}$ has been first presented in \cite{dozier2007analysis}, and
is based on the study of the properties of function $m_{N}(z)$ which, since it
is a Stieltjès transform, is holomorphic on $\mathbb{C}\backslash
\mathcal{S}_{N}$ and real-valued on $\mathbb{R}\backslash\mathcal{S}_{N}$. In
order to characterize $\mathcal{S}_{N}$, we will also consider the function
$w_N(z)$, introduced in \cite{dozier2007analysis}, 
defined from $m_{N}(z)$ as follows
\begin{equation}
w_{N}(z)=z\left(  1+\sigma^{2}c_{N}m_{N}(z)\right)  ^{2}-\sigma^{2}%
(1-c_{N})(1+\sigma^{2}c_{N}m_{N}(z)). \label{eq:def-w}%
\end{equation}
It will be seen later on that the function $w_{N}(z)$ has very interesting
properties that will be crucial for the derivations in this paper. In
particular, we will show in the following that the support of $\mu_{N}$, namely
$\mathcal{S}_{N}$, is in fact equal to the support of the imaginary part of
$w_{N}(z)$ when $z$ approaches the real axis. Thanks to this fact, we will be
able to characterize the support $\mathcal{S}_{N}$ by studying the properties
of $w_{N}(z)$ for $z$ on the real axis.

The next proposition provides some preliminary properties of $m_{N}(z)$ and
$w_{N}(z)$ that will become useful in the following sections. Most of these
properties are established in \cite{dozier2007analysis}. We will denote by
$f_{N}(w)$ the function on $\mathbb{C}-\left\{  \gamma_{1},\ldots,\gamma
_{M}\right\}  $ defined by%
\[
f_{N}(w)=\frac{1}{M}\mathrm{Tr}\left[  \left(  \mathbf{B}_{N}\mathbf{B}%
_{N}^{H}-w\mathbf{I}_{M}\right)  ^{-1}\right]
\]
which coincides with the Stieltjès transform of the eigenvalue distribution
$\nu_{N}(\mathrm{d}\lambda)=\frac{1}{M}\sum_{k=1}^{M}\delta(\lambda-\gamma
_{k})$ associated with the signal matrix $\mathbf{B}_{N}\mathbf{B}_{N}^{H}$.

\begin{proposition}
\label{prop:m}The following properties hold:

\begin{enumerate}
\item \label{enu:O_supp} The condition $c_{N}<1$ implies that $0$ does not
belong to $\mathcal{S}_{N}.${}

\item \label{enu:lim_m_R} For each $x\in\mathbb{R}$, $\lim_{z\in\mathbb{C}%
_{+},z\rightarrow x}m_{N}(z)$ exists, and will be denoted $m_{N}(x)$. The
function $m_{N}(z)$ thus defined is continuous on $\mathbb{C}_{+}%
\cup\mathbb{R}$, and continuously differentiable on $\mathbb{C}_{+}%
\cup\mathbb{R}-\partial\mathcal{S}_{N}$. Moreover, for each $x\in\mathbb{R}$,
$\lim_{z\in\mathbb{C}_{-},z\rightarrow x}m_{N}(z)$ exists, and is equal to
$(m_{N}(x))^{\ast}$. The measure $\mu_{N}$ is absolutely continuous, its
density is $\frac{1}{\pi}\mathrm{Im}(m_{N}(x))$, and the interior
$\mathrm{Int}(\mathcal{S}_{N})$ of $\mathcal{S}_{N}$ is given by
\begin{equation}
\mathrm{Int}(\mathcal{S}_{N})=\{x>0:\mathrm{Im}(m_{N}(x))>0\}
\label{eq:expre-intsupp}%
\end{equation}

\item \label{enu:derivee_w} For each $x\in\mathbb{R}$, $\lim_{z\in
\mathbb{C}_{+},z\rightarrow x}w_{N}(z)$ exists, and is still denoted by
$w_{N}(x)$. The function $z\rightarrow w_{N}(z)$ is continuous on
$\mathbb{C}_{+}\cup\mathbb{R}$, and is continuously differentiable on
$\mathbb{C}_{+}\cup\mathbb{R}-\partial\mathcal{S}_{N}$. Moreover,
$w_{N}(x)=x\left(  1+\sigma^{2}c_{N}m_{N}(x)\right)  ^{2}-\sigma^{2}%
(1-c_{N})(1+\sigma^{2}c_{N}m_{N}(x))$. Finally, $\lim_{z\in\mathbb{C}%
_{-},z\rightarrow x}w_{N}(z)=w_{N}(x)^{\ast}$.

\item \label{enu:image_w_R} $w_{N}(x)$ does not belong to the set
$\{\gamma_{1}^{(N)},\ldots,\gamma_{M}^{(N)}\}$ if $x\in\mathbb{R}%
-\mathcal{S}_{N}.$

\item \label{enu:Imw_positive} $\mathrm{Im}\left[  w_{N}(z)\right]  >0$ if
$\mathrm{Im}z>0$.

\item \label{enu:Re_m} $\mathrm{Re}\left[  1+c_{N}\sigma^{2}m_{N}(z)\right]
>0$ for each $z\in\mathbb{C}$.

\item \label{enu:mx_sol_can_eq}For any $x\in\mathbb{R}-\partial\mathcal{S}%
_{N}$, the function $m_{N}(x)$ is solution of the equation in
(\ref{eq:canonique-1})

\item \label{enu:wx_sol_can_eq}For any $x\in\mathbb{R}-\partial\mathcal{S}%
_{N}$, the function $w_{N}(x)$ is a solution of the equation%
\begin{equation}
\phi_{N}(w_{N}(x))=x \label{eq:equation-Phi}%
\end{equation}
where $\phi_{N}(w)$ is defined by
\begin{equation}
\phi_{N}(w)=w\;(1-c_{N}\sigma^{2}f_{N}(w))^{2}+(1-c_{N})\sigma^{2}%
(1-c_{N}\sigma^{2}f_{N}(w)) \label{eq:expre-phi}%
\end{equation}

\end{enumerate}
\end{proposition}%

\begin{IEEEproof}
Property \ref{enu:O_supp} is not established in \cite{dozier2007analysis}, and
is proved in Appendix \ref{appendix:proof_zero_not_belong}. As for Property
\ref{enu:lim_m_R}, the existence of the limit of $m_{N}(x+\mathrm{i}y)$ is
proved in \cite{dozier2007analysis} for $x\neq0$ because
\cite{dozier2007analysis} did not assume that $c_{N}<1$. However, Property
\ref{enu:O_supp} implies immediately that the limit exists if $x=0$ because
$m_{N}(z)$ is holomorphic in a neighborhood of the origin. The continuity and
the differentiability of $x\rightarrow m_{N}(x)$ is established in
\cite{dozier2007analysis} on $\mathbb{R}^{\ast}$ and $\mathbb{R}^{\ast
}\backslash\partial\mathcal{S}_{N}$ respectively, but it also holds on
$\mathbb{R}$ and $\mathbb{R}\backslash\partial\mathcal{S}_{N}$ by Property
\ref{enu:O_supp} and the fact that $m_{N}(z)$ is holomorphic $\mathbb{C}%
\backslash\mathcal{S}_{N}$. Since $m_{N}(z)$ is the Stieltjès transform of a
positive measure, it is clear that $m_{N}(z^{\ast})$ coincides with
$m_{N}^{\ast}(z)$. This implies immediately that $\lim_{y<0,y\rightarrow
0}m_{N}(x+\mathrm{i}y)=m_{N}^{\ast}(x)$. Finally, (\ref{eq:expre-intsupp}) is
a direct consequence of the continuity of $x\rightarrow m_{N}(x)$. Property
\ref{enu:derivee_w} follows directly from Property \ref{enu:lim_m_R}.
Properties \ref{enu:image_w_R} and \ref{enu:Imw_positive} are established in
\cite{dozier2007analysis}. As for Property \ref{enu:Re_m}, it was initially
proven in \cite{dozier2007analysis} for $z\in\mathbb{C}^{\ast}$, but it can
be shown easily that it
holds for $z=0$ using Property \ref{enu:O_supp} 
as well as the proof of Lemma 2-1 of \cite{dozier2007analysis}.
Finally, \cite{dozier2007analysis} established that $m_{N}(x)$ is solution
of \eqref{eq:canonique-1} if $x\in\mathrm{int}(\mathcal{S}_{N})$. This also
holds if $x\in\mathbb{C}\backslash\mathcal{S}_{N}$ because by Properties
\ref{enu:image_w_R} and \ref{enu:Re_m}, the right hand side of
(\ref{eq:canonique-1}) is holomorphic on $\mathbb{C}\backslash\mathcal{S}_{N}%
$. Since $m_{N}(z)$ is itself holomorphic on $\mathbb{C}\backslash
\mathcal{S}_{N}$, the equality in \eqref{eq:canonique-1} must hold not only on
$\mathbb{C}\backslash\mathbb{R}_{+}$ but also on $\mathbb{C}\backslash
\mathcal{S}_{N}$. Recalling that $\mathcal{S}_{N}$ is a closed set, all this
implies that $m_{N}(x)$ is solution of equation \eqref{eq:canonique-1} for
$x\in\mathbb{R}\backslash\partial\mathcal{S}_{N}$.

Let us finally establish Property \ref{enu:wx_sol_can_eq}. Thanks to
Properties \ref{enu:Re_m} and \ref{enu:mx_sol_can_eq} and to (\ref{eq:canonique-1}), we can write%
\begin{equation}
\frac{m_{N}(x)}{1+\sigma^{2}c_{N}m_{N}(x)}=f_{N}(w_{N}%
(x))\label{eq:canonical_simplified}%
\end{equation}
for each $x\in\mathbb{R}\backslash\partial\mathcal{S}_{N}$. This last equality
can be rewritten as
\begin{equation}
1-\sigma^{2}c_{N}f_{N}(w_{N}(x))=\frac{1}{1+\sigma^{2}c_{N}m_{N}%
(x)}\label{equation:link_delta_f_1}%
\end{equation}
where the right hand side is well defined thanks to Property \ref{enu:Re_m}.
Now, plugging (\ref{equation:link_delta_f_1}) into (\ref{eq:def-w}), we obtain
that, for $x\in\mathbb{R}\backslash\partial\mathcal{S}_{N}$, $w_{N}(x)$ is a
solution of the equation
\begin{equation}
\phi_{N}(w)=x\label{eq:equation-Phi0}%
\end{equation}
where function $\phi_{N}(w)$ is defined in (\ref{eq:expre-phi}). In other
words, the function $w_{N}(x)$ satisfies (\ref{eq:equation-Phi}) for each
$x\in\mathbb{R}\backslash\partial\mathcal{S}_{N}$.
\end{IEEEproof}%

Proposition \ref{prop:m} is establishing the fact that both $m_{N}(z)$ and
$w_{N}(z)$ are well defined when $z$ approaches the real axis, and that
$m_{N}(x)$ and $w_{N}(x)$ can be determined as one of the solutions to
(\ref{eq:canonique-1}) and (\ref{eq:equation-Phi}) respectively for any
$x\in\mathbb{R}\backslash\partial\mathcal{S}_{N}$. In the next section we will
establish some properties that characterize $w_{N}(x)$ out of the set of all
the solutions of (\ref{eq:equation-Phi}), and this will in turn help us in the
characterization of the support $\mathcal{S}_{N}$.

\section{An alternative characterization of $\mathcal{S}_{N}$}

\label{section:support}

In this section we will provide a characterization of the support
$\mathcal{S}_{N}$ as a simpler alternative to the study provided in
\cite{dozier2007analysis}. It must be pointed out that
\cite{dozier2007analysis} assumed that the eigenvalue distribution of matrix
$\mathbf{B}_{N}\mathbf{B}_{N}^{H}$ converges to a limit distribution
$\nu_{\infty}(\mathrm{d}\lambda)$, and showed that $\mu_{N}$ converges towards
a probability distribution $\mu_{\infty}$. Its Stieltjès transform $m_{\infty
}$ is solution of (\ref{eq:canonique-1}), but in which the discrete measure
$\nu_{N}(\mathrm{d}\lambda)=\frac{1}{M}\sum_{j=1}^{M}\delta(\lambda-\gamma
_{k}^{(N)})$ is replaced by measure $\nu_{\infty}(\mathrm{d}\lambda)$, i.e.
\[
m_{\infty}(z)=\int\left[  -z(1+\sigma^{2}c_{N}m_{N}(z))+\sigma^{2}%
(1-c_{N})+\frac{\lambda}{1+\sigma^{2}c_{N}m_{N}(z))}\right]  ^{-1}%
\;\nu_{\infty}(\mathrm{d}\lambda).
\]
In \cite{dozier2007analysis}, a detailed analysis of the support
$\mathcal{S}_{\infty}$ of $\mu_{\infty}$ was presented. The corresponding
results provide of course a characterization of $\mathcal{S}_{N}$ by replacing
the general probability distribution $\nu_{\infty}(\mathrm{d}\lambda)$ by the
discrete measure $\nu_{N}(\mathrm{d}\lambda)=\frac{1}{M}\sum_{j=1}^{M}%
\delta(\lambda-\gamma_{k}^{(N)})$. However, we show in the following that it
is possible to reformulate the results of \cite{dozier2007analysis} in a more
explicit manner by taking into account immediately that $\frac{1}{M}\sum
_{j=1}^{M}\delta(\lambda-\gamma_{k}^{(N)})$ is a discrete measure. We hope
that the following analysis, based on quite elementary technics, is easier to
follow than the general approach of \cite{dozier2007analysis}.

Our approach is based on the study of the function $w_{N}(z)$ that has been
introduced in \eqref{eq:def-w}. We have established in Proposition
\ref{prop:m} that $w_{N}(x)$ is well defined in the real axis, and that it can
be expressed as one of the roots of the polynomial equation in
(\ref{eq:equation-Phi}). Let us now see how this function can help us in the
characterization of the support $\mathcal{S}_{N}$.

\begin{proposition}
\label{prop:w} The function $w_{N}(z)$ defined in \eqref{eq:def-w} satisfies
the following properties:

\begin{enumerate}
\item \label{item: w_on_supp} $\mathrm{Int}\left(  \mathcal{S}_{N}\right)
=\{x\in\mathbb{R}_{+}:\mathrm{Im}\{w_{N}(x)\}>0\}$

\item \label{item:w_increasing} $w_{N}^{\prime}(x)>0$, for $x\in
\mathbb{R}\backslash\mathcal{S}_{N}$.

\item \label{item:Re_1-f} $1-\sigma^{2}c_{N}f_{N}(w_{N}(x))>0\quad\forall
x\in\mathbb{R}\backslash\mathcal{S}_{N}.$
\end{enumerate}
\end{proposition}

%

\begin{IEEEproof}%
See Appendix \ref{appendix:proof_properties_w}.%
\end{IEEEproof}%

\begin{remark}
\label{remark:derivativew}By taking derivatives with respect to $x$ on both
sides of the equation $\phi_{N}(w_{N}(x))=x$, we see that $w_{N}^{\prime
}(x)\phi_{N}^{\prime}(w_{N}(x))=1$ holds for $x\in\mathbb{R}-\partial
\mathcal{S}_{N}$. Property \ref{item:w_increasing} of the above proposition is
thus equivalent to
\begin{equation}
\phi_{N}^{\prime}(w_{N}(x))>0\;\mathrm{if}\;x\in\mathbb{R}\backslash
\mathcal{S}_{N}. \label{eq:signe-phi'(w)}%
\end{equation}

\end{remark}

Property \ref{item: w_on_supp} in Proposition \ref{prop:w} is basically stating the fact that the interior
of the support $\mathcal{S}_{N}$ coincides the region of values
of\ $\mathbb{R}_{+}$ for which the imaginary part of $w_{N}(x)$ is strictly
positive. Hence, it suffices to study the behavior of $\mathrm{Im}\left[
w_{N}(x)\right]  $ in order to characterize the interior of the support
$\mathcal{S}_{N}$. On the other hand, we know from Property
\ref{enu:wx_sol_can_eq} in Proposition \ref{prop:m} that, for any
$x\in\mathbb{R}\backslash\partial\mathcal{S}_{N}$, $w_{N}(x)$ is one of the
solutions to the polynomial equation in (\ref{eq:equation-Phi}). Proposition
\ref{prop:w} is helping us to identify which one of the roots is in fact
$w_{N}(x)$. More specifically, we will later show that:

\begin{itemize}
\item If $x\in\mathrm{Int}\left(  \mathcal{S}_{N}\right)  $, then $w_{N}(x)$
will be the unique root of (\ref{eq:equation-Phi}) with positive imaginary
part\footnote{The existence and unicity of such root will be established in
what follows. }, thanks to Property \ref{item: w_on_supp}.

\item If $x\in\mathbb{R}\backslash\mathcal{S}_{N}$, then $w_{N}(x)$ will be
the unique root of (\ref{eq:equation-Phi}) such that Properties
\ref{item:w_increasing} and \ref{item:Re_1-f} hold.
\end{itemize}

In order to establish the fact that these properties completely determine the
value of $w_{N}(x)$ out of the set of roots of the equation in
(\ref{eq:equation-Phi}), we need to study the form of the function $\phi_{N}$
\ in (\ref{eq:expre-phi}) more closely. The analysis of the roots of the
corresponding equation in\ (\ref{eq:equation-Phi}) will allow us to determine
the intervals of $\mathbb{R}$ for which $w_{N}(x)$ is real-valued and the
intervals in which it has a strictly positive imaginary part.

\subsection{Characterization of the function $\phi_{N}(w)$}

\label{section:propertiesphi}In the following, we assume that the $K$ non-zero
eigenvalues of the matrix $\B_{N}\B_{N}^{H}$, namely $\left\{  \gamma
_{M-K+1}^{(N)},\ldots,\gamma_{M}^{(N)}\right\}  $, have multiplicity $1$.
Under this hypothesis, the equation in (\ref{eq:equation-Phi}) is in fact
equivalent to a polynomial equation of degree $2(K+1)$. This can be readily
seen by using the expression of $f_{N}(w)$ in (\ref{eq:expre-phi}), so that we
can express $\phi_{N}(w)$ as sums of quotients of polynomials in $w$, i.e.
\begin{align}
\phi_{N}(w) &  =w\;\left(  1+\sigma^{2}\frac{M-K}{M}\frac{c_{N}}{w}-\sigma
^{2}\frac{c_{N}}{M}\sum_{m=M-K+1}^{M}\frac{1}{\gamma_{m}-w}\right)
^{2}\nonumber\\
&  +(1-c_{N})\sigma^{2}\left(  1+\sigma^{2}\frac{M-K}{M}\frac{c_{N}}{w}%
-\sigma^{2}\frac{c_{N}}{M}\sum_{m=M-K+1}^{M}\frac{1}{\gamma_{m}-w}\right)
.\label{eq:alternative_phi}%
\end{align}
Hence, multiplying both sides of equation $\phi_{N}(w)=x$ by $w%
{\displaystyle\prod\nolimits_{m=M-K+1}^{M}}
\left(  \gamma_{m}-w\right)  ^{2}$ we end up with a polynomial equation of
degree $2(K+1)$. If certain eigenvalues of $\B_{N}\B_{N}^{H}$ are multiple,
$\phi_{N}(w)=x$ will be a polynomial equation of degree $2(\overline{K}+1)$
where $\overline{K}$ represents the number of distinct non zero eigenvalues of
$\B_{N}\B_{N}^{H}$. The following results can thus be immediately adapted by
replacing $K$ by $\overline{K}$. The assumption $K=\overline{K}$ allows to
avoid the introduction of new notations representing the distinct eigenvalues
of $\B_{N}\B_{N}^{H}$ in the forthcoming analysis.

\subsubsection{\textbf{Zeros of }$\phi_{N}(w)$}

It is easily seen that the function $\phi_{N}$ has exactly $2K+2$ different
real zeros, which will be denoted\ as $z_{0}^{(N)-}<z_{0}^{(N)+}<\ldots
<z_{K}^{(N)-}<z_{K}^{(N)+}$. An elementary analysis of the function $\phi_{N}$
determines the position of these zeros, as well as the behavior of the
function $\phi_{N}(w)$ in their neighborhood:\

\begin{itemize}
\item The lowest couple of zeros are located on the negative real axis, namely
$z_{0}^{(N)-},z_{0}^{(N)+}\in\left]  -\infty,0\right[  $. Furthermore, the
function $\phi_{N}$ is increasing at $z_{0}^{(N)-}$ and decreasing at
$z_{0}^{(N)+}$, namely $\phi_{N}^{\prime}\left(  z_{0}^{(N)-}\right)  >0$ and
$\phi_{N}^{\prime}\left(  z_{0}^{(N)+}\right)  <0$, where $\phi_{N}^{\prime}$
denotes the derivative of $\phi_{N}$.

\item The next couple of zeros are located between zero and the first positive
eigenvalue of $\B_{N}\B_{N}^{H}$, i.e. $z_{1}^{(N)-},z_{1}^{(N)+}\in\left]
0,\gamma_{M-K+1}^{(N)}\right[  $, and it turns out that the function $\phi
_{N}$ is decreasing at $z_{1}^{(N)-}$ and increasing at $z_{1}^{(N)+}$, namely
$\phi_{N}^{\prime}\left(  z_{1}^{(N)-}\right)  <0$ and $\phi_{N}^{\prime
}\left(  z_{1}^{(N)+}\right)  >0.$

\item Each one of the remaining couples of zeros is located between two
positive eigenvalues of $\B_{N}\B_{N}^{H}$ , i.e. $z_{k}^{(N)-},z_{k}%
^{(N)+}\in\left]  \gamma_{M-K+k-1}^{(N)},\gamma_{M-K+k}^{(N)}\right[  $,
$\forall\ k=2,\ldots,K$, and the function $\phi_{N}$ is always decreasing at
the first zero and increasing at the second, i.e. $\phi_{N}^{\prime}\left(
z_{k}^{(N)-}\right)  <0$ and $\phi_{N}^{\prime}\left(  z_{k}^{(N)+}\right)
>0$, $\forall\ k=2,\ldots,K$.
\end{itemize}

In order to obtain these results, one only needs to factor $\phi_{N}(w)$ as
the product of two terms, namely
\begin{equation}
\phi_{N}(w)=\left[  1-c_{N}\sigma^{2}f_{N}(w)\right]  \left[  w\;(1-c_{N}%
\sigma^{2}f_{N}(w))+(1-c_{N})\sigma^{2}\right]  \label{eq:phi_2terms}%
\end{equation}
and therefore $\phi_{N}(w)=0$ if and only if one of these two terms is zero.
Out of the $2K+2$ zeros of the function $\phi_{N}(w)$, a total of $K+1$ are
the zeros of the first term in (\ref{eq:phi_2terms}). More formally:

\begin{itemize}
\item The second zero, namely $z_{0}^{(N)+}$, is solution of the equation
$1-\sigma^{2}c_{N}f_{N}(w)=0.$

\item The zeros $z_{k}^{(N)-}$ for $k=1,\ldots,K$ are the solutions of the
equation $1-\sigma^{2}c_{N}f_{N}(w)=0.$
\end{itemize}

This allows us to differentiate between intervals of the real axis where
$1-\sigma^{2}c_{N}f_{N}(w)>0$ and intervals where $1-\sigma^{2}c_{N}%
f_{N}(w)\leq0$, namely

\begin{itemize}
\item The function $1-\sigma^{2}c_{N}f_{N}(w)$ is positive on the intervals
\begin{equation}
\left]  -\infty,z_{0}^{+}\right[  \text{,\quad}\left\{  \left]  \gamma
_{M-K+k-1}^{(N)},z_{k}^{(N)-}\right[  \right\}  _{k=1,\ldots,K}\text{,\quad
}\left]  \gamma_{M}^{(N)},+\infty\right[  .\label{eq:intervals_zi}%
\end{equation}

\end{itemize}

\begin{figure}[h]
\centering
\par
\includegraphics[width=15cm]{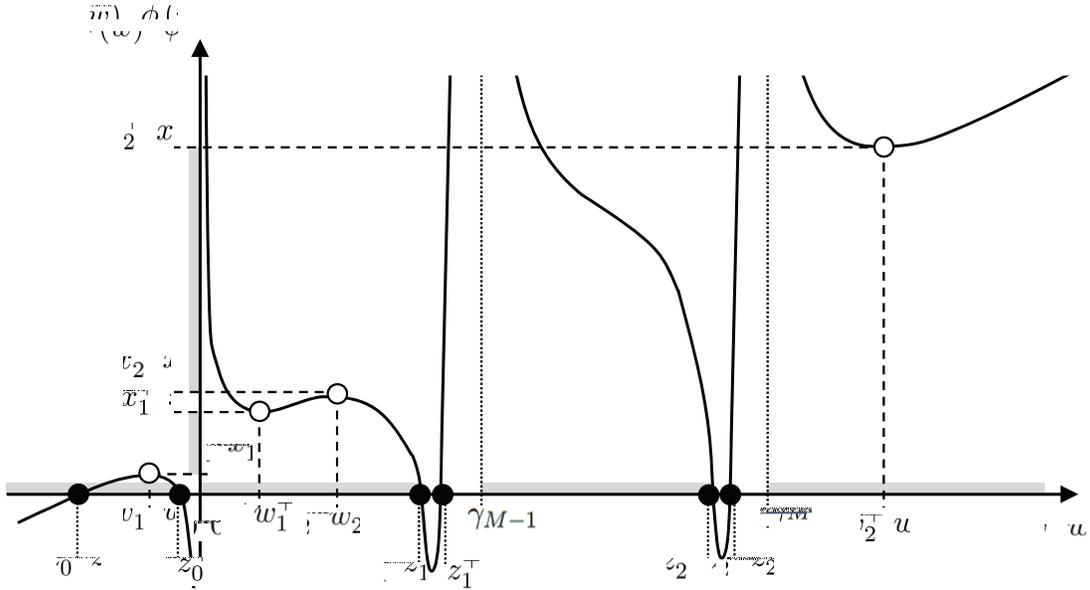}\caption{Typical
representation of $\phi_{N}\left(  w\right)  $ as a function of $w$ for $K=2$
and $Q=2$ (we drop the dependence on $N$ for clarity). The shaded region in
the horizontal axis represents the set of points for which $1-\sigma^{2}%
c_{N}f_{N}(w)>0$. The shaded region in the vertical axis represents
$\mathcal{S}_{N}$.}%
\label{figure:support}%
\end{figure}

This last fact is important, because we know from Property \ref{item:Re_1-f}
of Proposition \ref{prop:w} that, when $x$ does not belong to the support
$\mathcal{S}_{N}$, the solution of the equation $\phi_{N}(w)=x$ corresponding
to $w_{N}(x)$ will be such that $1-\sigma^{2}c_{N}f_{N}(w_{N}(x))>0$, and
therefore will be located inside of one of these intervals. In Figure
\ref{figure:support} we give a typical representation of function $\phi
_{N}(w)$ in a situation where $K=2$ (we drop the dependence on $N$ in all
quantities in the figure to simplify the representation). The function
$\phi_{N}(w)$ presents horizontal asymptotes at $w=0$ and also at the values
of the positive eigenvalues of $\B_{N}\B_{N}^{H}$, namely $\left\{
\gamma_{M-K+1}^{(N)},\ldots,\gamma_{M}^{(N)}\right\}  $. The region of the
horizontal axis where $1-\sigma^{2}c_{N}f_{N}(w)>0$ is shaded in grey.

\subsubsection{\textbf{Local extrema and monotonicity intervals of }$\phi
_{N}(w)$}

Next, we investigate the local extrema of the function $\phi_{N}$. The
following proposition summarizes the most interesting properties of the
positive local extrema.

\begin{proposition}
\label{property:phi_extrema}

\begin{enumerate}
\item \label{item:nb_positive_extrema} The function $\phi_{N}$ admits $2Q$
positive local extrema counting multiplicities (with $1\leq Q\leq K+1$) whose
preimages, denoted $w_{1}^{(N)-}<0<w_{1}^{(N)+}\leq w_{2}^{(N)-}\ldots\leq
w_{Q}^{(N)-}<w_{Q}^{(N)+},$ belong to the set $\{w\in\mathbb{R}:1-\sigma
^{2}c_{N}f_{N}(w)>0\}$

\item If we denote by $x_{k}^{(N)-}=\phi_{N}\left(  w_{k}^{(N)-}\right)  $ and
$x_{k}^{(N)+}=\phi_{N}\left(  w_{k}^{(N)+}\right)  $ these positive extrema,
then
\begin{equation}
0<x_{1}^{(N)-}<x_{1}^{(N)+}\leq x_{2}^{(N)-}\ldots\leq x_{Q}^{(N)-}%
<x_{Q}^{(N)+} \label{eq:extrema-croissants}%
\end{equation}

\item \label{item:location_extrema} Each eigenvalue $\gamma_{l}^{(N)}$ of
$\B_{N}\B_{N}^{H}$ belongs to one and only one of the intervals $\left]
w_{q}^{(N)-},w_{q}^{(N)+}\right[  $, $q=1\ldots Q$.

\item \label{item:croissance-phi} The function $\phi_{N}$ is increasing on the
intervals $\left]  -\infty,w_{1}^{(N)-}\right]  $, $\left\{  \left[
w_{q}^{(N)+},w_{q+1}^{(N)-}\right]  \right\}  _{q=1,Q-1}$, and $\left[
w_{Q}^{(N)+},+\infty\right]  $. Moreover,
\begin{gather*}
\phi_{N}\left(  \left]  -\infty,w_{1}^{(N)-}\right]  \right)  =\left]
-\infty,x_{1}^{(N)-}\right] \\
\phi_{N}\left(  \left[  w_{q}^{(N)+},w_{q+1}^{(N)-}\right]  \right)  =\left[
x_{q}^{(N)+},x_{q+1}^{(N)-}\right]  \text{ for each }q=1,\ldots,Q-1\text{,
and}\\
\phi_{N}\left(  \left[  w_{Q}^{(N)+},+\infty\right[  \right)  =\left[
x_{Q}^{(N)+},+\infty\right[  .
\end{gather*}

\end{enumerate}
\end{proposition}%

\begin{IEEEproof}%
Except for the inequalities in (\ref{eq:extrema-croissants}), which are proved
in Appendix \ref{appendix:proof_inequalities_x}, the statements of Proposition
\ref{property:phi_extrema} follow directly from an elementary analysis of the
function $\phi_{N}$.
\end{IEEEproof}%

\begin{figure}[h]
\centering
\par
\includegraphics[width=15cm]{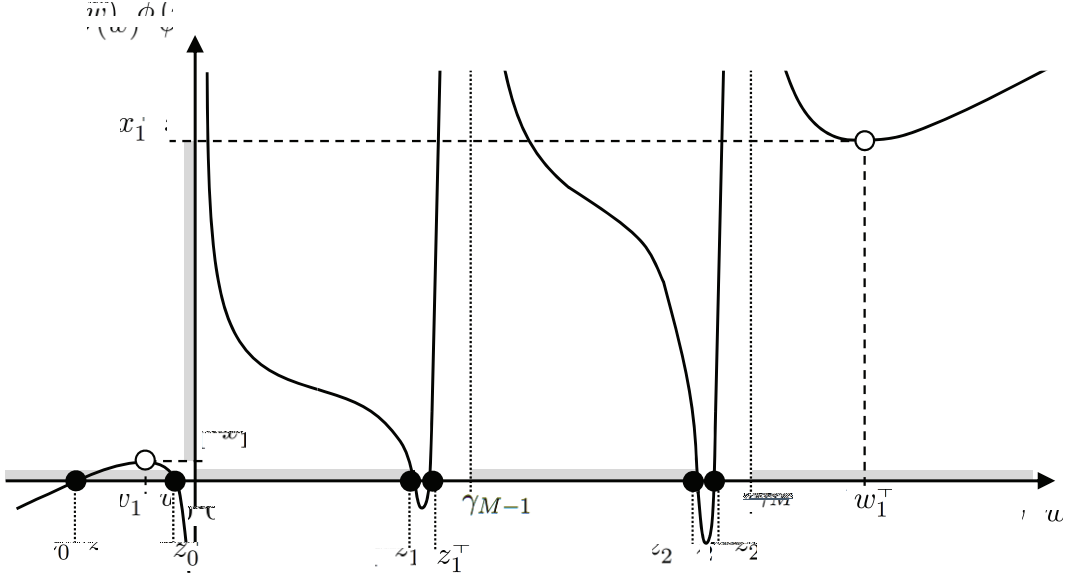}\caption{Typical
representation of $\phi_{N}\left(  w\right)  $ as a function of $w$ for $K=2$
and $Q=1$ (we drop the dependence on $N$ for clarity). The shaded region in
the horizontal axis represents the set of points for which $1-\sigma^{2}%
c_{N}f_{N}(w)>0$. The shaded region in the vertical axis represents
$\mathcal{S}_{N}$.}%
\label{figure:supporthighc}%
\end{figure}

\begin{figure}[h]
\centering
\par
\includegraphics[width=15cm]{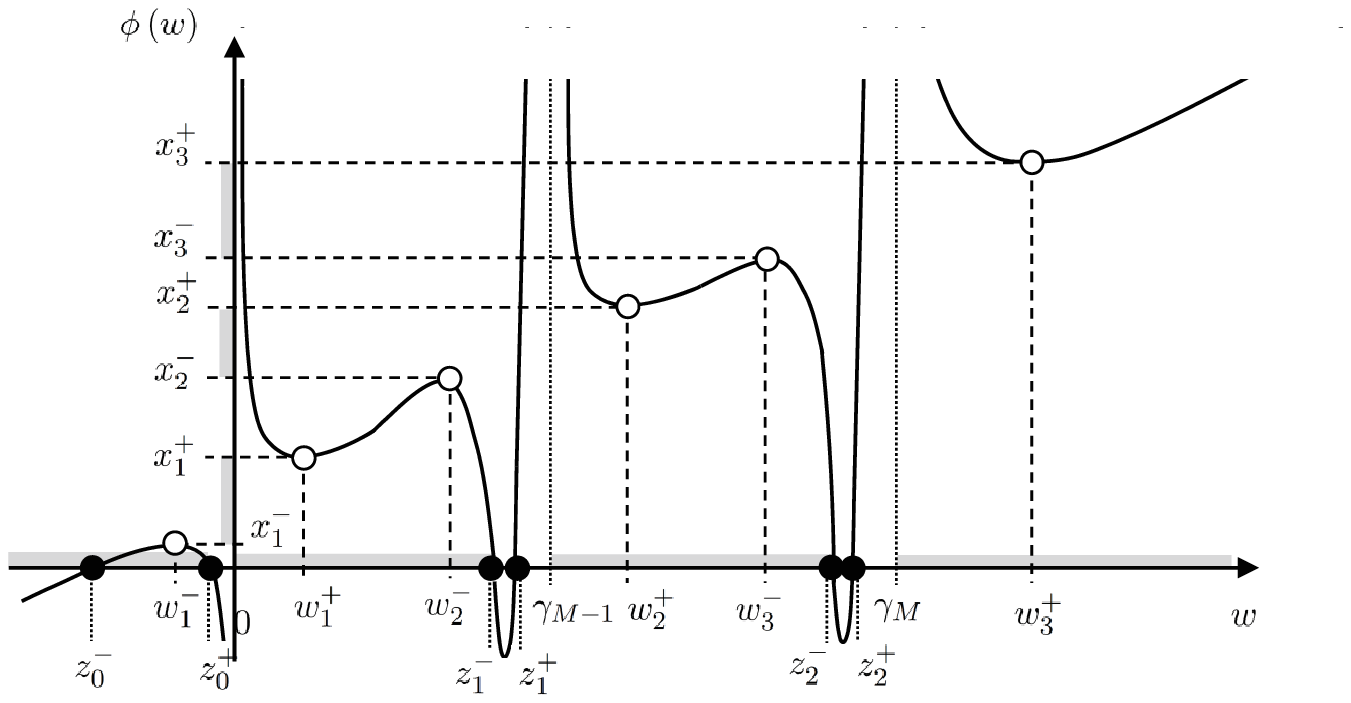}\caption{Typical
representation of $\phi_{N}\left(  w\right)  $ as a function of $w$ for $K=2$
and $Q=3$ (we drop the dependence on $N$ for clarity). The shaded region in
the horizontal axis represents the set of points for which $1-\sigma^{2}%
c_{N}f_{N}(w)>0$. The shaded region in the vertical axis represents
$\mathcal{S}_{N}$.}%
\label{figure:supportlowc}%
\end{figure}

We see from Proposition \ref{property:phi_extrema} that the local extrema
always appear in groups of two, and the actual number of extremum couples
($Q$) will generally depend on $\sigma^{2}$, $c_{N}$ and on the positive
eigenvalues of the matrix $\B_{N}\B_{N}^{H}$. For example, in the situation
represented in Figure \ref{figure:support}, the number of positive local
extrema was equal to four, which implies that $Q=2$. In Figures
\ref{figure:supporthighc} and \ref{figure:supportlowc} we depict other
equivalent examples of $\phi_{N}$, for which we had $Q=1$ and $Q=3$
respectively. 

\subsection{Characterization of $w_{N}(x)$ out of the roots of $\phi_{N}%
(w)=x$}

\label{section:discussion_w_select}We know from Proposition \ref{prop:m} that
$w_{N}(x)$ for real valued $x$ will be a solution of the equation $\phi
_{N}(w)=x$. In this section, we will characterize which one of these roots is
actually $w_{N}(x)$. First of all, observe that, since the equation $\phi
_{N}(w)=x\,\ $is equivalent to a polynomial equation of degree $2(K+1)$, the
number of solutions (counting multiplicities) will always be equal to
$2(K+1)$. Out of these solutions, we can graphically find the real-valued ones
by exploring the crossings between the graph of $\phi_{N}(w)$ and a horizontal
line at $x$. This is further illustrated in Figure \ref{fig:findingroots}. By
the properties of the function $\phi_{N}(w)\,$presented in Section
\ref{section:propertiesphi}, we can clearly differentiate between two
different situations:

\begin{itemize}
\item If$\ x\notin\bigcup_{k=1}^{Q}\left[  x_{k}^{(N)-},x_{k}^{(N)+}\right]
$, it is easily shown that the equation $\phi_{N}(w)=x$ presents exactly $2(K+1)$ different
real-valued solutions (cf. upper horizontal line in Figure
\ref{fig:findingroots}). Since the original equation has degree $2(K+1)$,
there are no complex-valued solutions. In particular, $w_{N}(x)$ will be real-valued.

\item If $x\in\bigcup_{k=1}^{Q}\left]  x_{k}^{(N)-},x_{k}^{(N)+}\right[  $,
in what follows, it will be  shown  that the equation $\phi_{N}(w)=x$ has exactly $2K$ different real-valued solutions
(cf. lower horizontal line in Figure \ref{fig:findingroots}). 
This implies
that there is a couple of complex conjugated solutions to the equation
$\phi_{N}(w)=x$. \begin{figure}[h]
\centering
\par
\includegraphics[width=15cm]{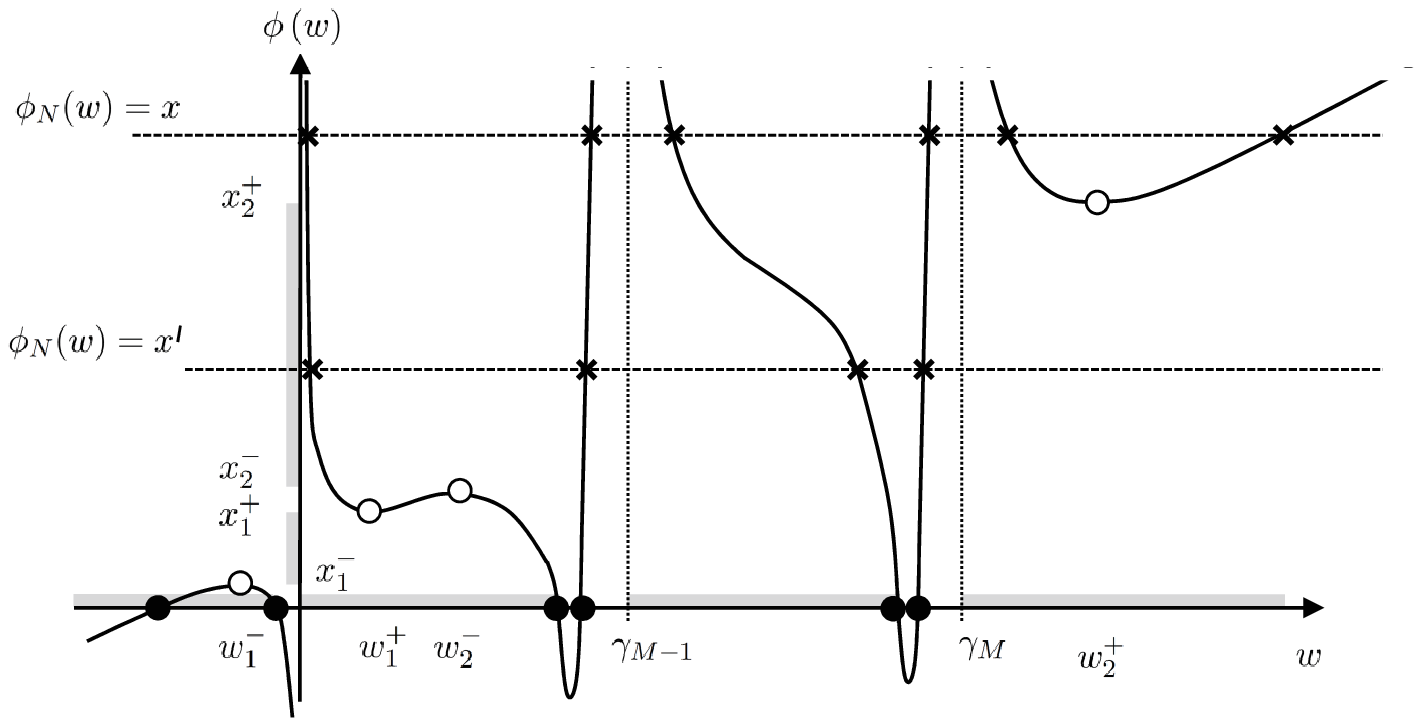}\caption{One can find
the real-valued solutions to $\phi_{N}\left(  w\right)  =x$ by examining the
crossings of the graph $\phi_{N}\left(  w\right)  $ with a horizontal line at
$x$. In this particular example, where $K=2$, we see that $\phi_{N}\left(
w\right)  =x$ presents $2(K+1)=6$ real-valued solutions, whereas $\phi
_{N}\left(  w\right)  =x^{\prime}$ has only $4$ real-valued solutions (plus a
couple of complex conjugated ones). }%
\label{fig:findingroots}%
\end{figure}
\end{itemize}

Let us now see how we can completely characterize $w_{N}(x)$ in these two
different situations:

\subsubsection{\textbf{Case }$x\in\mathbb{R}\backslash\bigcup_{k=1}^{Q}\left[
x_{k}^{(N)-},x_{k}^{(N)+}\right]  $}

From (\ref{eq:signe-phi'(w)}) and Property \ref{item:Re_1-f} of Proposition
\ref{prop:w}, we know that $w_{N}(x)$ is a root of the equation $\phi
_{N}(w)=x$ such that $\phi_{N}^{\prime}(w_{N}(x))>0$ and that $1-\sigma
^{2}c_{N}f_{N}(w_{N}(x))>0$. We now prove that this completely characterizes
$w_{N}(x)$ out of the set of all roots of $\phi_{N}(w)=x$, in the sense that
there is only one root of $\phi_{N}(w)=x$ that has these two properties. We
first consider the case $x<x_{1}^{(N)-}$. By Property
\ref{item:croissance-phi} of Proposition \ref{property:phi_extrema}, $\phi
_{N}$ is an increasing one to one correspondence from $\left]  -\infty
,w_{1}^{(N)-}\right[  $ onto $\left]  -\infty,x_{1}^{(N)-}\right[  $. Its
inverse $\phi_{N}^{-1}$ is thus a well defined increasing function from
$\left]  -\infty,x_{1}^{(N)-}\right[  $ onto $\left]  -\infty,w_{1}%
^{(N)-}\right[  $. We claim that $w_{N}(x)$ coincides with $\phi_{N}^{-1}(x)$.
Indeed, observe that since $\phi_{N}^{-1}(x)< w_{1}^{(N)-}$, we automatically have
$\phi_{N}^{\prime}(\phi_{N}^{-1}(x))>0$ and that $1-\sigma^{2}c_{N}f_{N}%
(\phi_{N}^{-1}(x))>0$. On the other hand, the behavior of $\phi_{N}$
established in Propositions \ref{prop:w} and \ref{property:phi_extrema}
implies that the other real-valued solutions of $\phi_{N}(w)=x$ do not satisfy
either $1-\sigma^{2}c_{N}f_{N}(w)>0$ or $\phi_{N}^{\prime}(w)>0$ (see further
Figures \ref{figure:support} to \ref{figure:supportlowc}). Therefore,
$w_{N}(x)$ can be expressed as $\phi_{N}^{-1}(x)$, and is the only root of
$\phi_{N}(w)=x$ such that $1-\sigma^{2}c_{N}f_{N}(w)>0$ and $\phi
_{N}^{^{\prime}}(w)>0$.

The above analysis can be extended if $x$ belongs to $\left]  x_{k}%
^{(N)+},x_{k+1}^{(N)-}\right[  $ for $k=1,\ldots,Q-1$ or if $x>x_{Q}^{(N)+}$.
Indeed, observe first that $\phi_{N}$ is a bijection between $\left]
w_{k}^{(N)+},w_{k+1}^{(N)-}\right[  $ and $\left]  x_{k}^{(N)+},x_{k+1}%
^{(N)-}\right[  $ for $k=1,\ldots,Q-1$ and between $\left]  w_{Q}%
^{(N)+},+\infty\right[  $ and $\left]  x_{Q}^{(N)+},+\infty\right[  $. Hence,
$\phi_{N}^{-1}$ is well defined on $\left]  x_{k}^{(N)+},x_{k+1}%
^{(N)-}\right[  $ for $k=1,\ldots,Q-1$ and on $\left]  x_{Q}^{(N)+}%
,+\infty\right[  $. Thanks to the form of the function $\phi_{N}$, we see that
$\phi_{N}^{-1}(x)$ is the only root that verifies $1-\sigma^{2}c_{N}%
f_{N}(w)>0$ and $\phi_{N}^{\prime}(w)>0$ (see further Figures
\ref{figure:support} to \ref{figure:supportlowc}), and this implies that
$w_{N}(x)=\phi_{N}^{-1}(x)$. Since $w_{N}\left(  x\right)  $ is continuous on
$\mathbb{R}$, we also get that $w_{N}(x_{k}^{(N)-})=w_{k}^{(N)-}$ as well as
$w_{N}(x_{k}^{(N)+})=w_{k}^{(N)+}$ for $k=1,\ldots,Q$.

\subsubsection{\textbf{Case} $x\in\bigcup_{k=1}^{Q}\left]  x_{k}^{(N)-}%
,x_{k}^{(N)+}\right[  $}

In this situation, we establish that the equation  $\phi_{N}(w)=x$ has exactly $2K$ real-valued
solutions, plus a couple of complex conjugated ones, and that 
$w_{N}(x)$ is equal to the complex-value root with strictly positive imaginary
part. We can reason from the behavior of $\phi_{N}$ that the polynomial
equation $\phi_{N}(w)=x$ has at least $2K$ real-valued solutions located in
the intervals $\left]  \gamma_{M-K+l-1}^{(N)},z_{l}^{-}\right[  $ and $\left]
z_{l}^{+},\gamma_{M-K+l}^{(N)}\right[  $ for $l=1,\ldots,K$. We however note
that none of them can satisfy both $\phi_{N}^{^{\prime}}(w)\geq0$ and
$1-\sigma^{2}c_{N}f_{N}(w)>0$. Therefore, $w_{N}(x)$ cannot coincide with one
of these solutions. Assume that the two remaining solutions of the equation are real. 
$w_N(x)$ of course coincides with one of these two solutions. The properties of function
$\phi_{N}$ as well as (\ref{eq:extrema-croissants}) imply the existence of two
extrema of $\phi_{N}$ , denoted by $x_{\ast}<x_{\ast}^{\prime}$ such that
$x\in\left]  x_{\ast},x_{\ast}^{\prime}\right[  $. Moreover, by (\ref{eq:extrema-croissants}), the two extra
solutions must belong to an interval $\left]  z_{l}^{+},\gamma_{M-K+l}%
^{(N)}\right[  $ for $l=1,\ldots,Q-1$. Consequently, these two solutions
satisfy $1-\sigma^{2}c_{N}f_{N}(w)<0$, and cannot coincide with $w_{N}(x)$,
which leads us to contradiction. Therefore, the two remaing solutions are 
complex conjugate, and $w_N(x)$ coincides the solution with strictly imaginary
part.

\subsection{Characterization of the support $\mathcal{S}_{N}$ }
As the interior of ${\cal S}_N$ coincides with $\{ x \in \mathbb{R}_{+}, \mathrm{Im}(w_N(x)) > 0 \}$
(see Property \ref{item: w_on_supp} of Proposition \ref{prop:w}),
we have shown the following Theorem. 
\begin{theorem}
\label{theorem:support} The support $\mathcal{S}_{N}$ is given by
\begin{equation}
\mathcal{S}_{N}=\bigcup_{k=1}^{Q}\left[  x_{k}^{(N)-},x_{k}^{(N)+}\right]  .
\label{eq:support_theorem}%
\end{equation}

\end{theorem}%

The above analysis shows that $x_1^{(N)-} < x_1^{(N)+} \leq x_2^{(N)-} < \ldots < x_{Q-1}^{(N)+}\leq x_{Q}^{(N)-} < x_Q^{(N)+}$ 
coincide with the set of all positive extrema of $\phi_N$. Theorem \ref{theorem:support} thus establishes a very simple method to determine
the support $\mathcal{S}_{N}$. First, one needs to determine all the local extrema
of $\phi_{N}\left(  w\right) $, namely the solutions to the polynomial
equation $\phi_{N}^{\prime}\left(  w\right)  =0$. The solutions will
be $\left\{  w_{1}^{(N)-},w_{1}^{(N)+},\ldots,w_{Q}^{(N)-},w_{Q}^{(N)+}%
\right\}  $ with possible repetitions if one of these roots has multiplicity
two,\ plus $K$ additional ones (it is easily seen that $\phi_N$ has exactly $K$ negative local minima). By evaluating the function $\phi_{N}$ at these
points, and selecting those for which $\phi_{N}$ is positive, we are
determining the values $\left\{  x_{1}^{(N)-},x_{1}^{(N)+},\ldots,x_{Q}%
^{(N)-},x_{Q}^{(N)+}\right\}  $ that characterize the support in
(\ref{eq:support_theorem}). Observe that the support $\mathcal{S}_{N}$ is a
disjoint reunion of compact intervals, which will be referred to as clusters.
Each of these clusters $\left[  x_{q}^{(N)-},x_{q}^{(N)+}\right]  $ will be
associated to an interval of the type $\left[  w_{q}^{(N)-},w_{q}%
^{(N)+}\right]  $, $q=1\ldots Q$, in the sense that $x_{q}^{(N)-}=\phi
_{N}\left(  w_{q}^{(N)-}\right)  $ and $x_{q}^{(N)+}=\phi_{N}\left(
w_{q}^{(N)+}\right)  $. On the other hand, we can also clearly see that a
specific eigenvalue$\ \gamma_{k}^{(N)},$ $k=1,\ldots,M$, always belongs to
one, and only one of the intervals $\left[  w_{q}^{(N)-},w_{q}^{(N)+}\right]
$. This motivates the following definition.

\begin{definition}
We say that the eigenvalue $\gamma_{k}^{(N)}$, $k=1,\ldots,M$, of the matrix
$\mathbf{B}_{N}\mathbf{B}_{N}^{H}$ is associated with the cluster $\left[
x_{q}^{(N)-},x_{q}^{(N)+}\right]  $ if $\gamma_{k}^{(N)}\in\left[
w_{q}^{(N)-},w_{q}^{(N)+}\right]  $.
\end{definition}

Observe that this is not a one-to-one correspondence, in the sense that
multiple consecutive eigenvalues of $\mathbf{B}_{N}\mathbf{B}_{N}^{H}$ may be
associated with the same cluster. For instance, in Figure
\ref{figure:supporthighc} the three eigenvalues ($0$, $\gamma_{M-1}^{(N)}$ and
$\gamma_{M}^{(N)}$) are associated with the same eigenvalue cluster, while in
Figure \ref{figure:supportlowc} each eigenvalue is associated with its own
different cluster.

The first cluster $[x_{1}^{(N)-},x_{1}^{(N)+}]$ plays a special role because
it is always associated with the eigenvalue $0$ of matrix $\mathbf{B}%
_{N}\mathbf{B}_{N}^{H}$. As seen below, the main results of this paper will be
valid under the assumption that the strictly positive eigenvalues of
$\mathbf{B}_{N}\mathbf{B}_{N}^{H}$ are not associated to the cluster
$[x_{1}^{(N)-},x_{1}^{(N)+}]$. Intuitively, this means that the eigenvalues
corresponding to the noise subspace are separated from the eigenvalues of the
signal subspace. Both Figure \ref{figure:support} and Figure
\ref{figure:supportlowc} satisfy this property, but not Figure
\ref{figure:supporthighc}.

More rigorously, we assume from now on that the following hypotheses hold.

\begin{assumption}
\label{assumption:exact_sep_1} $\exists N_{0}\in\mathbb{N}$ such that $\forall
N\in\mathbb{N},N\geq N_{0}$, the non zero eigenvalues $\left\{  \gamma
_{k}^{(N)}\right\}  _{k=M-K+1,\ldots,M}$ of $\mathbf{B}_{N}\mathbf{B}_{N}^{H}$
are not associated to the first cluster $[x_{1}^{(N)-},x_{1}^{(N)+}]$.
\end{assumption}

\begin{assumption}
\label{assumption:exact_sep_2} $\exists t_{1}^{-}>0,t_{1}^{+},t_{2}^{-}%
\in\mathbb{R}$ independent of $N$ such that
\begin{equation}
t_{1}^{-}<\inf_{N\geq N_{0}}\left\{  x_{1}^{(N)-}\right\}  <\sup_{N\geq N_{0}%
}\left\{  x_{1}^{(N)+}\right\}  <t_{1}^{+}<t_{2}^{-}<\inf_{N\geq N_{0}%
}\left\{  x_{2}^{(N)-}\right\}  \quad\forall N\geq N_{0}.
\label{eq:separation}%
\end{equation}

\end{assumption}

These two assumptions imply that for each $N\geq N_{0}$, the eigenvalue $0$ of
$\mathbf{B}_{N}\mathbf{B}_{N}^{H}$ belongs to the interval $\left]
w_{1}^{(N)-},w_{1}^{(N)+}\right[  $ and thus to $\left]  w_{N}(t_{1}%
^{-}),w_{N}(t_{1}^{+})\right[  $ because $w_{N}(t_{1}^{-})<w_{1}^{(N)-}$ and
$w_{N}(t_{1}^{+})>w_{1}^{(N)+}$. Similarly, the non zero eigenvalues $\left\{
\gamma_{M-K+l}^{(N)}\right\}  _{l=1,\ldots,K}$ of $\mathbf{B}_{N}%
\mathbf{B}_{N}^{H}$ satisfy $\gamma_{M-K+l}^{(N)}>w_{N}(t_{2}^{+})$. \newline

\section{Convergence and localization of the sample eigenvalues}

\label{section:exactseparation}The previous results are related to the
properties of the limit deterministic distribution $\mu_{N}$. The almost sure 
convergence of $\hat{\mu}_{N}-\mu_{N}$ towards 0 does not mean by itself that
the eigenvalues of $\hat{\mathbf{R}}_{N}$ belong almost surely to
$\mathcal{S}_{N}$, or to an interval containing $\mathcal{S}_{N}$. As one may
imagine, it is important to be able to locate the eigenvalues $(\hat{{\lambda
}}_{k}^{(N)})_{k=1,\ldots,M}$ of matrix $\hat{\mathbf{R}}_{N}$ with respect to
$\mathcal{S}_{N}$ for $N$ large enough. Bai and Silverstein established in
\cite{bai1998no}, \cite{bai1999exact} powerful related results in the context
of correlated zero-mean, possibly non Gaussian, random matrices. In the
following, we establish similar results for the Information plus Noise model.
However, the mathematical approach we use in the present paper has no
connection with the techniques used in \cite{bai1998no}, \cite{bai1999exact}
also valid in the non Gaussian case. Since ${\boldsymbol{\Sigma}}$ is assumed
Gaussian, we rather adapt to the Information plus Noise model the ideas
developed in \cite{capitaine2009largest} in the context of Gaussian Wigner
matrices. We prove in the following two theorems which are believed to be of
independent interest.

\begin{theorem}
\label{theo:no-eigenvalue}Assume that there exists a positive quantity
$\epsilon>0$, two real values $a,b\in\mathbb{R}$, and an integer $N_{0}$ such
that
\begin{equation}
\left]  a-\epsilon,b+\epsilon\right[  \cap\mathcal{S}_{N}=\varnothing
\qquad\forall N\in\mathbb{N},N\geq N_{0} \label{eq:abinterS}%
\end{equation}
where $\mathcal{S}_{N}$ denotes the support of $\mu_{N}$. Then, with
probability one, no eigenvalue of $\hat{\mathbf{R}}_{N}$ appears in $[a,b]$
for all $N$ large enough.
\end{theorem}

\begin{theorem}
\label{theo:exact-separation} If Assumptions \ref{assumption:exact_sep_1} and
\ref{assumption:exact_sep_2} hold, then, for all $N$ large enough, with
probability one,
\begin{align}
&  \hat{\lambda}_{1}^{(N)},\ldots,\hat{\lambda}_{M-K}^{(N)}\in\left]
t_{1}^{-},t_{1}^{+}\right[ \\
&  \hat{\lambda}_{M-K+1}^{(N)}>t_{2}^{-}%
\end{align}

\end{theorem}

Although Assumptions \ref{assumption:exact_sep_1} and
\ref{assumption:exact_sep_2} depend on the deterministic distributions
$\mu_{N}$, Theorem \ref{theo:exact-separation} shows that almost surely, the
smallest $M-K$ eigenvalues of $\hat{\mathbf{R}}_{N}$ are always separated from
the others for all $N$ large enough.

\subsection{Proof of Theorem \ref{theo:no-eigenvalue}}

We first state the following proposition, the proof of which is demanding, and is detailed in Appendix
\ref{appendix:proof_polynomial_bounding}. The result will play a fundamental
role in the proof of Theorem \ref{theo:no-eigenvalue}.

\begin{proposition}
\label{proposition:decomposition} $\forall z\in\mathbb{C}\backslash
\mathbb{R}_{+}$, we have for $N$ large enough,
\[
\mathbb{E}\left[  \frac{1}{M}\mathrm{Tr}\left[  \Q_{N}(z)\right]  \right]
=\frac{1}{M}\mathrm{Tr}\left[  \T_{N}(z)\right]  +\frac{1}{N^{2}}\chi_{N}(z)
\]
with $\chi$ is analytic in $\mathbb{C}-\mathbb{R}_{+}$ and satisfies
\begin{equation}
\left\vert \chi_{N}(z)\right\vert \leq(|z|+C)^{k}\mathrm{P}(|\mathrm{Im}%
(z)|^{-1}) \label{eq:inegalite-chi}%
\end{equation}
for each $z \in \mathbb{C}_{+}$ where $C$ is a constant, $k$ is an integer independent of $N$ and $\mathrm{P}$
is a polynomial with positive coefficients independent of $N$.
\end{proposition}

We now follow \cite{haagerup2005new} and \cite{capitaine2009largest} and prove 
the Lemma: 

\begin{lemma}
\label{le:phi(Q-T)} Let $\phi$ be a compactly supported real-valued smooth function
defined on $\mathbb{R}$, i.e. $\phi\in\mathcal{C}_{c}^{\infty}(\mathbb{R}%
,\mathbb{R})$. Then\footnote{By applying the function $\phi$ to a Hermitian
matrix, we implicitly represent the action of $\phi$ on the corresponding
eigenvalues.},
\begin{equation}
\mathbb{E}\left[  \frac{1}{M}\mathrm{Tr}\left[  \phi\left(
{\boldsymbol{\Sigma}_{N}}{\boldsymbol{\Sigma}_{N}}^{H}\right)  \right]
\right]  -\int_{\mathcal{S}_{N}}\phi(\lambda) \mathrm{d} \mu_{N}(\lambda
)=\mathcal{O}(\frac{1}{N^{2}}) \label{eq:phi(Q-T)}%
\end{equation}

\end{lemma}%

\begin{IEEEproof}%
We first note that, by Property \ref{enu:inversion_phi} in Lemma
\ref{property:stieltjes}, we can write
\[
\mathbb{E}\left[  \frac{1}{M}\mathrm{Tr}\left[  \phi\left(
\boldsymbol{\boldsymbol{\Sigma}}_{N}\boldsymbol{\boldsymbol{\Sigma}}_{N}%
^{H}\right)  \right]  \right]  =\frac{1}{\pi}\lim_{y\downarrow0}%
\ \mathrm{Im}\left\{  \int_{\mathbb{R}_{+}}\phi(x)\mathbb{E}\left[  \frac
{1}{M}\mathrm{Tr}\left[  \Q_{N}(x+\mathrm{i}y)\right]  \right]  \mathrm{d}%
x\right\}
\]
as well as
\[
\left[  \int_{\mathcal{S}_{N}}\phi(\lambda) \mathrm{d}\mu_{N}(
\lambda)\right]  =\frac{1}{\pi}\lim_{y\downarrow0}\ \mathrm{Im}\left\{
\int_{\mathbb{R}_{+}}\phi(x)\left[  \frac{1}{M}\mathrm{Tr}\left[
\T_{N}(x+\mathrm{i}y)\right]  \right]  \mathrm{d}x\right\}
\]
Therefore, using Proposition \ref{proposition:decomposition}, we can express
the right hand side of (\ref{eq:phi(Q-T)}) as
\begin{equation}
\mathbb{E}\left[  \frac{1}{M}\mathrm{Tr}\left[  \phi\left(
{\boldsymbol{\Sigma}_{N}}{\boldsymbol{\Sigma}_{N}}^{H}\right)  \right]
\right]  -\int_{\mathcal{S}_{N}}\phi(\lambda)\mu_{N}(\mathrm{d}\lambda
)=\frac{1}{N^{2}}\frac{1}{\pi}\lim_{y\downarrow0}\ \mathrm{Im}\left\{
\int_{\mathbb{R}_{+}}\phi(x)\chi_{N}(x+\mathrm{i}y) \, dx\right\}
\label{eq:egalite-utile}%
\end{equation}
Since the function $\chi_{N}(z)$ satisfies the inequality
\eqref{eq:inegalite-chi}, the Appendix of \cite{capitaine2007freeness} implies
that
\[
\limsup_{y\downarrow0}\left\vert \int_{\mathbb{R}}\varphi(x)\chi
_{N}(x+\mathrm{i}y)\mathrm{d}x\right\vert \leq C<+\infty
\]
where $C$ is a constant independent of $N$. Hence, \eqref{eq:egalite-utile}
readily implies \eqref{eq:phi(Q-T)}.
\end{IEEEproof}%

In order to establish Theorem \ref{theo:no-eigenvalue}, we consider a function
$\psi\in\mathcal{C}_{c}^{\infty}(\mathbb{R},\mathbb{R})$ satisfying $0\leq
\psi\leq1$ and
\[
\psi(\lambda)=%
\begin{cases}
1\quad & \mathrm{for}\quad\lambda\in\lbrack a,b]\\
0\quad & \mathrm{for}\quad\lambda\in\mathbb{R}-\left]  a-\epsilon
,b+\epsilon\right[
\end{cases}
\]
Condition (\ref{eq:abinterS}) implies that $\int_{\mathcal{S}_{N}}\psi
(\lambda) \mathrm{d} \mu_{N}(\lambda)=0$ if $N$ is large enough. Therefore,
(\ref{eq:phi(Q-T)}) implies that
\[
\mathbb{E}\left[  \frac{1}{M}\mathrm{Tr}\left[  \psi\left(
\boldsymbol{\boldsymbol{\Sigma}}_{N}\boldsymbol{\boldsymbol{\Sigma}}_{N}%
^{H}\right)  \right]  \right]  =\mathcal{O}\left(  \frac{1}{N^{2}}\right)  .
\]
We now establish that
\begin{equation}
\mathrm{Var}\left[  \frac{1}{M}\mathrm{Tr}\left[  \psi\left(
\boldsymbol{\boldsymbol{\Sigma}}_{N}\boldsymbol{\boldsymbol{\Sigma}}_{N}%
^{H}\right)  \right]  \right]  =\mathcal{O}\left(  \frac{1}{N^{4}}\right)
\label{eq:Varpsi}%
\end{equation}
In order to prove (\ref{eq:Varpsi}), we use the Nash-Poincaré inequality
\cite{hachem08, pastur05, Chatterjee04, capitaine2009largest} which implies that
\begin{equation}
\mathrm{Var}\left[  \frac{1}{M}\mathrm{Tr}\left[  \psi\left(
\boldsymbol{\Sigma}_{N}\boldsymbol{\Sigma}_{N}^{H}\right)  \right]  \right]
\leq\frac{\sigma^{2}}{N}\sum_{i,j}\mathbb{E}\left[  \left\vert \frac{\partial
}{\partial\mathbf{W}_{ij}}\left[  \frac{1}{M}\mathrm{Tr}\left[  \psi\left(
\boldsymbol{\Sigma}_{N}\boldsymbol{\Sigma}_{N}^{H}\right)  \right]  \right]
\right\vert ^{2}+\left\vert \frac{\partial}{\partial\mathbf{{W}}_{ij}^{\ast}%
}\left[  \frac{1}{M}\mathrm{Tr}\left[  \psi\left(  \boldsymbol{\Sigma}%
_{N}\boldsymbol{\Sigma}_{N}^{H}\right)  \right]  \right]  \right\vert
^{2}\right]  \label{eq:Nash-Poincare}%
\end{equation}
where $\mathbf{W}_{ij}$ denotes the ($i,j$)th entry of matrix $\mathbf{W}$
defined in (\ref{eq:def-normalisation}). Now, applying e.g. \cite[Lemma
4.6]{haagerup2005new} we can readily see that
\begin{align}
\frac{\partial}{\partial\mathbf{{W}}_{ij}}\left[  \frac{1}{M}\mathrm{Tr}%
\left[  \psi\left(  \boldsymbol{\Sigma}_{N}\boldsymbol{\Sigma}_{N}^{H}\right)
\right]  \right]   &  =  \frac{1}{M}\left[  {\boldsymbol{\Sigma}}_{N}^{H}%
\ \psi^{\prime}\left(  \boldsymbol{\Sigma}_{N}\boldsymbol{\Sigma}_{N}%
^{H}\right)  \right]  _{j,i}\label{eq:derivee}\\
\frac{\partial}{\partial\mathbf{{W}}_{ij}^{\ast}}\left[  \frac{1}%
{M}\mathrm{Tr}\left[  \psi\left(  \boldsymbol{\Sigma}_{N}\boldsymbol{\Sigma
}_{N}^{H}\right)  \right]  \right]   &  =  \frac{1}{M}\left[  \psi^{\prime
}\left(  \boldsymbol{\Sigma}_{N}\boldsymbol{\Sigma}_{N}^{H}\right)
\boldsymbol{\Sigma}_{N}\right]  _{i,j}%
\end{align}
where $\psi^{\prime}$ denotes the derivative of $\psi$. Consequently, the sum
on the right hand side of (\ref{eq:Nash-Poincare}) can be written as
\begin{multline*}
\sum_{i,j}\mathbb{E}\left[  \left\vert \frac{\partial}{\partial\mathbf{W}%
_{ij}}\left[  \frac{1}{M}\mathrm{Tr}\left[  \psi\left(  \boldsymbol{\Sigma
}_{N}\boldsymbol{\Sigma}_{N}^{H}\right)  \right]  \right]  \right\vert
^{2}+\left\vert \frac{\partial}{\partial\mathbf{{W}}_{ij}^{\ast}}\left[
\frac{1}{M}\mathrm{Tr}\left[  \psi\left(  \boldsymbol{\Sigma}_{N}%
\boldsymbol{\Sigma}_{N}^{H}\right)  \right]  \right]  \right\vert ^{2}\right]
=\\
=\frac{2}{M^{2}}\mathbb{E}\left[  \mathrm{Tr}\left[  \left[  \psi^{\prime
}\left(  \boldsymbol{\Sigma}_{N}\boldsymbol{\Sigma}_{N}^{H}\right)  \right]
^{2}{\boldsymbol{\Sigma}}_{N}{\boldsymbol{\Sigma}}_{N}^{H}\right]  \right]  .
\end{multline*}
This yields
\begin{equation}
\mathrm{Var}\left[  \frac{1}{M}\mathrm{Tr}\left[  \psi\left(
\boldsymbol{\Sigma}_{N}\boldsymbol{\Sigma}_{N}^{H}\right)  \right]  \right]
\leq C\frac{1}{N^{2}}\mathbb{E}\left[  \frac{1}{M}\mathrm{Tr}\left[  \left[
\psi^{\prime}\left(  \boldsymbol{\Sigma}_{N}\boldsymbol{\Sigma}_{N}%
^{H}\right)  \right]  ^{2}\boldsymbol{\Sigma}_{N}\boldsymbol{\Sigma}_{N}%
^{H}\right]  \right]  \label{eq:proche}%
\end{equation}
for some constant $C$ independent of $N$. Next, consider the function
$h(\lambda)$, defined as $h(\lambda)=\lambda\left[  \psi^{\prime}%
(\lambda)\right]  ^{2}$, which clearly belongs to $\mathcal{C}_{c}^{\infty
}(\mathbb{R},\mathbb{R})$. Lemma \ref{le:phi(Q-T)} implies that
\[
\mathbb{E}\left[  \frac{1}{M}\mathrm{Tr}\left[  \left[  \psi^{\prime}\left(
\boldsymbol{\Sigma}_{N}\boldsymbol{\Sigma}_{N}^{H}\right)  \right]
^{2}\boldsymbol{\Sigma}_{N}\boldsymbol{\Sigma}_{N}^{H}\right]  \right]
=\int_{\mathcal{S}_{N}}h(\lambda) \mathrm{d} \mu_{N}(\lambda)+\mathcal{O}\left(  \frac
{1}{N^{2}}\right)  .
\]
But it is clear from (\ref{eq:abinterS}) that $\int_{\mathcal{S}_{N}}%
h(\lambda) \mathrm{d} \mu_{N}(\lambda)=0$ if $N$ is large enough. Therefore,
(\ref{eq:proche}) gives (\ref{eq:Varpsi}).

We are now in position to complete the proof of Theorem
\ref{theo:no-eigenvalue} as in \cite{capitaine2009largest}. Applying the
classical Markov inequality together with the above results, we can write (for
$N$ large enough)
\begin{multline}
\mathbb{P}\left(  \frac{1}{M}\mathrm{Tr}\left[  \psi\left(
\boldsymbol{\boldsymbol{\Sigma}}_{N}\boldsymbol{\boldsymbol{\Sigma}}_{N}%
^{H}\right)  \right]  >\frac{1}{N^{4/3}}\right)  \leq N^{8/3}\mathbb{E}\left[
\left\vert \frac{1}{M}\mathrm{Tr}\left[  \psi\left(
\boldsymbol{\boldsymbol{\Sigma}}_{N}\boldsymbol{\boldsymbol{\Sigma}}_{N}%
^{H}\right)  \right]  \right\vert ^{2}\right] \\
=N^{8/3}\left(  \left\vert \mathbb{E}\left[  \frac{1}{M}\mathrm{Tr}\left[
\psi\left(  \boldsymbol{\boldsymbol{\Sigma}}_{N}\boldsymbol{\boldsymbol{\Sigma
}}_{N}^{H}\right)  \right]  \right]  \right\vert ^{2}+\mathrm{Var}\left[
\frac{1}{M}\mathrm{Tr}\left[  \psi\left(  \boldsymbol{\boldsymbol{\Sigma}}%
_{N}\boldsymbol{\boldsymbol{\Sigma}}_{N}^{H}\right)  \right]  \right]
\right)  =\mathcal{O}\left(  \frac{1}{N^{4/3}}\right)
\label{equation:Prob_N43}%
\end{multline}
Then, by Borel-Cantelli lemma, for $N$ large enough, we have with probability
one,
\[
\frac{1}{M}\mathrm{Tr}\left[  \psi\left(  \boldsymbol{\boldsymbol{\Sigma}}%
_{N}\boldsymbol{\boldsymbol{\Sigma}}_{N}^{H}\right)  \right]  \leq\frac
{1}{N^{4/3}}%
\]
By the very definition of $\psi$, the number of eigenvalues of $\hat
{\mathbf{R}}_{N}=\boldsymbol{\boldsymbol{\Sigma}}_{N}%
\boldsymbol{\boldsymbol{\Sigma}}_{N}^{H}$ in $[a,b]$ is upper-bounded by
$\mathrm{Tr}\left[  \psi(\boldsymbol{\boldsymbol{\Sigma}}_{N}%
\boldsymbol{\boldsymbol{\Sigma}}_{N}^{H})\right]  $ and is therefore a
$\mathcal{O}(N^{-\frac{1}{3}})$ with probability one. Since this number has to
be an integer, we deduce that for $N$ large enough, there is no eigenvalue in
$[a,b]$. This completes the proof of Theorem \ref{theo:no-eigenvalue}.

\subsection{Proof of Theorem \ref{theo:exact-separation}}
\begin{figure}[h]
\centering
\par
\includegraphics[width=15cm]{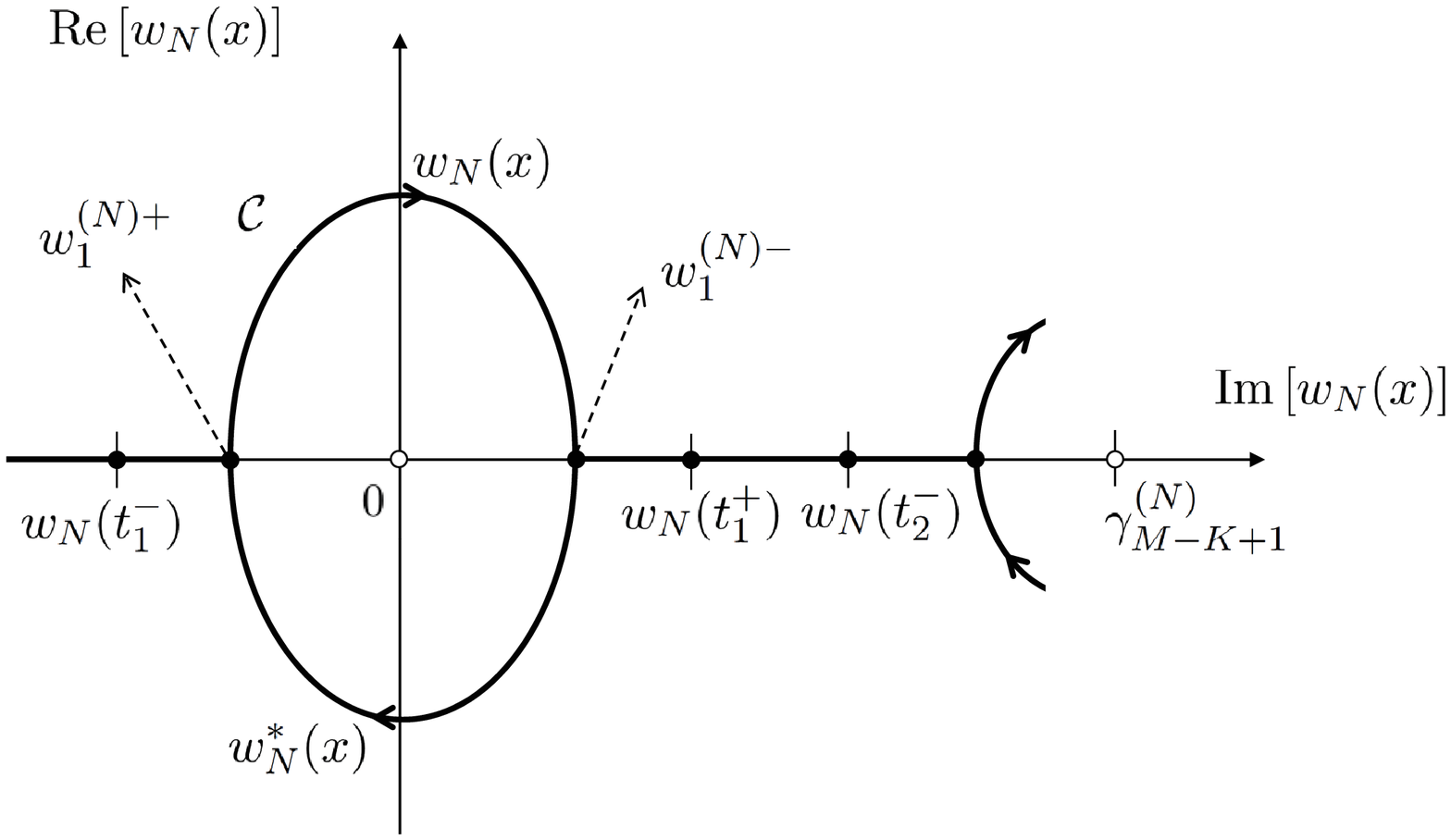}\caption{Representation of
the contour $\mathcal{C}$ on the complex plane. }%
\label{figure:contour}%
\end{figure}

The approach we use to establish Theorem \ref{theo:exact-separation}
differs from the method of \cite{capitaine2009largest} which is 
inspired by \cite{bai1999exact}. 
The first part our proof is similar to the proof of Theorem
\ref{theo:no-eigenvalue}, and thus we will omit certain details. For the
second part, we will need a certain result that we summarize in the following proposition:

\begin{proposition}
\label{prop:contour}Consider the curve $\mathcal{C}$ defined by the complex
valued function $w_{N}(x)$ in (\ref{eq:def-w}) on the complex plane as $x$
moves from $t_{1}^{-}$ to $t_{1}^{+}$, concatenated with the function
$w_{N}^{\ast}(x)$ as $x$ moves back from $t_{1}^{+}$ to $t_{1}^{-}$, namely \
\begin{equation}
\mathcal{C}=\left\{  w_{N}(x):x\in\left[  t_{1}^{-},t_{1}^{+}\right]
\right\}  \cup\left\{  w_{N}^{\ast}(x):x\in\left[  t_{1}^{-},t_{1}^{+}\right]
\right\}  .\label{definition:contour_C}%
\end{equation}
This is a closed curve that encloses the points of $\left]  w_{1}^{-}%
,w_{1}^{+}\right[  $ (see further Figure \ref{figure:contour}). Let $\psi(z)$
be a function holomorphic in a neighborhood of $\mathcal{C}$. Then, the
contour integral $\int_{\mathcal{C}^{-}}\psi(\lambda)\,\mathrm{d}\lambda$ is
well defined by
\begin{equation}
\oint_{\mathcal{C}^{-}}\psi(\lambda)\,\mathrm{d}\lambda=2\mathrm{i~Im}\left[
\int_{[t_{1}^{-},t_{1}^{+}]}\psi(w_{N}(x))w_{N}^{\prime}(x)\,\mathrm{d}%
x\right]  .\label{eq:defintegraleC}%
\end{equation}
where $w_{N}^{\prime}(z)$ denotes the derivative of $w_{N}(z)$ and where the
symbol $\mathcal{C}^{-}$ means that $\mathcal{C}$ is oriented clockwise.
\newline Finally, let $\xi\in\mathbb{R}$ a point that does not belong to
$[w_{N}(t_{1}^{-}),w_{1}^{(N)-}]\cup\lbrack w_{1}^{(N)+},w_{N}(t_{1}^{+})]$.
Then,
\[
\mathrm{Ind}_{\mathcal{C}}(\xi) = \frac{1}{2 i \pi} \int_{{\cal C}_{-}} \frac{d \lambda}{\xi - \lambda} = \left\{
\begin{array}
[c]{cc}%
1 & \mathrm{if\quad}\xi\in\left]  w_{1}^{(N)-},w_{1}^{(N)+}\right[  \\
0 & \mathrm{if\quad}\xi<w_{N}(t_{1}^{-})\text{\quad}\mathrm{or}\text{\quad}%
\xi>w_{N}(t_{1}^{+}),
\end{array}
\right.
\]

\end{proposition}%

\begin{IEEEproof}%
According to the discussion in Section \ref{section:discussion_w_select}, if
$x\in\left[  t_{1}^{-},x_{1}^{(N)-}\right]  $, then $w_{N}(x)$ is real-valued,
and increases from $w_{N}(t_{1}^{-})$ to $w_{N}\left(  x_{1}^{(N)-}\right)
=w_{1}^{(N)-}$. For $x\in\left]  x_{1}^{(N)-},x_{1}^{(N)+}\right[  $, the
point $w_{N}(x)$ belongs to $\mathbb{C}_{+}$. Finally, if $x\in\left[
x_{1}^{(N)+},t_{1}^{+}\right]  $, $w_{N}(x)$ is again real-valued, and
increases from $w_{N}\left(  x_{1}^{(N)+}\right)  =w_{1}^{(N)+}$ to
$w_{N}(t_{1}^{+})$. The contour $\mathcal{C}$ is therefore well defined and
encloses the points of $\left]  w_{1}^{(N)-},w_{1}^{(N)+}\right[  $.

Let us now prove (\ref{eq:defintegraleC}). Observe that the function
$x\rightarrow w_{N}(x)$ is not exactly a piecewise continuously differentiable
function on $\left[  t_{1}^{-},t_{1}^{+}\right]  $ because $|w_{N}^{\prime
}(x)|$ increases without bound when $x\rightarrow x_{1}^{(N)-},x_{1}^{(N)+}$. To see that $w_{N}(x)$ can indeed be
used as a valid parametrization of $\mathcal{C}$, we need to see that the
integral in (\ref{eq:defintegraleC}) is well defined. It is thus necessary to
study the behavior of $w_{N}^{\prime}$ around the points$\ \left\{
x_{1}^{(N)-},x_{1}^{(N)+}\right\}  $. The following lemma is an immediate
consequence of the analysis of the behavior of the density of measure $\mu
_{N}$ near a point of $\partial\mathcal{S}_{N}$ provided in
\cite{dozier2007analysis} (see Appendix \ref{section:proof_lemma_domination}
for a proof).

\begin{lemma}
\label{lemma:domination}There exists neighborhoods $\mathcal{V}\left(
x_{1}^{(N)-}\right)  $ and $\mathcal{V}\left(  x_{1}^{(N)+}\right)  $ of
$x_{1}^{(N)-}$ and $x_{1}^{(N)+}$ such that
\begin{equation}
\left\vert w_{N}^{\prime}(x+\mathrm{i}y)\right\vert \leq\frac{C}%
{\sqrt{\left\vert x-x_{1}^{(N)-}\right\vert }}%
\;\mbox{for $y \geq  0$, $x+\mathrm{i}y \in {\cal V}(x_1^{(N)-})$, and $x \neq x_1^{(N)-}$}\label{eq:domination-}%
\end{equation}
and
\begin{equation}
\left\vert w_{N}^{\prime}(x+\mathrm{i}y)\right\vert \leq\frac{C}%
{\sqrt{\left\vert x-x_{1}^{(N)+}\right\vert }}%
\;\mbox{for $y  \geq 0$, $x+\mathrm{i}y \in {\cal V}(x_1^{(N)+})$ and  $x \neq x_1^{(N)+}$}\label{eq:domination+}%
\end{equation}

\end{lemma}

In particular, Lemma \ref{lemma:domination} implies that $\int_{[t_{1}%
^{-},t_{1}^{+}]}|\psi(w_{N}(x))||w_{N}^{\prime}(x)|\,dx<+\infty$ so that the
right hand side of (\ref{eq:defintegraleC}) is well defined. The reader may
check that it is possible to use the usual results related to integrals over
piecewise continuously differentiable contours. In particular, as
$\mathrm{Im}(w_{N}(x))>0$ if $x\in\left]  x_{1}^{(N)-},x_{1}^{(N)+}\right[  $,
the index of a point $\xi\in\mathbb{R}$ which does not belong to $\left[
w_{N}(t_{1}^{-}),w_{1}^{(N)-}\right]  \cup\left[  w_{1}^{(N)+},w_{N}(t_{1}%
^{+})\right]  $ 
is equal to $1$ is $\xi \in\left]  w_{1}^{(N)-},w_{1}^{(N)+}\right[  $ and to $0$
if either $\xi<w_{N}(t_{1}^{-})$ or $w>w_{N}(t_{1}^{+})$.
\end{IEEEproof}

Proposition \ref{prop:contour} is basically pointing out that the function
$w_{N}(x)$ defines a valid parametrization of a contour that will not
intersect with any eigenvalue of $\mathbf{B}_{N}\mathbf{B}_{N}^{H}$.
Furthermore, Assumptions \ref{assumption:exact_sep_1} and
\ref{assumption:exact_sep_2} imply that
\begin{equation}
\mathrm{Ind}_{\mathcal{C}}(0)=1\label{eq:indice0}%
\end{equation}
and
\begin{equation}
\mathrm{Ind}_{\mathcal{C}}(\gamma_{M-K+l})=0\label{eq:indiceeigenvalues}%
\end{equation}
for $l=1,\ldots,K$. This means that the contour will only enclose the zero
eigenvalue, and none of the positive eigenvalues of $\mathbf{B}_{N}%
\mathbf{B}_{N}^{H}$, which will be of crucial importance in the following
development. Figure \ref{figure:contour} gives a schematic representation of
the form of the contour $\mathcal{C}$.   

Having introduced the result in Proposition \ref{prop:contour}, we are now in
the position of establishing the proof of Theorem \ref{theo:exact-separation}.
Let $\psi\in\mathcal{C}_{c}^{\infty}(\mathbb{R},\mathbb{R})$ such that
$0\leq\psi\leq1$ and
\[
\psi(\lambda)=%
\begin{cases}
1 & \forall\lambda\in\lbrack t_{1}^{-},t_{1}^{+}]\\
0 & \forall\lambda\in\mathbb{R}-[t_{1}^{-}-\epsilon,t_{1}^{+}+\epsilon]
\end{cases}
\]
with $\epsilon$ chosen in such a way that $t_{1}^{+}+\epsilon<t_{2}^{-}$.
Since $\psi\in\mathcal{C}_{c}^{\infty}(\mathbb{R},\mathbb{R})$, we can use
Lemma \ref{le:phi(Q-T)} to get
\[
\mathbb{E}\left[  \frac{1}{M}\mathrm{Tr}\left[  \psi\left(
\boldsymbol{\boldsymbol{\Sigma}}_{N}\boldsymbol{\boldsymbol{\Sigma}}_{N}%
^{H}\right)  \right]  \right]  =\int_{\mathbb{R}_{+}}\psi(\lambda
)\mathrm{d}\mu_{N}(\lambda)+\mathcal{O}\left(  \frac{1}{N^{2}}\right)  .
\]
Assumptions \ref{assumption:exact_sep_1} and \ref{assumption:exact_sep_2}
imply that
\[
\int_{\mathbb{R}_{+}}\psi(\lambda)\mathrm{d}\mu_{N}(\lambda)=\mu_{N}\left(
\left[  x_{1}^{(N)-},x_{1}^{(N)+}\right]  \right)  =\mu_{N}([t_{1}^{-}%
,t_{1}^{+}])
\]
for $N$ large enough. This leads to
\[
\mathbb{E}\left[  \frac{1}{M}\mathrm{Tr}\left[  \psi\left(
\boldsymbol{\boldsymbol{\Sigma}}_{N}\boldsymbol{\boldsymbol{\Sigma}}_{N}%
^{H}\right)  \right]  \right]  =\mu_{N}([t_{1}^{-},t_{1}^{+}])+\mathcal{O}%
\left(  \frac{1}{N^{2}}\right)
\]
As established in \eqref{eq:Varpsi}, we also have
\[
\mathrm{Var}\left[  \frac{1}{M}\mathrm{Tr}\left[  \psi\left(
\boldsymbol{\Sigma}_{N}\boldsymbol{\Sigma}_{N}^{H}\right)  \right]  \right]
=\mathcal{O}\left(  \frac{1}{N^{4}}\right)
\]
because $\mathrm{supp}(\psi^{\prime})\cap\mathcal{S}_{N}=\varnothing$ for $N$
large enough. Therefore, using again the proof of theorem
\ref{theo:no-eigenvalue} (inequality \eqref{equation:Prob_N43}), we get that
\begin{equation}
\frac{1}{M}\mathrm{Tr}\left[  \psi\left(  \boldsymbol{\boldsymbol{\Sigma}}%
_{N}\boldsymbol{\boldsymbol{\Sigma}}_{N}^{H}\right)  \right]  -\mu_{N}\left(
\left[  t_{1}^{-},t_{1}^{+}\right]  \right)  =\mathcal{O}\left(  \frac
{1}{N^{4/3}}\right)  \quad\mathrm{a.s.}\label{equation:exact_separation_final}%
\end{equation}
Let us now find a closed form expression for $\mu_{N}([t_{1}^{-},t_{1}^{+}])$.
Noting that $\mu_{N}$ is absolutely continuous with density $\frac{1}{\pi
}\mathrm{Im}(m_{N}(x))$, we can write
\[
\mu_{N}\left(  \left[  t_{1}^{-},t_{1}^{+}\right]  \right)  =\frac{1}{\pi
}\mathrm{Im}\left[  \int_{t_{1}^{-}}^{t_{1}^{+}}m_{N}(x)\mathrm{d}x\right]  .
\]
By expressing the Stieltjès transform as $m_{N}(x)=\frac{f_{N}(w_{N}%
(x))}{1-\sigma^{2}c_{N}f_{N}(w_{N}(x))}$ (see further
(\ref{equation:link_delta_f_1})), $\mu_{N}\left(  \left[  t_{1}^{-},t_{1}%
^{+}\right]  \right)  $ can be written as
\[
\mu_{N}\left(  \left[  t_{1}^{-},t_{1}^{+}\right]  \right)  =\frac{1}{\pi
}\mathrm{Im}\left[  \int_{t_{1}^{-}}^{t_{1}^{+}}\frac{f_{N}(w_{N}%
(x))}{1-\sigma^{2}c_{N}f_{N}(w_{N}(x))}\mathrm{d}x\right]
\]
In order to express $\mu_{N}\left(  \left[  t_{1}^{-},t_{1}^{+}\right]
\right)  $ in terms of an integral over the contour $\mathcal{C}$, we can use
the relation $w_{N}^{\prime}(x)\phi_{N}^{\prime}(w_{N}(x))=1$ for each
$x\in\mathbb{R}-\partial\mathcal{S}_{N}$ (see further (\ref{eq:equation-Phi}%
)). Now, using Proposition \ref{prop:contour}, we see that
\begin{multline}
\mu_{N}\left(  \left[  t_{1}^{-},t_{1}^{+}\right]  \right)  =\frac{1}{\pi
}\mathrm{Im}\left[  \int_{\left[  t_{1}^{-},t_{1}^{+}\right]  }\frac
{f_{N}(w_{N}(x))\phi_{N}^{\prime}(w_{N}(x))}{1-\sigma^{2}c_{N}f_{N}(w_{N}%
(x))}w_{N}^{\prime}(x)\mathrm{d}x\right]  =\frac{1}{2\pi\mathrm{i}}%
\oint_{\mathcal{C}^{-}}\frac{f_{N}(\lambda)\phi_{N}^{\prime}(\lambda
)}{1-\sigma^{2}c_{N}f_{N}(\lambda)}\mathrm{d}\lambda\\
=\frac{1}{2\pi\mathrm{i}}\oint_{\mathcal{C}^{-}}f_{N}(\lambda)\frac
{\;(1-c_{N}\sigma^{2}f_{N}(\lambda))^{2}-2c_{N}\sigma^{2}\lambda f_{N}%
^{\prime}(\lambda)\;(1-c_{N}\sigma^{2}f_{N}(\lambda))-c_{N}\sigma^{4}%
(1-c_{N})f_{N}^{\prime}(\lambda))}{1-\sigma^{2}c_{N}f_{N}(\lambda
)}\,\mathrm{d}\lambda\label{eq:expre-masse}%
\end{multline}
The integrand of the right hand side of (\ref{eq:expre-masse}) is a
meromorphic function. The contour integral can be thus evaluated using the
residue theorem. The poles of the integrand are the eigenvalues of
$\mathbf{B}_{N}\mathbf{B}_{N}^{H}$ as well as the solutions of the equation
$1-\sigma^{2}c_{N}f_{N}(\lambda)=0$. This equation has $K+1$ real-valued
solutions that we have denoted $z_{0}^{(N)+}$, and $\left\{  z_{l}%
^{(N)-}\right\}  _{l=1,\ldots,K}$ (see further Figures \ref{figure:support} to
\ref{figure:supportlowc}). Assumptions \ref{assumption:exact_sep_1} and
\ref{assumption:exact_sep_2} imply that only the poles $\left\{  0\right\}  $
and $\left\{  z_{0}^{(N)+}\right\}  $ of the integrand are in fact enclosed by
$\mathcal{C}$. Using the residue theorem, and after some straightforward
calculations, we obtain a closed form for the above integral, namely
\[
\mu_{N}\left(  \left[  t_{1}^{-},t_{1}^{+}\right]  \right)  =\frac{M-K}%
{M}\alpha_{1}^{(N)}+\frac{1}{M}\sum_{k=M-K+1}^{M}\alpha_{k}^{(N)}%
\]
with
\begin{align}
\alpha_{1}^{(N)} &  =\frac{N-K}{M-K}\left(  1-\frac{\sigma^{2}c_{N}}{M}%
\sum_{l=M-K+1}^{M}\frac{1}{\gamma_{l}^{(N)}}\right)  +\frac{\sigma^{2}%
(1-c_{N})}{z_{0}^{(N)+}}\\
\alpha_{k}^{(N)} &  =\left(  1-\frac{K}{N}\right)  \frac{\sigma^{2}}%
{\gamma_{k}^{(N)}}+\frac{\sigma^{2}(1-c_{N})}{z_{0}^{(N)+}-\gamma_{k}^{(N)}}%
\end{align}
Therefore, we can write
\begin{align}
\mu_{N}\left(  \left[  t_{1}^{-},t_{1}^{+}\right]  \right)   &  =\frac{N-K}%
{M}+\sigma^{2}\frac{1-c_{N}}{M}\left(  \frac{M-K}{M}\frac{1}{z_{0}^{(N)+}%
}+\sum_{k=M-K+1}^{M}\frac{1}{z_{0}^{(N)+}-\gamma_{k}^{(N)}}\right)  \\
&  =\frac{N-K}{M}-\sigma^{2}\left(  1-c_{N}\right)  f_{N}\left(  z_{0}%
^{(N)+}\right)
\end{align}
but, using the fact that $1-\sigma^{2}c_{N}f_{N}(z_{0}^{(N)+})=0$, we obtain
that $\mu_{N}\left(  \left[  t_{1}^{-},t_{1}^{+}\right]  \right)  =\frac
{M-K}{M}$. Inserting this into \eqref{equation:exact_separation_final}, we
get
\[
\mathrm{Tr}\left[  \psi\left(  \boldsymbol{\boldsymbol{\Sigma}}_{N}%
\boldsymbol{\boldsymbol{\Sigma}}_{N}^{H}\right)  \right]  -(M-K)=\mathcal{O}%
\left(  \frac{1}{N^{1/3}}\right)
\]
with probability $1$. Moreover, thanks to theorem \ref{theo:no-eigenvalue}, no
eigenvalue of $\boldsymbol{\boldsymbol{\Sigma}}_{N}%
\boldsymbol{\boldsymbol{\Sigma}}_{N}^{H}$ appears in $[t_{1}^{-}%
-\epsilon,t_{1}^{-}]\cup\lbrack t_{1}^{+},t_{1}^{+}+\epsilon]$ almost surely
for $N$ large enough. Therefore, almost surely for $N$ large enough,
$\mathrm{Tr}\left[  \psi(\boldsymbol{\boldsymbol{\Sigma}}_{N}%
\boldsymbol{\boldsymbol{\Sigma}}_{N}^{H})\right]  $ coincides with the number
of eigenvalues of $\boldsymbol{\boldsymbol{\Sigma}}_{N}%
\boldsymbol{\boldsymbol{\Sigma}}_{N}^{H}$ contained in the interval $\left]
t_{1}^{-},t_{1}^{+}\right[  $. This number is thus equal to $M-K$. These
eigenvalues are moreover the $M-K$ smallest ones: otherwise the smallest
eigenvalue of $\boldsymbol{\boldsymbol{\Sigma}}_{N}%
\boldsymbol{\boldsymbol{\Sigma}}_{N}^{H}$ would belong to $\left[  0,t_{1}%
^{-}\right]  $, a contradiction by Theorem \ref{theo:no-eigenvalue}. Finally,
Theorem \ref{theo:no-eigenvalue} again implies that $\hat{\lambda}%
_{M-K+1}^{(N)}>t_{2}^{-}$. This completes the proof of Theorem
\ref{theo:exact-separation}.

\section{Consistent estimation of the localization function}

\label{section:estimationeigenvalues}We now present a consistent estimator
$\eta_{N}=\mathbf{b}_{N}^{H}{\boldsymbol{\Pi}}_{N}\mathbf{b}_{N}$ of the
subspace method localization function. Here, $\mathbf{b}_{N}$ represents a
$M$--dimensional deterministic vector, and we assume that $\sup_{N}%
\Vert\mathbf{b}_{N}\Vert<\infty$. The new consistent estimator presented in
this section can be seen as an extension of the work in
\cite{mestre2008modified}, which implicitely assumes that the useful signals are
Gaussian random i.i.d. sequences. In order to simplify the notation, we drop the
dependence on $N$ from all the sample eigenvalues and sample eigenvectors.

\begin{theorem}
\label{proposition:MUSIC} Under Assumptions \ref{assumption:exact_sep_1} and
\ref{assumption:exact_sep_2}, we have with probability one,
\[
\hat{\eta}_{N}^{new}-\eta_{N}\longrightarrow0
\]
where $\hat{\eta}_{N}^{new}$ is defined by
\begin{equation}
\hat{\eta}_{N}^{new}=\sum_{k=1}^{M}\hat{\xi}_{k}{}\mathbf{b}_{N}^{{H}}%
\hat{\mathbf{e}}_{k}\hat{\mathbf{e}}_{k}^{H}\mathbf{b}_{N}
\label{definition:music_G_estimator}%
\end{equation}
Here, the coefficients $\left\{  \hat{\xi}_{k}\right\}  _{k=1,\ldots,M-K}$ are
given by
\begin{equation}
\hat{\xi}_{k}=1+\frac{\sigma^{2}c_{N}}{M}\sum_{l=M-K+1}^{M}\frac{\hat{\lambda
}_{k}+\hat{\lambda}_{l}}{(\hat{\lambda}_{k}-\hat{\lambda}_{l})^{2}}+\sigma
^{2}(1-c_{N})\sum_{l=M-K+1}^{M}\left(  \frac{1}{\hat{\lambda}_{k}-\hat
{\lambda}_{l}}-\frac{1}{\hat{\lambda}_{k}-\hat{\omega}_{l}}\right)
\label{definition:coef_xi_1}%
\end{equation}
and $\left\{  \hat{\xi}_{k}\right\}  _{k=M-K+1,\ldots,M}$ by
\begin{equation}
\hat{\xi}_{k}=-\frac{\sigma^{2}c_{N}}{M}\sum_{l=1}^{M-K}\frac{\hat{\lambda
}_{k}+\hat{\lambda}_{l}}{(\hat{\lambda}_{k}-\hat{\lambda}_{l})^{2}}-\sigma
^{2}(1-c_{N})\sum_{l=1}^{M-K}\left(  \frac{1}{\hat{\lambda}_{k}-\hat{\lambda
}_{l}}-\frac{1}{\hat{\lambda}_{k}-\hat{\omega}_{l}}\right)
\label{definition:coef_xi_k}%
\end{equation}
and where $\left\{  \hat{\omega}_{l}\right\}  _{l=1,\ldots,M}$ represent the
solutions (arranged in increasing order) of the equation%
\begin{equation}
1+\frac{\sigma^{2}c_{N}}{M}\mathrm{Tr}\left[  \left(  \boldsymbol{\Sigma}%
_{N}\boldsymbol{\Sigma}_{N}^{H}-x\mathbf{I}_{M}\right)  ^{-1}\right]  =0.
\label{eq:equations_w}%
\end{equation}

\end{theorem}

We remark that the consistent estimator is a linear combination of the terms
$\left(  {\mathbf{b}}_{N}^{H}\mathbf{\hat{e}}_{k}\mathbf{\hat{e}}_{k}%
^{H}{\mathbf{b}}_{N}\right)  _{k=1,\ldots,M}$. In contrast to the traditional
estimator $\eta_{trad}=\sum_{k=1}^{M-K}{\mathbf{b}}_{N}^{H}\mathbf{\hat{e}%
}_{k}\mathbf{\hat{e}}_{k}^{H}{\mathbf{b}}_{N}$, it contains contributions of
both the noise subspace and the signal subspace. We also note that the
assumptions \ref{assumption:exact_sep_1} and \ref{assumption:exact_sep_2} and
Theorem \ref{theo:exact-separation} are intuitively important because the
various sums on the right hand side of (\ref{definition:coef_xi_1}) and
(\ref{definition:coef_xi_k}) remain bounded: in (\ref{definition:coef_xi_1})
and (\ref{definition:coef_xi_k}), the terms $\left\vert \hat{\lambda}_{k}%
-\hat{\lambda}_{l}\right\vert $ are greater than $t_{2}^{-}-t_{1}^{+}$, and it
will be shown that a similar property holds for the terms $\left\vert
\hat{\lambda}_{k}-\hat{\omega}_{l}\right\vert $.

\begin{remark}
It is worth pointing out that whenever the number of samples is forced to be
much larger than the observation dimension ($N>>M\,$\ or equivalently
$c_{N}\rightarrow0$), the proposed estimator converges to the classical sample
eigenvector estimate. This can be readily seen by taking the limit as
$c_{N}\rightarrow0$ in the coefficients of (\ref{definition:coef_xi_1}) and
(\ref{definition:coef_xi_k}) and noticing that $\hat{\omega}_{l}%
\rightarrow\hat{\lambda}_{l}$ when $c_{N}\rightarrow0$. Hence, as
$c_{N}\rightarrow0$ we have $\hat{\xi}_{k}\rightarrow1$ for $k=1,\ldots,M-K$,
\ and $\hat{\xi}_{k}\rightarrow0$ for $k=M-K+1,\ldots,M\,$, implying that
$\hat{\eta}_{N}^{new} - \hat{\eta}_{N}^{trad} \rightarrow 0$. This shows that the
proposed estimator is in fact a generalization of the classical sample
eigenvector estimate.
\end{remark}

The remaining of this section is devoted to presenting the main points of the
proof of Theorem \ref{proposition:MUSIC}. The starting point consists in
remarking that Assumptions \ref{assumption:exact_sep_1} and
\ref{assumption:exact_sep_2} imply that
\[
\eta_{N}=\frac{1}{2\pi{\mathrm{i}}}\oint_{\mathcal{C}^{-}}{{}{\mathbf{b}}%
_{N}^{H}\left(  \mathbf{B}_{N}\mathbf{B}_{N}^{H}-\lambda\mathbf{I}_{M}\right)
^{-1}\ \mathbf{b}_{N}\mathrm{d}\lambda}%
\]
where $\mathcal{C}^{-}$ is the closed path defined by
(\ref{definition:contour_C}). This leads to
\begin{multline}
\eta_{N}=\frac{1}{2\pi{\mathrm{i}}}\int_{t_{1}^{-}}^{t_{1}^{+}}{\mathbf{b}%
}_{N}^{H}\left(  \mathbf{B}_{N}\mathbf{B}_{N}^{H}-w_{N}(x)\mathbf{I}%
_{M}\right)  ^{-1}\mathbf{b}_{N}w_{N}^{\prime}(x)\mathrm{d}x+\\
-\frac{1}{2\pi{\mathrm{i}}}\int_{t_{1}^{-}}^{t_{1}^{+}}{\mathbf{b}}_{N}%
^{H}\left(  \mathbf{B}_{N}\mathbf{B}_{N}^{H}-w_{N}^{\ast}(x)\mathbf{I}%
_{M}\right)  ^{-1}{\mathbf{b}_{N}}\left(  {w_{N}^{\prime}(x)}\right)  ^{\ast
}\mathrm{d}x=\\
\frac{1}{\pi}\mathrm{Im}\left(  \int_{t_{1}^{-}}^{t_{1}^{+}}{\mathbf{b}}%
_{N}^{H}\left(  \mathbf{B}_{N}\mathbf{B}_{N}^{H}-w_{N}(x)\mathbf{I}%
_{M}\right)  ^{-1}{\mathbf{b}_{N}}w_{N}^{\prime}(x)\mathrm{d}x\right)  .
\end{multline}
Let $g_{N}(x+\mathrm{i}y)={{\mathbf{b}}_{N}^{H}}\left(  \mathbf{B}%
_{N}\mathbf{B}_{N}^{H}-w_{N}(x+\mathrm{i}y)\mathbf{I}_{M}\right)
^{-1}{{\mathbf{b}}_{N}}w_{N}^{\prime}(x+\mathrm{i}y)$. The function
$y\rightarrow g_{N}(x+\mathrm{i}y)$ is continuous on $\mathbb{R}_{+}$ for each
$x\in\mathbb{R}\backslash\partial\mathcal{S}_{N}$ thanks to Proposition
\ref{prop:m}. Lemma \ref{lemma:domination} and the dominated convergence
theorem imply that
\begin{align}
\eta_{N} &  =\lim_{y\downarrow0}\frac{1}{\pi}\mathrm{Im}\left(  \int
_{t_{1}^{-}}^{t_{1}^{+}}{\mathbf{b}}_{N}^{H}\left(  \mathbf{B}_{N}%
\mathbf{B}_{N}^{H}-w_{N}(x+\mathrm{i}y)\mathbf{I}_{M}\right)  ^{-1}%
{\mathbf{b}_{N}}w_{N}^{\prime}(x+\mathrm{i}y)\mathrm{d}x\right)  \\
&  =\lim_{y\downarrow0}\left[  \frac{1}{2\pi\mathrm{i}}\oint_{\partial
\mathcal{R}_{y}^{-}}g_{N}(z)\mathrm{d}z-\frac{1}{2\pi}\int_{-y}^{y}g_{N}%
(t_{1}^{-}+\mathrm{i}h)\mathrm{d}h+\frac{1}{2\pi}\int_{-y}^{y}g_{N}(t_{1}%
^{+}-\mathrm{i}h)\mathrm{d}h\right]
\end{align}
where $\partial\mathcal{R}_{y}^{-}$ is the boundary (clockwise oriented) of
the rectangle $\mathcal{R}_{y}$ defined for $y>0$ by
\begin{equation}
\mathcal{R}_{y}=\left\{  u+\mathrm{i}v:u\in\lbrack t_{1}^{-},t_{1}^{+}%
],v\in\lbrack-y,y]\right\}  .\label{definition:rectangle}%
\end{equation}
Notice that the last two integrands vanish as $y\downarrow0$ (since the
function $v\mapsto g_{N}(t_{1}^{-}+\mathrm{i}v)$ is continuous on $[-y,y]$),
and thus\textbf{ }
\[
\eta_{N}=\lim_{y\downarrow0}\frac{1}{2\pi\mathrm{i}}\oint_{\partial
\mathcal{R}_{y}^{-}}g_{N}(z)\mathrm{d}z.
\]
Moreover, since $g_{N}(z)$ is holomorphic in $\mathbb{C}\backslash [x_1^{(N)-},x_1^{(N)+}] $,
the value of the contour integral does not depend on $y>0$, and therefore the
limit can be dropped, namely
\[
\eta_{N}=\frac{1}{2\pi\mathrm{i}}\oint_{\partial\mathcal{R}_{y}^{-}}%
g_{N}(z)\mathrm{d}z.
\]
Using the equality $(1+\sigma^{2}cm_{N}(z))(\mathbf{B}_{N}\mathbf{B}_{N}%
^{H}-w_{N}(z)\mathbf{I}_{M})^{-1}=\mathbf{T}_{N}(z)$, which follows easily
from the definition in (\ref{eq:def-T}), we can write
\[
g_{N}(z)={\mathbf{b}}_{N}^{H}\mathbf{T}_{N}(z){\mathbf{b}_{N}\frac
{w_{N}^{\prime}(z)}{1+\sigma^{2}cm_{N}(z)}.}%
\]
Now, the key point of the proof is based on the observation that $g_{N}(z)$
can be estimated consistently from the elements of matrix $\hat{\mathbf{R}}_{N}$.
We recall that $\hat{m}_{N}(z)$ is defined by
\begin{equation}
\hat{m}_{N}(z)=\frac{1}{M}\mathrm{Tr}\left[  \mathbf{Q}_{N}(z)\right]
=\frac{1}{M}\sum_{k=1}^{M}\frac{1}{\hat{\lambda}_{k}-z}\label{eq:def-mhat}%
\end{equation}
and we define $\hat{w}_{N}(z)$ as the function obtained by replacing function $m_{N}(z)$
with $\hat{m}_N(z)$ in the definition of $w_{N}(z)$, i.e.
\begin{equation}
\hat{w}_{N}(z)=z\left(  1+\sigma^{2}c_{N}\hat{m}_{N}(z)\right)  ^{2}%
-\sigma^{2}(1-c_{N})\left(  1+\sigma^{2}c_{N}\hat{m}_{N}(z)\right)
\label{eq:def-what}%
\end{equation}
We define the corresponding random asymptotic equivalent of $g_{N}(z)$ by
\[
\hat{g}_{N}(z)={\mathbf{b}}_{N}^{H}\mathbf{Q}_{N}(z){\mathbf{b}}%
_{N}{\frac{\hat{w}_{N}^{\prime}(z)}{1+\sigma^{2}c_{N}\hat{m}_{N}(z)}.}%
\]
Observe from the definition of $\hat{m}_N$ and of ${\bf Q}_N$ that the function $\hat{g}_N$ is
meromorphic with poles at $\hat{\lambda}_{1}$,\ldots,$\hat{\lambda}_{M}$ and
at $\hat{\omega}_{1}$,\ldots,$\hat{\omega}_{M}$, the $M$ real-valued solutions
to the polynomial equation (of degree $M$) $1+\sigma^{2}c_{N}\hat{m}_{N}%
(x)=0$. In the following, it is important to locate the
$(\hat{\omega}_{l})_{l=1,\ldots,M}$.

\begin{lemma}
\label{lemma:poles_localization} For $N$ large enough, with probability one
\begin{align}
&  \hat{\lambda}_{1},\ldots,\hat{\lambda}_{M-K},\hat{\omega}_{1},\ldots
,\hat{\omega}_{M-K}\in]t_{1}^{-},t_{1}^{+}[\\
&  \hat{\lambda}_{M-K+1},\ldots,\hat{\lambda}_{M},\hat{\omega}_{M-K+1}%
,\ldots,\hat{\omega}_{M}\;\mbox{are greater than $t_2^{-}$}
\end{align}

\end{lemma}%

Theorem \ref{theo:convergence-hatmu} implies that almost surely,
$g_{N}(z)-\hat{g}_{N}(z)\rightarrow0$ on $\partial\mathcal{R}_{y}%
\backslash\{t_{1}^{-},t_{1}^{+}\}$. In order to be able to use the dominated convergence theorem, 
we first state the following inequalities proven in Appendix \ref{sec:borne-g-hatg}: there exists $N_0 \in \mathbb{N}$ 
such that 
\begin{equation}
\label{eq:borne-g}
\sup_{N \geq N_0} \sup_{z \in \partial\mathcal{R}_{y}} |g_N(z)| < +\infty
\end{equation}
and 
\begin{equation}
\label{eq:borne-hatg}
\sup_{N \geq N_0} \sup_{z \in \partial\mathcal{R}_{y}} |\hat{g}_N(z)| < +\infty
\end{equation}
almost surely. The dominated convergence theorem thus implies that 
\[
\left\vert \frac{1}{2\pi\mathrm{i}}\oint_{\partial\mathcal{R}_{y}^{-}}%
g_{N}(z)-\hat{g}_{N}(z)\mathrm{d}z\right\vert \longrightarrow0\qquad
\mathrm{a.s.}%
\]
We now establish that the integral
\[
\hat{\tilde{\eta}}_{N}^{new}=\frac{1}{2\pi\mathrm{i}}\oint_{\partial
\mathcal{R}_{y}^{-}}\hat{g}_{N}(z)\mathrm{d}z
\]
is equal to $\hat{\eta}_{N}^{new}$ defined by
\eqref{definition:music_G_estimator}. This can be shown using residue Theorem.

Lemma \ref{lemma:poles_localization} implies that for $N$ large enough
\[
\hat{\tilde{\eta}}_{N}^{new}=\sum_{k=1}^{M-K}\left[  \mathrm{Ind}%
_{\partial\mathcal{R}_{y}^{-}}\left(  \hat{\lambda}_{k}\right)  \mathrm{Res}%
\left(  \hat{g}_{N},\hat{\lambda}_{k}\right)  +\mathrm{Ind}_{\partial
\mathcal{R}_{y}^{-}}\left(  \hat{\omega}_{k}\right)  \mathrm{Res}\left(
\hat{g}_{N},\hat{\omega}_{k}\right)  \right]
\]
where $\mathrm{Res}(\hat{g}_{N},\lambda)$ denotes the residue of function
$\hat{g}_{N}$ at point $\lambda$.

In order to evaluate these residues, we first remark that
\[
\mathbf{b}_{N}^{H}\mathbf{Q}_{N}(z)\mathbf{b}_{N}=\sum_{k=1}^{M}%
\frac{\mathbf{b}_{N}^{H}\hat{\mathbf{e}}_{k}\hat{\mathbf{e}}_{k}^{H}%
\mathbf{b}_{N}}{\hat{\lambda}_{k}-z}%
\]
$\hat{g}_{N}\left(  z\right)  $ can thus be written as
\[
\hat{g}_{N}\left(  z\right)  =\sum_{k=1}^{M}\mathbf{b}_{N}^{H}\hat{\mathbf{e}%
}_{k}\hat{\mathbf{e}}_{k}^{H}\mathbf{b}_{N}\left[  \hat{\alpha}_{k}%
(z)+\hat{\beta}_{k}(z)+\hat{\gamma}_{k}(z)\right]
\]
where we have defined 
\begin{gather}
\hat{\alpha}_{k}(z)=\frac{1+\sigma^{2}c_{N}\hat{m}_{N}(z)}{\hat{\lambda}%
_{k}-z}\\
\hat{\beta}_{k}(z)=\frac{2\sigma^{2}c_{N}z\hat{m}_{N}^{\prime}(z)}%
{\hat{\lambda}_{k}-z}\\
\hat{\gamma}_{k}(z)=-\sigma^{4}c_{N}(1-c_{N})\frac{\hat{m}_{N}^{\prime}%
(z)}{\left(  \hat{\lambda}_{k}-z\right)  \left(  1+\sigma^{2}c_{N}\hat{m}%
_{N}(z)\right)  }%
\end{gather}
and consequently with probability one for $N$ large enough
\[
\hat{\tilde{\eta}}_{N}^{new}=-\sum_{k=1}^{M}{}\mathbf{b}_{N}^{H}%
\hat{\mathbf{e}}_{k}\hat{\mathbf{e}}_{k}^{H}\mathbf{b}_{N}\sum_{m=1}%
^{M-K}\left[  \mathrm{Res}\left(  \hat{\alpha}_{k},\hat{\lambda}_{m}\right)
+\mathrm{Res}\left(  \hat{\beta}_{k},\hat{\lambda}_{m}\right)  +\mathrm{Res}%
\left(  \hat{\gamma}_{k},\hat{\lambda}_{m}\right)  +\mathrm{Res}\left(
\hat{\gamma}_{k},\hat{\omega}_{m}\right)  \right]  .
\]
Classical residue calculus gives
\begin{align}
\mathrm{Res}\left(  \hat{\alpha}_{k},\hat{\lambda}_{m}\right)   &  =\left\{
\begin{array}
[c]{ccc}%
-\frac{\sigma^{2}c_{N}}{M}\frac{1}{\hat{\lambda}_{k}-\hat{\lambda}_{m}} & \  &
k\neq m\\
-\left(  1+\sigma^{2}c_{N}\frac{1}{M}\sum_{\substack{i=1\\i\neq k}}^{M}%
\frac{1}{\hat{\lambda}_{i}-\hat{\lambda}_{k}}\right)   & \  & k=m
\end{array}
\right.  \\
\mathrm{Res}\left(  \hat{\beta}_{k},\hat{\lambda}_{m}\right)   &  =\left\{
\begin{array}
[c]{ccc}%
\frac{2\sigma^{2}c_{N}}{M}\frac{\hat{\lambda}_{k}}{\left(  \hat{\lambda}%
_{k}-\hat{\lambda}_{m}\right)  ^{2}} & \  & k\neq m\\
-\frac{2\sigma^{2}c_{N}}{M}\sum_{\substack{i=1\\i\neq k}}^{M}\frac
{\hat{\lambda}_{k}}{\left(  \hat{\lambda_{i}}-\hat{\lambda}_{k}\right)  ^{2}}
& \  & k=m
\end{array}
\right.  \\
\mathrm{Res}\left(  \hat{\gamma}_{k},\hat{\lambda}_{m}\right)   &  =\left\{
\begin{array}
[c]{ccc}%
\sigma^{2}\left(  1-c_{N}\right)  \frac{1}{\hat{\lambda}_{k}-\hat{\lambda}%
_{m}} & \  & k\neq m\\
-M\frac{1-c_{N}}{c_{N}}\left(  1+\frac{\sigma^{2}c_{N}}{M}\sum
_{\substack{i=1\\i\neq k}}^{M}\frac{1}{\hat{\lambda_{i}}-\hat{\lambda}_{k}%
}\right)   & \  & k=m
\end{array}
\right.  \\
\mathrm{Res}\left(  \hat{\gamma}_{k},\hat{\omega}_{m}\right)   &  =-\sigma
^{2}\frac{1-c_{N}}{\hat{\lambda}_{k}-\hat{\omega}_{m}}.
\end{align}
Next, we define $\hat{\xi}_{k}$ as 
\[
\hat{\xi}_{k}=-\sum_{m=1}^{M-K}\mathrm{Res}\left(  \hat{\alpha}_{k}%
,\hat{\lambda}_{m}\right)  +\mathrm{Res}\left(  \hat{\beta}_{k},\hat{\lambda
}_{m}\right)  +\mathrm{Res}\left(  \hat{\gamma}_{k},\hat{\lambda}_{m}\right)
+\mathrm{Res}\left(  \hat{\gamma}_{k},\hat{\omega}_{m}\right)  .
\]
We obtain, for $k=1,\ldots,M-K$ 
\begin{align}
\hat{\xi}_{k} &  =1-\frac{\sigma^{2}c_{N}}{M}\sum_{i=M-K+1}^{M}\frac{1}%
{\hat{\lambda}_{k}-\hat{\lambda}_{i}}+\frac{2\sigma^{2}c_{N}}{M}\sum
_{i=M-K+1}^{M}\frac{\hat{\lambda}_{k}}{\left(  \hat{\lambda}_{k}-\hat{\lambda
}_{i}\right)  ^{2}}+M\frac{1-c_{N}}{c_{N}}\\
&  +\sigma^{2}(1-c_{N})\left(  \sum_{\substack{i=1\\i\neq k}}^{M-K}\frac
{1}{\hat{\lambda}_{i}-\hat{\lambda}_{k}}-\sum_{\substack{i=1\\i\neq k}%
}^{M-K}\frac{1}{\hat{\omega}_{i}-\hat{\lambda}_{k}}+\sum_{\substack{i=1\\i\neq
k}}^{M}\frac{1}{\hat{\lambda}_{i}-\hat{\lambda}_{k}}\right)
\end{align}
and for $k=M-K+1,\ldots,M$
\begin{align}
\hat{\xi}_{k} &  =-\frac{\sigma^{2}c_{N}}{M}\sum_{i=1}^{M-K}\frac{1}%
{\hat{\lambda}_{i}-\hat{\lambda}_{k}}-\frac{2\sigma^{2}c_{N}}{M}\sum
_{i=1}^{M-K}\frac{\hat{\lambda}_{k}}{\left(  \hat{\lambda}_{k}-\hat{\lambda
}_{i}\right)  ^{2}}\\
&  +\sigma^{2}(1-c_{N})\sum_{i=1}^{M-K}\frac{\hat{\omega}_{i}-\hat{\lambda
}_{i}}{\left(  \hat{\lambda}_{i}-\hat{\lambda}_{k}\right)  \left(  \hat
{\omega}_{i}-\hat{\lambda}_{k}\right)  }.
\end{align}
To retrieve the final form of $\hat{\xi}_{k}$ given in the statement of the
theorem, we notice that
\[
1+\sigma^{2}c_{N}\frac{1}{M}\sum_{i=1}^{M}\frac{1}{\hat{\lambda}_{i}%
-\hat{\omega}_{k}}=0
\]
and use the following lemma proved in Appendix
\ref{section:proof_lemma_xi_transformation}:

\begin{lemma}
\label{lemma:xi_transformation} The following identity holds for any
$k=1\ldots M$
\[
\frac{1}{M}\sum_{\substack{i=1\\i\neq k}}^{M}\frac{1}{\hat{\lambda}_{i}%
-\hat{\omega}_{k}}=\frac{2}{M}\sum_{\substack{i=1\\i\neq k}}^{M}\frac{1}%
{\hat{\lambda}_{i}-\hat{\lambda}_{k}}-\frac{1}{M}\sum_{\substack{i=1\\i\neq
k}}^{M}\frac{1}{\hat{\omega}_{i}-\hat{\lambda}_{k}}%
\]

\end{lemma}

This establishes that $\hat{\tilde{\eta}}_{new}=\hat{\eta}_{new}$ and
completes the proof of Theorem \ref{proposition:MUSIC}.

\section{Numerical results}

\label{section:simu}

In this section, we compare the results provided by the traditional subspace estimate, the new
estimate \eqref{definition:music_G_estimator} (referred to in the figure as
the "conditional estimator"), and the improved estimate of \cite{mestre2008modified} derived under the assumption
that the source signals are i.i.d. sequences (referred to as the "unconditional estimator"). 

\begin{figure}[h!]
\centering
\par
\includegraphics[scale=0.4]{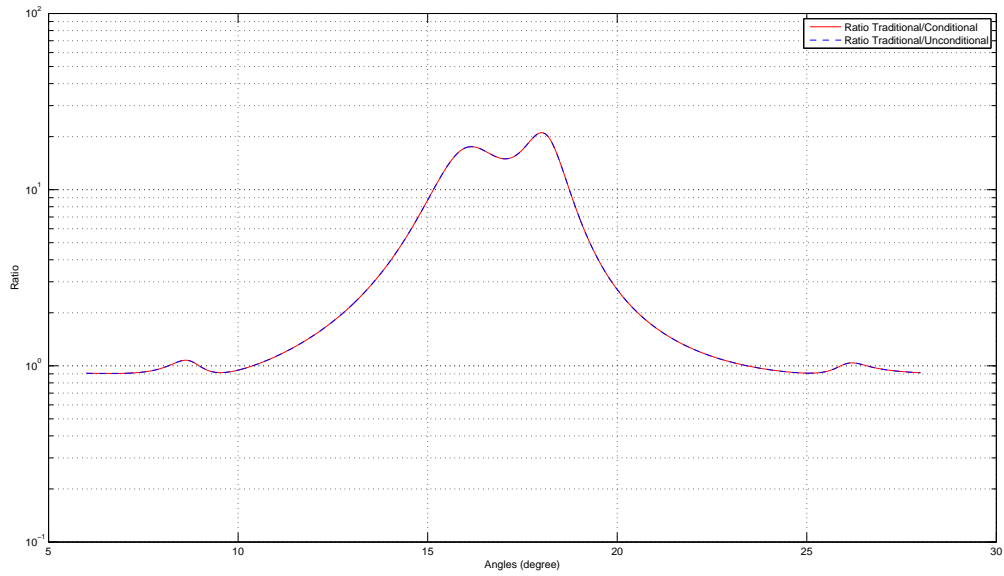}
\caption{Ratio (in dB) of the MSE of the traditional estimate of ${\bf a}(\theta)^{H} {\bs \Pi} {\bf a}(\theta)$ over the improved estimates vs angles.}%
\label{figure:ratiomsevsangles-exper1}%
\end{figure}

\begin{figure}[h!]
\centering
\par
\includegraphics[scale=0.4]{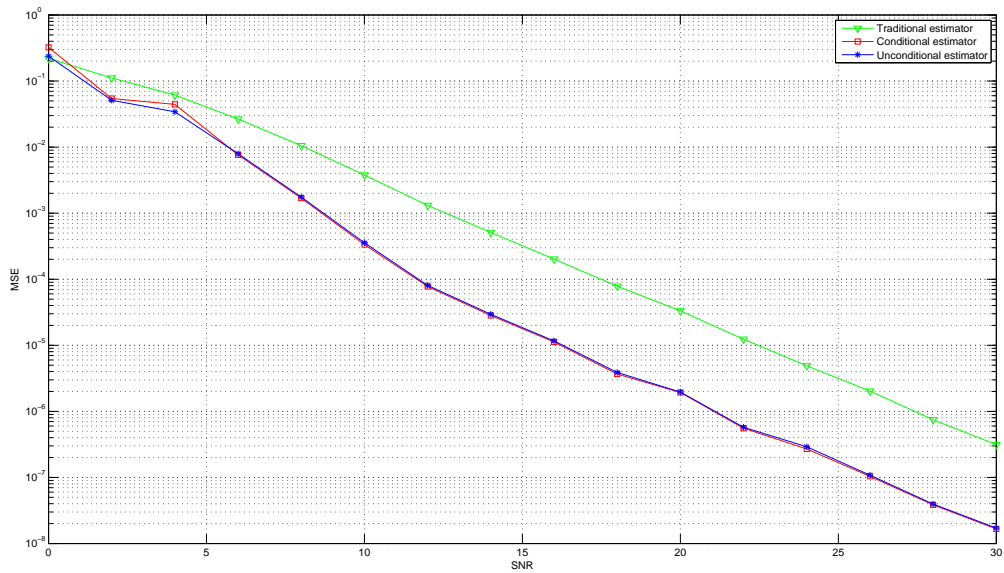}
\caption{Mean of the MSE of the estimates of  ${\bf a}(\theta_1)^{H} \bs{\Pi} {\bf a}(\theta_1)$
and  ${\bf a}(\theta_2)^{H} \bs{\Pi} {\bf a}(\theta_2)$. }%
\label{figure:msequadform-exper-1}%
\end{figure}

We consider a uniform linear array of antennas the elements of which are located at half the wavelenght. 
The steering vector ${\bf a}(\theta)$ is thus given by
\begin{align}
	\a(\theta) = \frac{1}{\sqrt{M}}\left[1,e^{i \pi \sin(\theta)},\ldots,e^{i (M-1) \pi \sin(\theta)}\right]^T
\end{align}
In the following numerical experiments, source signals are realizations of mutually independent unit variance AR(1) sequences 
with correlation coefficient $0.9$. In order to evaluate the performance of the various estimators, 
we use Monte Carlo simulations. The additive noise varies from trials to trials, but, for fixed $M$ and $N$, 
matrix ${\bf S}$ remains unchanged. Finally, unless otherwise stated, the cluster associated to 
the eigenvalue 0 of matrix ${\bf A} {\bf S}$ is assumed to be separated from the clusters corresponding to its non zero eigenvalues, i.e. 
for each $\sigma^2$, $M$ and $N$, it holds that
\begin{equation}
\label{eq:expre-sepcond-Nfini}
0  <  w_1^{(N)+} <  w_{2}^{(N)-}  < \gamma_{M-K+1}^{(N)}
\end{equation}
We finally mention that the estimate of \cite{mestre2008modified} is supposed to 
be unconsistent in the context of the following experiments because the source signals are not i.i.d. sequences. 
However, we will see that the performance of the conditional and the unconditional estimates
are quite close, a property which will need further work (see Remark \ref{rem:cond-uncond}).  
\\
\textbf{Experiment 1}:
We first consider two closely spaced sources, i.e.  $\theta_1=16^{\circ}$ and $\theta_2=18^{\circ}$.
The number of antennas is $M=20$ and the number of snapshots is $N=40$. The separation condition (\ref{eq:expre-sepcond-Nfini}) is verified if the SNR is larger than 10 dB.
 In order to evaluate the performance of the estimates of the localization 
function, for each improved estimator (conditional and unconditional), we  plot versus $\theta$ in figure \ref{figure:ratiomsevsangles-exper1}
the ratio of the MSE of the traditional estimator of 
${\bf a}(\theta)^{H} {\bs \Pi} {\bf a}(\theta)$ over the MSE of the improved estimator. The SNR is equal to $16$ dB. Figure  \ref{figure:ratiomsevsangles-exper1}
shows that the 2 improved estimates have nearly the same performance, and that they outperform significantly the traditional approach around the 2 angles. We however
notice that the 3 estimates have nearly the same performance if $\theta$ is far away from  $\theta_1=16^{\circ}$ and $\theta_2=18^{\circ}$. 
In order to evaluate more precisely the improvements provided by the conditional and the unconditional estimators around $\theta_1$ and
$\theta_2$, we plot vs SNR in figure \ref{figure:msequadform-exper-1} the mean of the MSEs of the estimates of ${\bf a}(\theta_1)^{H} \bs{\Pi} {\bf a}(\theta_1)$
and  ${\bf a}(\theta_2)^{H} \bs{\Pi} {\bf a}(\theta_2)$. 
\\

\begin{figure}[h!]
\centering
\par
\includegraphics[scale=0.7]{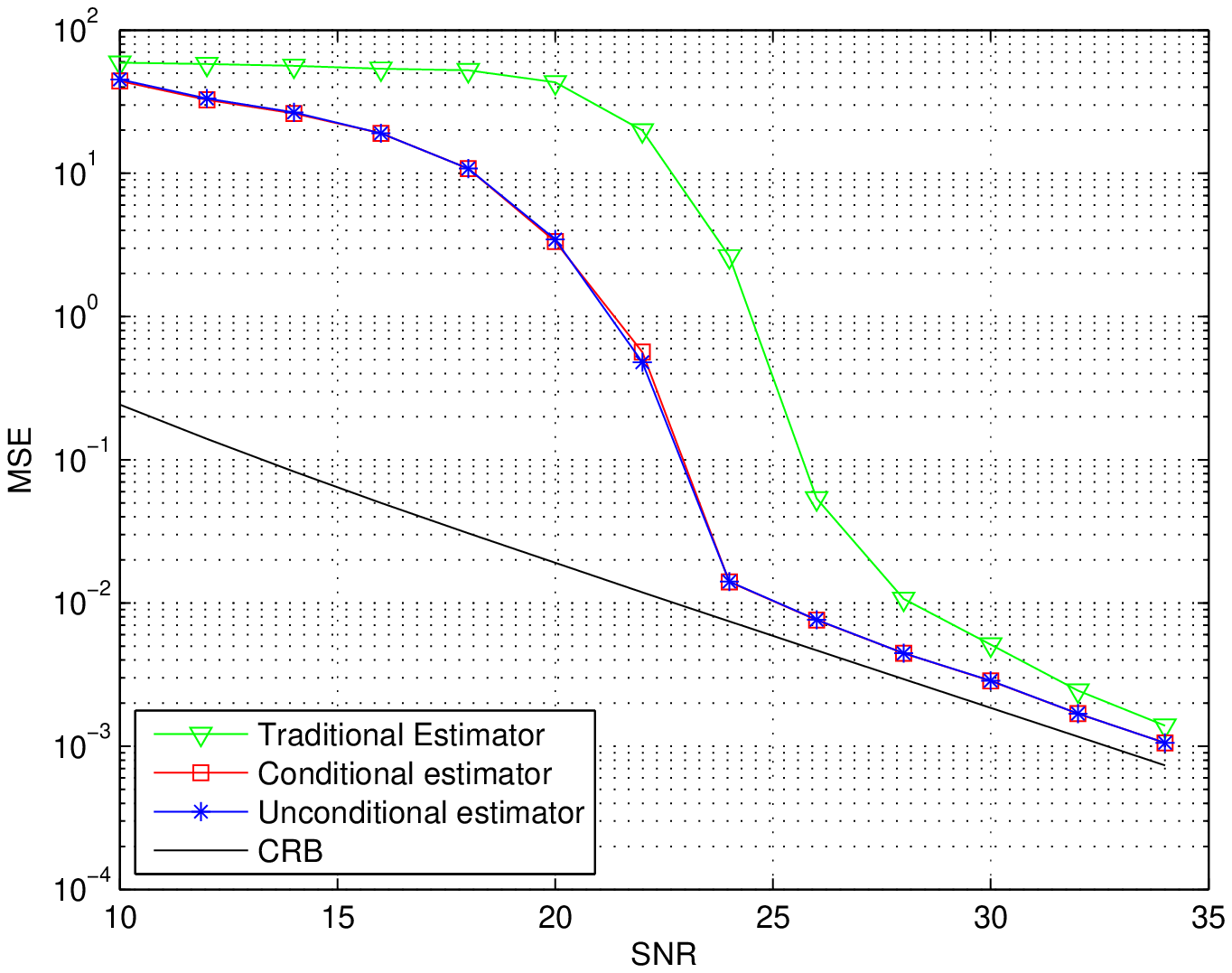}
\caption{Mean of the MSE of the angles estimates versus SNR}%
\label{figure:rel_st_dev_doa}%
\end{figure}

In figure \ref{figure:rel_st_dev_doa}, we plot for each method the mean of the MSE of the two estimated angles versus the SNR. The estimates of $\theta_1$ and $\theta_2$ are defined 
as the arguments of the two deepest local minima of the estimated localization function.  
The mean of the two Cramer-Rao bounds is also represented. The performance of the 2 improved estimates are again quite similar, and they provide an improvement of $4$ dB 
w.r.t the traditional estimator in the range 15dB-25dB. 
\\

\begin{figure}[h!]
\centering
\par
\includegraphics[scale=0.7]{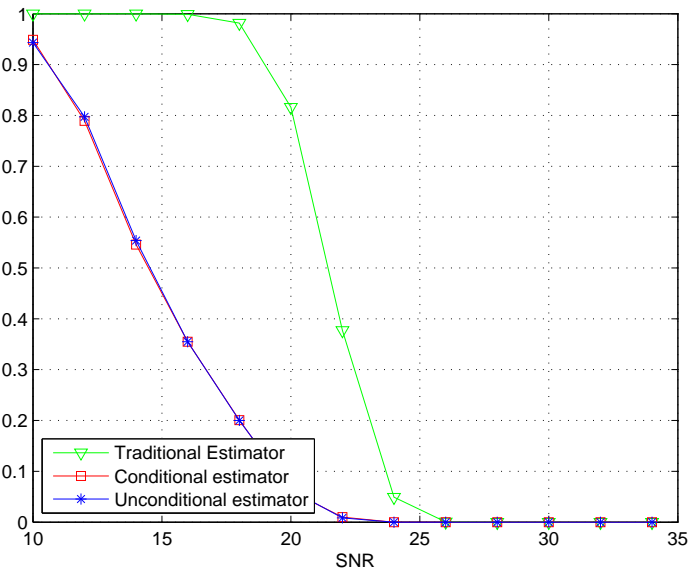}\caption{Outlier
Probability vs the SNR}%
\label{figure:proba}%
\end{figure}

We now plot the probability of outlier, i.e. the probability that one of the two estimated angles is separated from
the true one by more than half of the separation between the two true sources.
In figure \ref{figure:proba}, we compare the outlier probability of the three
approaches versus the SNR of the three estimators.
For a target probability of error of $0.5$, the 2 improved estimators provide a gain of $8$ dB over the traditional estimate. 

\begin{figure}[h!]
\centering
\par
\includegraphics[scale=0.3]{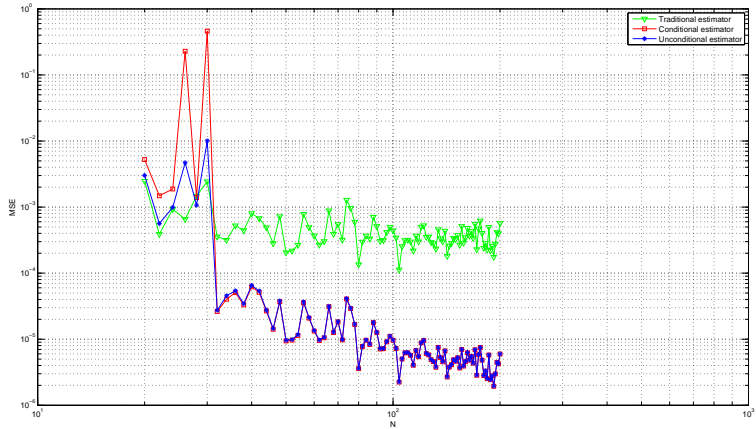}
\caption{MSE for the estimators of the localization function vs N}%
\label{figure:msevsN}%
\end{figure}

We finally evaluate the influence of $M$ and $N$  on the performance. $N$ varies from $20$ to $200$ while the ratio $c_N$ is kept constant to $0.5$, and 
SNR = 15 dB.
In figure \ref{figure:msevsN} we have plotted the mean of the MSEs on the estimates of  ${\bf a}(\theta_i)^{H} {\bs \Pi} {\bf a}(\theta_i)$ for $i=1,2$. The separation condition (\ref{eq:expre-sepcond-Nfini}) occurs for $N\geq32$. Figure \ref{figure:msevsN} illustrates clearly the unconsistency of that the traditional estimate. 

\begin{figure}[h!]
\centering
\par
\includegraphics[scale=0.3]{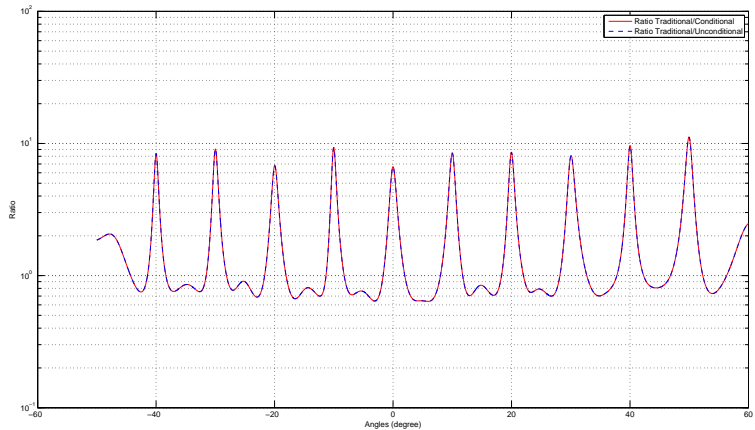}
\caption{Ratio (in dB) of the MSE of the traditional estimate of the localization function over the 
MSE of its improved estimates versus $\theta$}
\label{figure:ratiomse}%
\end{figure}

\textbf{Experiment 2:}
We now assume that the number of sources $K$ is of the same order of magnitude that $M$ and $N$, i.e. $K = 10$, $M=20, N=40$.  
The ten angles $(\theta_i)_{i=1, \ldots, 10}$ are equal to $\theta_i = -40^{\circ} + (i-1) 10^{\circ}$ for $i=1, \ldots, 10$. 
The separation condition holds if SNR is greater than 15 dB. 
We again plot versus $\theta$ in figure \ref{figure:ratiomse} the ratio of the MSE of the traditional estimator  of the localization function over the 
MSE of its conditional and unconditional estimators. SNR is equal to $16$ dB. Figure \ref{figure:ratiomse}
shows again that the performance improvement of the conditional and unconditional estimates is optimum 
around the angles  $(\theta_i)_{i=1, \ldots, 10}$. 

\begin{figure}[h!]
\centering
\par
\includegraphics[scale=0.3]{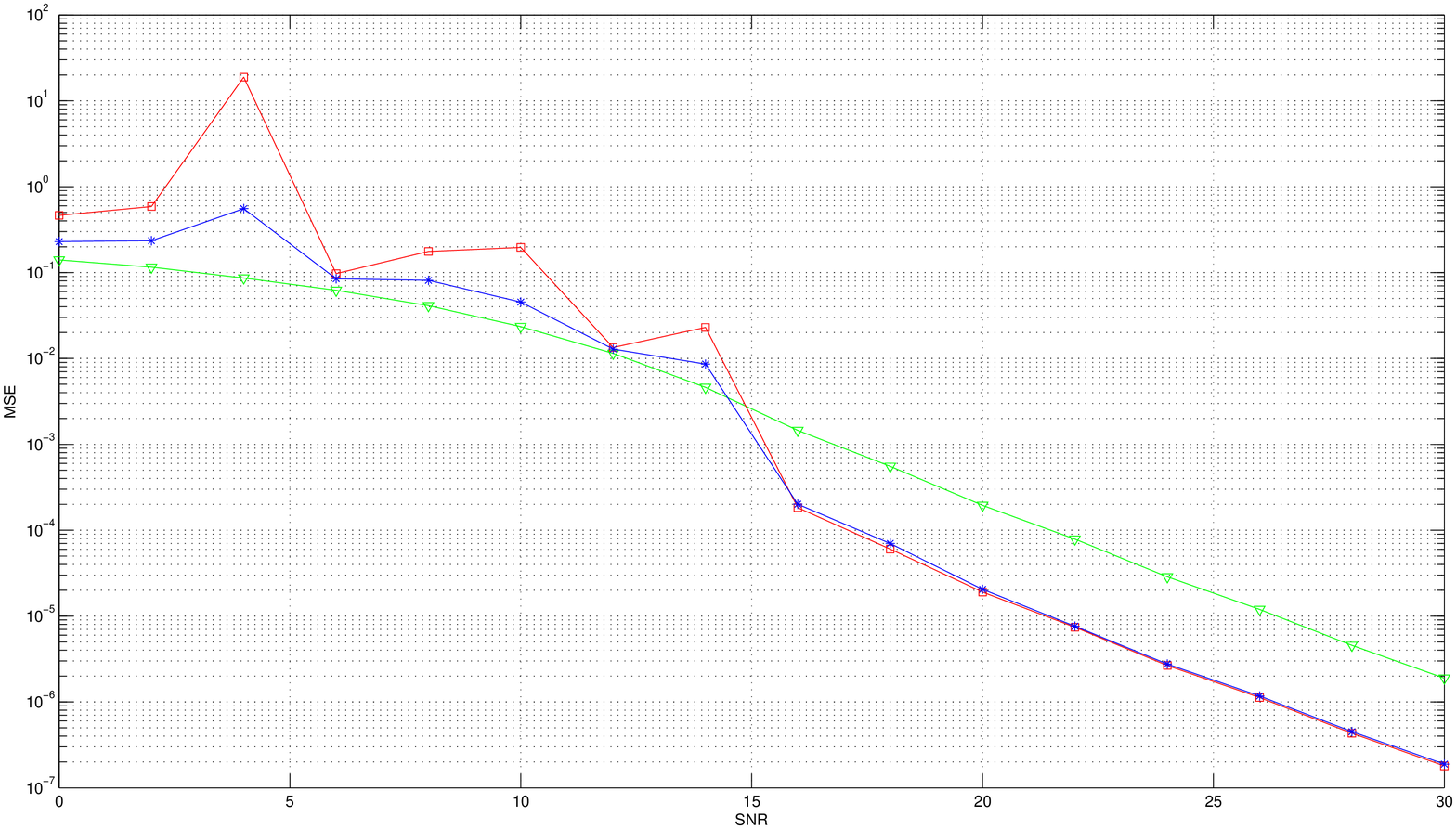}
\caption{Mean of the MSE of the estimates of  ${\bf a}(\theta_i)^{H} \bs{\Pi} {\bf a}(\theta_i)$
for $i=1, \ldots,10$ versus SNR }%
\label{figure:mseprojvssnr-10}%
\end{figure}

Figure \ref{figure:mseprojvssnr-10} represents the mean of the MSEs of the various estimates of ${\bf a}(\theta_i)^{H} {\bs \Pi} {\bf a}(\theta_i)$
for $i=1, \ldots, 10$ w.r.t. the SNR, and confirms the superiority of the 2 improved estimates when 
the separation condition (\ref{eq:expre-sepcond-Nfini}). 
We note that  

\begin{remark}
\label{rem:cond-uncond}
	All the previous plots clearly show that the conditional estimator outperforms the traditional one, while its difference with the unconditional one is negligible. 	
        This is a quite surprising fact. To explain this, we recall that the unconditional estimator has been derived in \cite{mestre2008modified} under the assumption 
        that matrix ${\bf S}_N$ is a Gaussian matrix with unit variance i.i.d. entries.  The unconditional estimator of \cite{mestre2008modified} is based on the observation
	that if  ${\bf S}_N$ is an i.i.d. Gaussian matrix, then the entries of $(\hat{\mathbf{R}}_N - z \mathbf{I})^{-1}$ have the same behaviour than the entries of 
        matrix $ \mathbf{T}_{N,iid}(z)$ defined by the following equation
	\begin{align}
		m_{N,iid}(z) &= \frac{1}{M}\mathrm{Tr}\ \mathbf{T}_{N,iid}(z) \notag \\
		 \mathbf{T}_{N,iid}(z) &= \left[\left(\mathbf{A}\mathbf{A}^H+\sigma^2 \mathbf{I}\right) (1-c_N-c_N z m_{N,iid}(z)) - z \mathbf{I}\right]^{-1} \notag
	\end{align}
	One can verify that the entries of $\mathbf{T}_N(z)$ defined by (\ref{eq:def-T}), which depend on $\mathbf{S}_N$, have the same asymptotic behaviour than the entries 
        of $\mathbf{T}_{N,iid}(z)$ 
        when ${\bf S}_N$ is a realization of an i.i.d. matrix. In this case, the conditional and unconditional estimators have of course the same behaviour. 
        If however $\mathbf{S}_N$ is not an i.i.d. matrix, then the entries of $(\hat{\mathbf{R}}_N - z \mathbf{I})^{-1}$ do not behave like 
        the entries of $\mathbf{T}_{N,iid}(z)$ so that the unconditional estimator should become unconsistent. The previous simulation results tend to indicate 
        that it is not the case. The explanation of this phenomenon is a topic for further researchs. 	
\end{remark}

\section{Conclusions}

\label{section:conclusions}This paper has considered the use of subspace
estimation algorithms in situations where the number of available samples and
the observation dimension are comparable in magnitude. We have considered the
information plus noise signal model, according to which the received signals
are deterministic unknowns whose empirical spatial correlation matrix is
low-rank. We have derived an estimator of the noise subspace of the spatial
correlation matrix that is consistent, not only when the number of samples
tends to infinity for a fixed observation dimension, but also when these two
quantities increase to infinity at the same rate. This guarantees that the
estimator will present a good performance even when these two quantities are
comparable in magnitude. In order to establish the consistency of the 
estimator, we have proven new results concerning the almost sure location 
of the eigenvalues of the sample covariance matrix of an Information plus 
Noise Gaussian model.  

\appendices

\section{Proof of Property \ref{enu:O_supp} of Proposition \ref{prop:m}}

\label{appendix:proof_zero_not_belong}
In order to establish that $0$ does not belong
to the support $\mathcal{S}_{N}$, we show that it exists $\epsilon>0$ for
which $\mu_N([0,x]) = 0$ for each $x \in ]0,\epsilon[)$.  In order to show this, we will make
us of the function $h(m,z)$ defined as%
\begin{equation}
h(m,z)=\frac{1}{M}\mathrm{Tr}\left[  -z(1+\sigma^{2}c_{N}m)\mathbf{I}%
_{M}+\sigma^{2}(1-c_{N})\mathbf{I}_{M}+\frac{\mathbf{B}_{N}\mathbf{B}_{N}^{H}%
}{1+\sigma^{2}c_{N}m}\right]  ^{-1}. \label{eq:def-h}%
\end{equation}
Observe that the equation $m=h(m,0)$ is equivalent to
\[
m=\frac{1}{M}\mathrm{Tr}\left[  \sigma^{2}(1-c_{N})\mathbf{I}_{M}%
+\frac{\mathbf{B}_{N}\mathbf{B}_{N}^{H}}{1+\sigma^{2}c_{N}m}\right]  ^{-1}.
\]
Now, the condition $c_{N}<1$ implies that the function $m\rightarrow \frac{h(m,0)}{m}$
is decreasing on $\mathbb{R}_{+}$. Therefore, the equation $m=h(m,0)$ has a unique strictly positive solution
denoted $m_{\ast}$. Next, we will check that
\begin{equation}
1-\left.  \frac{\partial h}{\partial m}\right\vert _{\left(  m_{\ast
},0\right)  }>0. \label{eq:partialh}%
\end{equation}
Indeed, observe that
\[
\left.  \frac{\partial h}{\partial m}\right\vert _{\left(  m_{\ast},0\right)
}=\frac{\sigma^{2}c_{N}}{1+\sigma^{2}c_{N}m_{\ast}}\;\frac{1}{M}%
\mathrm{Tr}\left[  \frac{\mathbf{B}_{N}\mathbf{B}_{N}^{H}}{1+\sigma^{2}%
c_{N}m_{\ast}}\left(  \sigma^{2}(1-c_{N})\mathbf{I}_{M}+\frac{\mathbf{B}%
_{N}\mathbf{B}_{N}^{H}}{1+\sigma^{2}c_{N}m_{\ast}}\right)  ^{-2}\right]
\]
so that
\[
\left.  \frac{\partial h}{\partial m}\right\vert _{\left(  m_{\ast},0\right)
}<\frac{\sigma^{2}c_{N}}{1+\sigma^{2}c_{N}m_{\ast}}\;\frac{1}{M}%
\mathrm{Tr}\left[  \sigma^{2}(1-c_{N})\mathbf{I}_{M}+\frac{\mathbf{B}%
_{N}\mathbf{B}_{N}^{H}}{1+\sigma^{2}c_{N}m_{\ast}}\right]  ^{-1}=\frac
{\sigma^{2}c_{N}m_{\ast}}{1+\sigma^{2}c_{N}m_{\ast}}<1
\]
as required. Hence, the implicit function theorem implies that there exists an
open disk centered at zero with radius $\eta>0$, i.e. $D(0,\eta)$, and a
unique function $\overline{m}(z)$, holomorphic on $D(0,\eta)$, satisfying
$\overline{m}(0)=m_{\ast}$ and such that
\begin{equation}
\overline{m}(z)=h(\overline{m}(z),z) \label{eq:implicite}%
\end{equation}
for $|z|<\eta$. Evaluating the successive derivatives of function $z\rightarrow
h(\overline{m}(z),z)$ at the origin, one can check that for each $l\geq0$,
$\overline{m}^{(l)}(0)$ is real-valued. Since $m_{\ast}>0$, there exists a
positive quantity $\epsilon$, $0<\epsilon\leq\eta$ such that $\overline{m}(x)$
is real-valued and $\overline{m}(x)>0$ if $x \in ]-\epsilon, \epsilon[$. On the other hand,
it can be readily checked that if $x<0$, the equation $m=h(m,x)$ has a unique
strictly positive solution. Now, for $x<0$, $m_{N}(x)$ is strictly positive, and satisfies this
equation. Therefore, it holds that $m_{N}(x)=\overline{m}(x)$ for
$-\epsilon<x<0$. Since the two functions $m_{N}$ and $\overline{m}$ are
holomorphic on $D(0,\epsilon)\backslash\left\{  \lbrack0,\epsilon
\lbrack\right\}  $ and coincide on a set of values with an accumulation point,
they must coincide on the whole domain of analicity, namely $D(0,\epsilon
)\backslash\left\{  \lbrack0,\epsilon\lbrack\right\}  $. We recall that for
$0\leq x<\epsilon$, $\mu_N([0,x])$ can be expressed as
\[
\mu_N([0,x]) = \frac{1}{\pi} \lim_{y \rightarrow 0, y>0} \int_0^{x} \mathrm{Im}(m_N(s+iy)) ds
\]
Therefore, 
\[
\mu_N([0,x]) = \frac{1}{\pi} \lim_{y \rightarrow 0, y>0} \int_0^{x} \mathrm{Im}(\overline{m}(s+iy)) ds
\]
As $\overline{m}$ is holomorphic on $D(0,\epsilon)$, the dominated convergence theorem implies that 
\[ 
\frac{1}{\pi} \lim_{y \rightarrow 0, y>0} \int_0^{x} \mathrm{Im}(\overline{m}(s+iy)) ds = \frac{1}{\pi} \int_0^{x} \mathrm{Im}(\overline{m}(s)) ds = 0
\]
because $\overline{m}(s) \in \mathbb{R}$ if $s \in [0, x]$. This establishes that $\mu_N([0,x]) = 0$. 


\section{Proof of Proposition \ref{prop:w}}

\label{appendix:proof_properties_w}

In order to prove Property \ref{item: w_on_supp}, we establish that
$\mathrm{Im}(w_{N}(x))>0$ if and only if $\mathrm{Im}(m_{N}(x))>0$. Assume
that $\mathrm{Im}(m_{N}(x))>0$, i.e. that $x\in\mathrm{Int}(\mathcal{S}_{N})$,
which in particular implies that $x>0$, and consider $z=x+\mathrm{i}y$ with
$y>0$. Equation \eqref{eq:canonique-1} can be written in terms of $w_{N}(z)$
as
\begin{equation}
\frac{m_{N}(z)}{1+c_{N}\sigma^{2}m_{N}(z)}=f_{N}(w_{N}%
(z)).\label{eq:canonique-bis}%
\end{equation}
Taking the imaginary part from both sides yields the identity
\[
\frac{\mathrm{Im}(m_{N}(z))}{|1+\sigma^{2}c_{N}m_{N}(z)|^{2}}=\mathrm{Im}%
(w_{N}(z))\frac{1}{M}\mathrm{Tr}\left[  (\mathbf{B}_{N}\mathbf{B}_{N}%
^{H}-w_{N}(z)\mathbf{I}_{M})^{-1}(\mathbf{B}_{N}\mathbf{B}_{N}^{H}-w_{M}%
^{\ast}(z)\mathbf{I}_{M})^{-1}\right]
\]
or equivalently,
\begin{align}
\mathrm{Im}(m_{N}(z)) &  =\mathrm{Im}(w_{N}(z))\left\vert 1+\sigma^{2}%
c_{N}m_{N}(z)\right\vert ^{2}\frac{1}{M}\mathrm{Tr}\left[  (\mathbf{B}%
_{N}\mathbf{B}_{N}^{H}-w_{N}(z))^{-1}(\mathbf{B}_{N}\mathbf{B}_{N}^{H}%
-w_{N}^{\ast}(z))^{-1}\right]  \\
&  =\mathrm{Im}(w_{N}(z))\frac{1}{M}\mathrm{Tr}\left[  \mathbf{T}%
_{N}(z)\mathbf{T}_{N}^{H}(z)\right]
\end{align}
It is shown in \cite{dozier2007analysis} (see Eq. (2.6)) that
\[
\frac{\sigma^{2}}{N}\mathrm{Tr}\left[  \mathbf{T}_{N}(z)\mathbf{T}_{N}%
^{H}(z)\right]  \leq\frac{1}{|z|}\leq\frac{1}{x}%
\]
which implies
\begin{equation}
\mathrm{Im}(m_{N}(z))\leq\mathrm{Im}(w_{N}(z))\frac{1}{\sigma^{2}c_{N}%
|x|}.\label{eq:Imm_N-versus-Imw_N}%
\end{equation}
If $y\rightarrow0$, we get that
\[
0<\mathrm{Im}(m_{N}(x))\leq\mathrm{Im}(w_{N}(x))\frac{1}{\sigma^{2}c_{N}|x|}%
\]
which implies that $\mathrm{Im}(w_{N}(x))>0$. Conversely, assume that
$\mathrm{Im}(w_{N}(x))>0$. Then, $m_{N}(x)$ cannot be real-valued, otherwise,
$w_{N}(x)=x(1+\sigma^{2}c_{N}m_{N}(x))^{2}-\sigma^{2}(1-c_{N})(1+\sigma
^{2}c_{N}m_{N}(x))$ would be also real-valued. \newline

Next, we prove Property \ref{item:w_increasing}. Since $x\rightarrow m_{N}(x)$
is differentiable on $\mathbb{R}-\partial\mathcal{S}_{N}$, $x\rightarrow
w_{N}(x)$ is differentiable on the same subset. By Property
\ref{enu:image_w_R} of Proposition \ref{prop:m}, $w_{N}(x)$ does not belong to
the spectrum of matrix $\mathbf{B}_{N}\mathbf{B}_{N}^{H}$ if $x\in
\mathbb{R}\backslash\mathcal{S}_{N}$. Therefore, the function $x\rightarrow
f_{N}(w_{N}(x))$ is differentiable for $x\in\mathbb{R}\backslash
\mathcal{S}_{N}$. Since (\ref{eq:canonique-bis}) holds on $x\in\mathbb{R}%
\backslash\mathcal{S}_{N}$, we can differentiate it with respect to $x$ on
$x\in\mathbb{R}\backslash\mathcal{S}_{N}$. This gives
\[
w_{N}^{\prime}(x)f_{N}^{\prime}(w_{N}(x))=\frac{m_{N}^{\prime}(x)}%
{(1+c_{N}\sigma^{2}m_{N}(x))^{2}}%
\]
for $x\in\mathbb{R}\backslash\mathcal{S}_{N}$. Now, observe that
$m_{N}^{\prime}(x)>0$ on $\mathbb{R}\backslash\mathcal{S}_{N}$ because
$m_{N}(z)$ is the Stieltjès transform of a probability measure carried by
$\mathcal{S}_{N}$. On the other hand, the function $f_{N}^{\prime}$ is of
course strictly positive on $\mathbb{R}$. This in turn
shows that $w_{N}^{\prime}(x)>0$ on $x\in\mathbb{R}\backslash\mathcal{S}_{N}$.
\newline

To establish the last property, we use \eqref{eq:canonique-1} at point
$x\in\mathbb{R}\backslash\mathcal{S}_{N}$, and get that
\begin{equation}
1-c_{N}\sigma^{2}f_{N}(w(x))=\frac{1}{1+c_{N}\sigma^{2}m_{N}(x)}.
\label{eq:expre-(1-f)}%
\end{equation}
The conclusion follows from the inequality $1+c_{N}\sigma^{2}m_{N}(x)>0$ if
$x\in$ $\mathbb{R}\backslash\mathcal{S}_{N}$ (see Proposition \ref{prop:m}).
\newline

\section{Proof of (\ref{eq:extrema-croissants}) in Proposition
\ref{property:phi_extrema}}

\label{appendix:proof_inequalities_x}

We consider $w_{1},w_{2}\in\left\{  w_{1}^{(N)-},w_{1}^{(N)+},\ldots
,w_{Q}^{(N)-},w_{Q}^{(N)+}\right\}  $, and denote by $\phi_{1}$ and $\phi_{2}$
the quantities $\phi_{N}(w_{1})$ and $\phi_{N}(w_{2})$ respectively. We define
$h_{n}=1-\sigma^{2}c_{N}f_{N}(w_{n})$ so that we can write $\phi_{n}%
=w_{n}h_{n}^{2}+\sigma^{2}(1-c_{N})h_{n}$, $n\in\left\{  1,2\right\}  $. Our
objective is to show that the quantity $\left(  \phi_{2}-\phi_{1}\right)
/\left(  w_{2}-w_{1}\right)  $ is always positive. Note that, by definition,
$w_{1}$ and $w_{2}$ are inflexion points of $\phi_{N}(w)$ such that $h_{1}%
\geq0$ and $h_{2}\geq0$.

Using direct substraction of the expressions of $\phi_{1}$ and $\phi_{2}$ we
can write%
\[
\frac{\phi_{2}-\phi_{1}}{w_{2}-w_{1}}=(h_{1}+h_{2})\frac{(w_{2}h_{2}%
-w_{1}h_{1})}{w_{2}-w_{1}}+\sigma^{2}(1-c_{N})\frac{h_{2}-h_{1}}{w_{2}-w_{1}%
}-h_{1}h_{2}%
\]
Consider now the following inequality
\begin{equation}
\frac{2}{M}\sum_{k=1}^{M}\frac{\gamma_{k}^{(N)}}{(\gamma_{k}^{(N)}%
-w_{1})(\gamma_{k}^{(N)}-w_{2})}\leq\frac{1}{M}\sum_{k=1}^{M}\frac{\gamma
_{k}^{(N)}}{(\gamma_{k}^{(N)}-w_{1})^{2}}+\frac{1}{M}\sum_{k=1}^{M}%
\frac{\gamma_{k}^{(N)}}{(\gamma_{k}^{(N)}-w_{2})^{2}}
\label{eq:inequality_squares}%
\end{equation}
which can be readily obtained by noting that
\[
\frac{1}{M}\sum_{k=1}^{M}\left(  \frac{\left(  \gamma_{k}^{(N)}\right)
^{1/2}}{(\gamma_{k}^{(N)}-w_{1})}-\frac{\left(  \gamma_{k}^{(N)}\right)
^{1/2}}{(\gamma_{k}^{(N)}-w_{2})}\right)  ^{2}\geq0.
\]
Using the definition of $h_{1}$ and $h_{2}$ we can readily write
\[
\frac{w_{2}h_{2}-w_{1}h_{1}}{w_{2}-w_{1}}=1-\frac{\sigma^{2}c_{N}}{M}%
\sum_{k=1}^{M}\frac{\gamma_{k}^{(N)}}{(\gamma_{k}^{(N)}-w_{1})(\gamma
_{k}^{(N)}-w_{2})},
\]
and hence the inequality in (\ref{eq:inequality_squares}) is giving us
\begin{multline}
\frac{\phi_{2}-\phi_{1}}{w_{2}-w_{1}}\geq(h_{1}+h_{2})\left[  1-\frac
{\sigma^{2}c_{N}}{2}\left(  f_{N}(w_{1})+f_{N}(w_{2})+w_{1}f_{N}^{\prime
}(w_{1})+w_{2}f_{N}^{\prime}(w_{2})\right)  \right]  +\\
-h_{1}h_{2}+\sigma^{2}(1-c_{N})\frac{h_{2}-h_{1}}{w_{2}-w_{1}}
\label{lemma_inf_pt_tmp1}%
\end{multline}
where $f_{N}^{\prime}(w)$ denotes the derivative of $f_{N}(w)$. Using again
the definition of $h_{1}$ and $h_{2}$, we can rewrite the last term of the
previous expression as
\[
\frac{h_{2}-h_{1}}{w_{2}-w_{1}}=-\frac{\sigma^{2}c_{N}}{2}\left[
f_{N}^{\prime}(w_{1})+f_{N}^{\prime}(w_{2})-\frac{1}{M}\sum_{k=1}^{M}%
\frac{(w_{2}-w_{1})^{2}}{(\gamma_{k}^{(N)}-w_{1})^{2}(\gamma_{k}^{(N)}%
-w_{2})^{2}}\right]  .
\]
By inserting this last equality into \eqref{lemma_inf_pt_tmp1} and replacing
$f_{N}(w_{1})$ with $\sigma^{-2}(1-h_{1})$, we obtain the expression
\begin{multline}
\frac{\phi_{2}-\phi_{1}}{w_{2}-w_{1}}\geq\frac{\sigma^{4}c_{N}(1-c_{N})}%
{2}\frac{1}{M}\sum_{k=1}^{M}\frac{(w_{2}-w_{1})^{2}}{(\gamma_{k}^{(N)}%
-w_{1})^{2}(\gamma_{k}^{(N)}-w_{2})^{2}}+\frac{h_{1}^{2}+h_{2}^{2}}{2}+\\
-\frac{\sigma^{4}c_{N}(1-c_{N})}{2}\left[  f_{N}^{\prime}(w_{1})+f_{N}%
^{\prime}(w_{2})\right]  -\frac{\sigma^{2}}{2}\frac{h_{1}+h_{2}}{\left(
w_{1}f_{N}^{\prime}(w_{1})+w_{2}f_{N}^{\prime}(w_{2})\right)  }.
\label{lemma_inf_pt_tmp2}%
\end{multline}
Now, both $w_{1}$ and $w_{2}$ are preimages of local extrema of $\phi_{N}$, so
that for $n=1,2$, we have $\phi_{N}^{\prime}(w_{n})=h_{n}^{2}-2\sigma^{2}%
w_{n}f_{N}^{\prime}(w_{n})h_{n}-\sigma^{4}(1-c_{N})f_{N}^{\prime}(w_{n})=0$.
Thus, we can write
\[
\frac{h_{1}^{2}+h_{2}^{2}}{2}=\sigma^{2}\left[  w_{1}h_{1}f_{N}^{\prime}%
(w_{1})+w_{2}h_{2}f_{N}^{\prime}(w_{2})\right]  +\frac{\sigma^{4}c_{N}%
(1-c_{N})}{2}\left[  f_{N}^{\prime}(w_{1})+f_{N}^{\prime}(w_{2})\right]
\]
and by inserting the last equality into \eqref{lemma_inf_pt_tmp2}, we obtain
\begin{equation}
\frac{\phi_{2}-\phi_{1}}{w_{2}-w_{1}}\geq\frac{\sigma^{4}c_{N}(1-c_{N})}%
{2}\frac{1}{M}\sum_{k=1}^{M}\frac{(w_{2}-w_{1})^{2}}{(\gamma_{k}^{(N)}%
-w_{1})^{2}(\gamma_{k}^{(N)}-w_{2})^{2}}+\frac{\sigma^{2}}{2}(h_{1}%
-h_{2})(w_{1}f_{N}^{\prime}(w_{1})-w_{2}f_{N}^{\prime}(w_{2})).
\label{lemma_inf_pt_tmp3}%
\end{equation}
Using again the fact that $\phi_{N}^{\prime}(w_{n})=0$, we can write
$w_{n}f_{N}^{\prime}(w_{n})=\frac{h_{n}}{2\sigma^{2}}-\frac{\sigma^{2}%
(1-c_{N})}{2}\frac{f_{N}^{\prime}(w_{n})}{h_{n}}$ and thus
\eqref{lemma_inf_pt_tmp3} becomes
\begin{multline}
\frac{\phi_{2}-\phi_{1}}{w_{2}-w_{1}}\geq\frac{\sigma^{4}c_{N}(1-c_{N})}%
{2}\frac{1}{M}\sum_{k=1}^{M}\frac{(w_{2}-w_{1})^{2}}{(\gamma_{k}^{(N)}%
-w_{1})^{2}(\gamma_{k}^{(N)}-w_{2})^{2}}+\frac{(h_{1}-h_{2})^{2}}{4}+\\
\quad-\frac{\sigma^{4}(1-c_{N})}{4}\left(  f_{N}^{\prime}(w_{1})-f_{N}%
^{\prime}(w_{2})\right)  +\frac{\sigma^{4}(1-c_{N})}{4}c_{N}\left[
\frac{h_{1}}{h_{2}}f_{N}^{\prime}(w_{2})+\frac{h_{2}}{h_{1}}f_{N}^{\prime
}(w_{1})\right]  \label{lemma_inf_pt_tmp4}%
\end{multline}
Clearly, we have
\[
\frac{1}{M}\sum_{k=1}^{M}\frac{(w_{2}-w_{1})^{2}}{(\gamma_{k}^{(N)}-w_{1}%
)^{2}(\gamma_{k}^{(N)}-w_{2})^{2}}-\left[  f_{N}^{\prime}(w_{1})+f_{N}%
^{\prime}(w_{2})\right]  =-\frac{2}{M}\sum_{k=1}^{M}\frac{1}{(\gamma_{k}%
^{(N)}-w_{1})(\gamma_{k}^{(N)}-w_{2})}%
\]
and thus by multiplying the previous equality with $h_{1}h_{2}$ and adding
$h_{2}^{2}f_{N}^{\prime}(w_{1})+h_{1}^{2}f_{N}^{\prime}(w_{2})$, we can also
write%
\begin{multline*}
h_{2}^{2}f_{N}^{\prime}(w_{1})+h_{1}^{2}f_{N}^{\prime}(w_{2})+\frac{1}{M}%
\sum_{k=1}^{M}\frac{h_{1}h_{2}(w_{2}-w_{1})^{2}}{(\gamma_{k}^{(N)}-w_{1}%
)^{2}(\gamma_{k}^{(N)}-w_{2})^{2}}+\\
-h_{1}h_{2}\left[  f_{N}^{\prime}(w_{1})+f_{N}^{\prime}(w_{2})\right]
=\frac{1}{M}\sum_{k=1}^{M}\left(  \frac{h_{2}}{\gamma_{k}^{(N)}-w_{1}}%
-\frac{h_{1}}{\gamma_{k}^{(N)}-w_{2}}\right)  ^{2}.
\end{multline*}
The left hand side of the previous equality appears in
\eqref{lemma_inf_pt_tmp4} as a common factor on the last two terms of the
right hand side of that equation. Hence, plugging it into
\eqref{lemma_inf_pt_tmp4}, we obtain
\begin{multline*}
\frac{\phi_{2}-\phi_{1}}{w_{2}-w_{1}}\geq\frac{\sigma^{4}c_{N}(1-c_{N})}%
{4}\frac{1}{M}\sum_{k=1}^{M}\frac{(w_{2}-w_{1})^{2}}{(\gamma_{k}^{(N)}%
-w_{1})^{2}(\gamma_{k}^{(N)}-w_{2})^{2}}+\\
+\frac{(h_{1}-h_{2})^{2}}{4}+\frac{\sigma^{4}c_{N}(1-c_{N})}{4h_{1}h_{2}}%
\frac{1}{M}\sum_{k=1}^{M}\left(  \frac{h_{2}}{\gamma_{k}^{(N)}-w_{1}}%
-\frac{h_{1}}{\gamma_{k}^{(N)}-w_{2}}\right)  ^{2}.
\end{multline*}
Finally, noting that all the terms of the above equation are non-negative, we
have established (\ref{eq:extrema-croissants}).

\section{Proof of Lemma \ref{lemma:domination}}

\label{section:proof_lemma_domination}

The proof of this Lemma is a direct consequence of \cite[Section
4]{dozier2007analysis}. Next, we provide some details on how to obtain
(\ref{eq:domination-}); the same procedure can be applied in order to obtain
(\ref{eq:domination+}). As in \cite{dozier2007analysis}, we define in this
section function $b_{N}(z)$ by $b_{N}(z)=1+\sigma^{2}c_{N}m_{N}(z)$ for
$z\in\mathbb{C}$, and denote by $b_{1}^{-}$ the quantity $b_{N}(x_{1}^{-})$
(note that we drop the dependence on $N$ in $x_{1}^{-}$). Since $x_{1}^{-}$
belongs to $\partial\mathcal{S}_{N}$, both $m_{N}(x_{1}^{-})$ and $b_{1}^{-}$
are real-valued. Proposition \ref{prop:m} thus implies that $z\rightarrow
b_{N}(z)$ is continuous at the point $x_{1}^{-}$. Similarly, $w_{N}(x_{1}%
^{-})=w_{1}^{-}$ is real-valued so that the function $z\rightarrow w_{N}(z)$
is also continuous at $x_{1}^{-}$.

Since $f_{N}^{\prime}(w_{1}^{-})>0$, there exists a neighborhood
$\mathcal{V}(w_{1}^{-})$ of $w_{1}^{-}$ on which $f_{N}$ is biholomorphic. For
$z\in\mathbb{C}_{+}\cup\mathbb{R}$, it follows from
(\ref{eq:canonical_simplified}) that we can write%
\begin{equation}
f_{N}(w_{N}(z))=\frac{m_{N}(z)}{1+\sigma^{2}c_{N}m_{N}(z)}=\frac{1}{\sigma
^{2}c_{N}}\left(  1-\frac{1}{b_{N}(z)}\right)  .\label{equation:f_at_w_z}%
\end{equation}
Since $w_{N}$ is continuous at $x_{1}^{-}$ and since $w_{N}(z)\in
\mathbb{C}_{+}$ if $z\in\mathbb{C}_{+}$ (see Property \ref{enu:Imw_positive}
of Proposition \ref{prop:m}), there exists a neighborhood $\mathcal{V}%
(x_{1}^{-})$ of $x_{1}^{-}$ such that
\[
w_{N}\left(  \mathcal{V}(x_{1}^{-})\cap\mathbb{C}_{+}\right)  \subset
\mathcal{V}(w_{1}^{-})\cap\mathbb{C}_{+}.
\]
Therefore, applying the holomorphic inverse of $f_{N}$, denoted as$\ f_{N}%
^{-1}$, to both sides of (\ref{equation:f_at_w_z}) we get, for any
$z\in\mathcal{V}(x_{1}^{-})\cap\mathbb{C}_{+}$,
\[
w_{N}(z)=f_{N}^{-1}\left(  \frac{1}{\sigma^{2}c_{N}}\left(  1-\frac{1}%
{b_{N}(z)}\right)  \right)  .
\]
Using the fact that $w_{N}(z)=zb_{N}^{2}(z)-\sigma^{2}(1-c_{N})b_{N}(z)$ and
solving with respect to $z$, we get that
\begin{equation}
z=Z_{N}\left(  b_{N}(z)\right)  \qquad z\in\mathcal{V}(x_{1}^{-}%
)\cap\mathbb{C}_{+}\label{eq:Z(b(z))=z}%
\end{equation}
where $Z_{N}$ is the function defined in an appropriate neighborhood of
$b_{1}^{-}$ by
\[
Z_{N}(b)=\frac{1}{b^{2}}f_{N}^{-1}\left(  \frac{1}{\sigma^{2}c_{N}}\left(
1-\frac{1}{b}\right)  \right)  +\frac{\sigma^{2}(1-c_{N})}{b}.
\]
Next, we recall the following result from \cite{dozier2007analysis}.

\begin{lemma}
\label{lemma:Psi}There exists a neighborhood $\mathcal{V}(b_{1}^{-})$ of
$b_{1}^{-}$ and a function $\Psi_{N}$, biholomorphic from $\mathcal{V}%
(b_{1}^{-})$ onto a neighborhood of the origin $\mathcal{V}(0)$ such that
$\forall b\in\mathcal{V}(b_{1}^{-})$
\[
Z_{N}(b)-x_{1}^{-}=\Psi_{N}^{2}(b).
\]

\end{lemma}

Since the function $b_{N}$ is continuous at the point $x_{1}^{-}$, and since
$b_{N}(z)\in\mathbb{C}_{+}$ if $z\in\mathbb{C}_{+}$ (which follows from the
definition of $b_{N}$), there exist two smaller neighborhoods $\mathcal{V}%
^{\prime}(x_{1}^{-})\subset\mathcal{V}(x_{1}^{-})$ and $\mathcal{V}^{\prime
}(b_{1}^{-})\subset\mathcal{V}(b_{1}^{-})$ of $x_{1}^{-}$ and $b_{1}^{-}$
respectively, such that
\[
b_{N}(z)\in\mathcal{V}^{\prime}(b_{1}^{-})\cap\mathbb{C}_{+}\qquad\forall
z\in\mathcal{V}^{\prime}\left(  x_{1}^{-}\right)  \cap\mathbb{C}_{+}%
\]
Therefore, using (\ref{eq:Z(b(z))=z}), we can write
\[
\left(  \Psi_{N}\left(  b_{N}(z)\right)  \right)  ^{2}=z-x_{1}^{-}%
\]
$\forall z\in\mathcal{V}^{\prime}(x_{1}^{-})\cap\mathbb{C}_{+}$. Let us now
choose, $\forall z\in\mathcal{V}^{\prime}(x_{1}^{-})\cap\mathbb{C}_{+}$,
\[
\Psi_{N}\left(  b_{N}(z)\right)  =\sqrt{z-x_{1}^{-}}%
\]
where $\sqrt{\left(  \text{\textperiodcentered}\right)  }$ represents any
determination of the complex square root that is holomorphic\footnote{This
property must hold for all possible choices of $\Psi_{N}$ because, by
definition, $\Psi_{N}$ is holomorphic on $\mathcal{V}(b_{1}^{-})$ and
$b_{N}(z)\in\mathbb{C}_{+}$ if $z\in\mathbb{C}_{+}$. Since $b_{N}(z)$ is
holomorphic on $\mathcal{V}^{\prime}(x_{1}^{-})\cap\mathbb{C}_{+}$, $\Psi
_{N}\left(  b_{N}(z)\right)  $ must be holomorphic on the same set.} on
$\mathbb{C}_{+}$ and such that $\sqrt{1}=1$ (the following reasoning applies
verbatim to the square root determination for which $\sqrt{1}=-1$). We denote
by $\Psi_{N}^{-1}$ the holomorphic inverse function of $\Psi_{N}$ defined
on $\mathcal{V}(0)$. We have
\[
b_{N}(z)=\Psi_{N}^{-1}\left(  \sqrt{z-x_{1}^{-}}\right)  \quad\forall
z\in\mathcal{V}^{\prime}(x_{1}^{-})\cap\mathbb{C}_{+}%
\]
Taking derivatives with respect to $z$ at both sides of the previous equality,
we obtain
\[
b_{N}^{\prime}(z)=\frac{1}{2\sqrt{z-x_{1}^{-}}}\left[  \Psi_{N}^{-1}\right]
^{\prime}\left(  \sqrt{z-x_{1}^{-}}\right)  .
\]
Now, since $\Psi_{N}^{-1}$ is holomorphic on $\mathcal{V}(0)$ by Lemma
\ref{lemma:Psi}, the function\ $\left[  \Psi_{N}^{-1}\right]  ^{\prime}$ will
be bounded on the same neighborhood of $0$ and thus we will have%
\[
\left\vert b_{N}^{\prime}(z)\right\vert \leq\frac{C}{\left\vert \sqrt
{z-x_{1}^{-}}\right\vert }%
\]
for some constant $C$ independent of $z$. Therefore, for $z=x+\mathrm{i}%
y\in\mathcal{V}^{\prime}(x_{1}^{-})\cap\mathbb{C}_{+}$, we can write%
\begin{equation}
\left\vert w_{N}^{\prime}(x+\mathrm{i}y)\right\vert =\left\vert b_{N}%
(z)^{2}+2zb_{N}^{\prime}(z)-\sigma^{2}(1-c_{N})b_{N}^{\prime}(z)\right\vert
\leq\frac{C}{\sqrt{\left\vert x-x_{1}^{-}+\mathrm{i}y\right\vert }}.
\end{equation}
The inequality
\[
\frac{C}{\sqrt{\left\vert x-x_{1}^{-}+\mathrm{i}y\right\vert }}\leq\frac
{C}{\sqrt{\left\vert x-x_{1}^{-}\right\vert }}%
\]
for $x\neq x_{1}^{-}$ completes the proof of (\ref{eq:domination-}) for $y>0$.
(\ref{eq:domination-}) for $y=0$ follows from the observation that
$w_{N}^{\prime}(x)=\lim_{y\downarrow0}w_{N}^{\prime}(x+\mathrm{i}y)$.

\section{Proof of Proposition \ref{proposition:decomposition}}

\label{appendix:proof_polynomial_bounding}%

In this section, we drop as much as possible the subscript $N$  for an easier reading.
In the following, $\mathrm{P}_1(|z|)$ and $\mathrm{P}_2(\frac{1}{|\mathrm{Im}(z)|})$ represent generic positive coefficients polynomials of
the variables $|z|$ and $\frac{1}{|\mathrm{Im}(z)|}$ whose mean feature is to be independent of $N$. The values
of $\mathrm{P}_1$ and $\mathrm{P}_2$ can change from one line to another. \\

We rely extensively on the results of the Appendix II of  \cite{dumont2009capacity} related to the properties 
of matrix $({\bf B} + {\bf D}^{1/2} {\bf W} \tilde{{\bf D}}^{1/2})({\bf B} + {\bf D}^{1/2} {\bf W} \tilde{{\bf D}}^{1/2})^{H}$
where ${\bf D}$ and $\tilde{{\bf D}}$ are deterministic diagonal matrix. We thus use \cite{dumont2009capacity} 
in the case where ${\bf D} = \sigma {\bf I}_M$ and $\tilde{{\bf D}} = \sigma {\bf I}_N$ which corresponds to 
the context of the present paper. In order to help the
reader, we use the same notations as in \cite{dumont2009capacity} all along this section. More precisely, we define
\begin{align}
        \delta(z) &= \sigma c m(z)
        \label{def:delta}
        \\
	\tilde{\delta}(z) &= \delta(z) - \sigma \frac{1-c}{z}
        \label{def:tildedelta}
	\\
        \alpha(z) &= \mathbb{E}\left[\frac{\sigma}{N}\Tr \Q(z)\right] 
        \label{def:alpha}
        \\
        \tilde{\alpha}(z) &= \alpha(z) - \sigma \frac{1-c}{z}
	\label{def:tildealpha}
\end{align}
We remark that $\alpha(z)$ is the Stieltjès transform of measure $c \sigma \omega$ 
where $\omega$ is the probability measure carried by $\mathbb{R}_{+}$ defined by 
\begin{equation}
\label{eq:def-omega}
\omega({\cal B}) = \mathbb{E}(\hat{\mu}({\cal B}))
\end{equation}
for each Borel set ${\cal B}$. We recall that $\hat{\mu}$ represents the empirical eigenvalue distribution 
of $\hat{{\bf R}}_N = \bs{\Sigma}_N \bs{\Sigma}_N^{H}$. 
Finally, it is easily seen that $\tilde{\delta}$ is the Stieltjès transform of measure $\sigma c \mu + \sigma (1-c) \delta_0$  ($\delta_0$ 
represents the Dirac distribution at 0), and that $\tilde{\alpha}(z)$, which can be expressed by 
\begin{equation}
\label{eq:expre-tildealpha}
\tilde{\alpha}(z) = \mathbb{E} \left[ \sigma \frac{1}{N} \mathrm{Tr} \tilde{{\bf Q}}(z) \right]
\end{equation}
where $\tilde{{\bf Q}}(z)$ is defined by 
\begin{equation}
\label{eq:def-tildeQ}
\tilde{{\bf Q}}(z) = \left( {\bs \Sigma}^{H}  {\bs \Sigma} - z {\bf I} \right)^{-1}
\end{equation}
coincides with the Stieltjès transform of measure 
$\sigma c  \omega + \sigma (1 -c ) \delta_0$. 

Matrix ${\bf T}(z)$ defined by (\ref{eq:def-T}) can be written as 
\[
{\bf T}(z) = \left[ -z(1+\sigma \tilde{\delta}(z)) {\bf I}_M + \frac{{\bf B}{\bf B}^{H}}{1+\sigma \delta(z)} \right]^{-1}
\]
and $\delta(z)$ is equal to 
\begin{equation}
\label{eq:canonique-delta}
\delta(z) = \sigma \frac{1}{N} \mathrm{Tr} {\bf T}(z)
\end{equation}
We also define matrix $\tilde{{\bf T}}(z)$ by 
\begin{equation}
\label{eq:def-tildeT}
\tilde{{\bf T}}(z) = \left[ -z(1+\sigma \delta(z)) {\bf I}_N + \frac{{\bf B}^{H}{\bf B}}{1+\sigma \tilde{\delta}(z)} \right]^{-1}
\end{equation}
and remark, after simple calculations, that
\begin{equation}
\label{eq:canonique-tildedelta}
\tilde{\delta}(z) = \sigma \frac{1}{N} \mathrm{Tr} \tilde{{\bf T}}(z)
\end{equation}
We finally denote by ${\bf R}(z)$ and $\tilde{{\bf R}}(z)$ the matrices defined by 
\begin{align}
\label{def:R}
{\bf R}(z) & =  \left[ -z(1+\sigma \tilde{\alpha}(z)) {\bf I}_M + \frac{{\bf B}{\bf B}^{H}}{1+\sigma \alpha(z)} \right]^{-1} \\
\label{def:tildeR}
\tilde{{\bf R}}(z) & =  \left[ -z(1+\sigma \alpha(z)) {\bf I}_N  + \frac{{\bf B}^{H}{\bf B}}{1+\sigma \tilde{\alpha}(z)} \right]^{-1}
\end{align}
Using Property \ref{enu:converse-stieljes} of Lemma \ref{property:stieltjes}, it is easily checked that 
functions $\left(-z(1+\sigma \delta(z))\right)^{-1}$, $\left(-z(1+\sigma \tilde{\delta}(z))\right)^{-1}$, 
$\left(-z(1+\sigma \alpha(z))\right)^{-1}$, $\left(-z(1+\sigma \tilde{\alpha}(z))\right)^{-1}$ 
are Stieltjès transforms of probability measures carried by $\mathbb{R}_{+}$. 
Proposition 5.1 of \cite{hachem2007deterministic} thus implies that matrix valued functions 
${\bf T}(z), \tilde{{\bf T}}(z), {\bf R}(z), \tilde{{\bf R}}(z)$ are holomorphic 
in $\mathbb{C}-\mathbb{R}_{+}$, coincide with the Stieltjès transforms of positive matrix valued measures 
carried by $\mathbb{R}_{+}$, the mass of which are equal to ${\bf I}$, and their spectral norms are bounded by
$\frac{1}{|\mathrm{Im}(z)|}$ on $\mathbb{C}_{+}$ (see \cite{hachem2007deterministic} for more details).  

We finally recall that matrices ${\bf Q}(z)$ and $\tilde{{\bf Q}}(z)$ satisfy $\| {\bf Q} \| \leq (\mathrm{Im}(z))^{-1}$ and 
 $\| \tilde{{\bf Q}} \| \leq (\mathrm{Im}(z))^{-1}$ for $z \in \mathbb{C}_{+}$ (see e.g. 
\cite{silverstein95, haagerup2005new, capitaine2009largest, hachem2007deterministic}).  \\

In order to establish Proposition \ref{proposition:decomposition}, we have first to study the term 
$$
\mathbb{E} \left( \frac{1}{N} \Tr {\bf Q}(z) \right) - \frac{1}{N} \Tr {\bf R}(z)
$$ 
\subsection{Study of $\mathbb{E} \left( \frac{1}{N} \Tr {\bf Q}(z) \right)
- \frac{1}{N} \Tr {\bf R}(z)$}

Let $\tilde{\tau}(z)$ and $\bs{\Delta}(z)$ defined by
\begin{align}
        \tilde{\tau}(z) = 
        \frac{- \sigma}{z \left(1 + \sigma \alpha(z)\right)} \left[1 - \frac{1}{N}\Tr\left(\frac{\B^H \mathbb{E}[\Q(z)] \B}{1 + \sigma \alpha (z)} \right) \right] 
        \label{def:tau_t}
\end{align}
and
\begin{align}
        \bs{\Delta}(z) &:= \bs{\Delta}_1(z) + \bs{\Delta}_2(z) + \bs{\Delta}_3(z)
        \label{def:Delta}
        \\
        \bs{\Delta}_1(z) &:= 
        - \frac{\sigma}{1+\sigma \alpha(z)}
        \mathbb{E}\left[\Q(z)\bs{\bs{\Sigma}} \bs{\bs{\Sigma}}^H \frac{\sigma}{N} \mathrm{Tr} \left(\Q(z) - \mathbb{E}[\Q(z)]\right)\right]
        \label{def:delta1}
        \\
        \bs{\Delta}_2(z) &:= 
        - \frac{\sigma^2}{1+\sigma \alpha(z)}\mathbb{E}\left[\left(\Q(z)-\mathbb{E}[\Q(z)]\right) \frac{\sigma}{N} \Tr \bs{\Sigma}^H \Q(z) \B\right] 
        \label{def:delta2}
        \\
        \bs{\Delta}_3(z) &:= 
        \frac{\sigma^2}{(1+\sigma \alpha(z))^2}\mathbb{E}\left[\Q(z)\right]
        \mathbb{E}\left[ \frac{\sigma}{N} \Tr \left(\Q(z)-\mathbb{E}[\Q(z)]\right) \frac{\sigma}{N} \Tr \bs{\Sigma}^H \Q(z) \B\right]
        \label{def:delta3}
\end{align}
As it will become apparent below, the entries of matrix ${\bs \Delta}(z)$ converge towards 0. 

It is proved in \cite{dumont2009capacity} that for each $z \in \mathbb{R}_{-}^{*}$, 
the following equality holds true
\begin{equation}
\label{eq:Delta_Q_1}
 \I_M + \bs{\Delta}(z) =
                \mathbb{E}\left[\Q(z)\right]\left(-z (1 + \sigma \tilde{\tau}(z))\I_M + \frac{\B\B^H}{1 + \sigma \alpha(z)} \right)
\end{equation}
As the lefthandside and the righthandside of (\ref{eq:Delta_Q_1})
are analytic on $\mathbb{C} - \mathbb{R}_{+}$, Eq.  (\ref{eq:Delta_Q_1}) 
holds not only on  $\mathbb{R}^{-}_{*}$, but on $\mathbb{C} - \mathbb{R}_{+}$. It is shown in 
 \cite{dumont2009capacity} that $\tilde{\alpha}(z) - \tilde{\tau}(z)$ converges towards 0  
for each $z \in \mathbb{C} - \mathbb{R}_{+}$ when $N \rightarrow +\infty$. 
The general  expression of $\tilde{\alpha}(z) - \tilde{\tau}(z)$ given in 
\cite{dumont2009capacity} is complicated. However, the simplicity of the model considered in 
this paper (matrices ${\bf D}$ and $\tilde{{\bf D}}$ in \cite{dumont2009capacity}
are reduced to $\sigma {\bf I}$) allows to derive the following Lemma. 
\begin{lemma}
\label{le:expression-tildealpha-tildetau}
For each $z \in \mathbb{C}-\mathbb{R}_{+}$, it holds that
\begin{equation}
\label{eq:expression-tildealpha-tildetau}
z \left( \tilde{\alpha}(z) - \tilde{\tau}(z) \right) = - \, \sigma \frac{1}{N} \mathrm{Tr} \bs{\Delta}(z)
\end{equation}
\end{lemma}
\begin{IEEEproof}
Multiplying \eqref{eq:Delta_Q_1} from both sides by $\sigma$ and taking the trace, we obtain
\begin{align}
        \sigma \, \frac{1}{N} \Tr\left(\frac{\B^H \mathbb{E}[\Q(z)]\B}{1 + \sigma \alpha(z)}\right)
        &=
        \sigma \, \frac{M}{N} + \sigma \, \frac{1}{N}\Tr \bs{\Delta}(z) + z \left( 1 + \sigma \tilde{\tau}(z)\right) \, \alpha(z)
\end{align}
From the definition of $\tilde{\tau}(z)$ (equation \eqref{def:tau_t}), we also have
\begin{align}
        \sigma \, \frac{1}{N} \Tr\left(\frac{\B^H \mathbb{E}[\Q(z)]\B}{1 + \sigma \alpha(z)}\right)
        &=
        z \tilde{\tau}(z)(1 + \sigma \alpha(z)) + \sigma
\end{align}
The two above equalities imply that 
\begin{align}
        \alpha(z) - \tilde{\tau}(z) = \frac{\sigma (1-c)}{z} - \frac{\sigma}{z} \frac{1}{N} \Tr \bs{\Delta}(z)
\end{align}
Using (\ref{def:tildealpha}), we get that
\begin{align}
        \tilde{\alpha}(z) - \tilde{\tau}(z)  = - \frac{\sigma}{z} \frac{1}{N} \Tr \bs{\Delta}(z)
\end{align}
and (\ref{eq:expression-tildealpha-tildetau}). 
\end{IEEEproof}
Writing the righthandside of (\ref{eq:Delta_Q_1}) as
$$
\mathbb{E}({\bf Q}(z)) {\bf R}(z)^{-1} +  z \sigma (\tilde{\alpha}(z) - \tilde{\tau}(z)) \mathbb{E}({\bf Q}(z))
$$
and using  (\ref{eq:expression-tildealpha-tildetau}), we obtain immediately that 
\begin{equation}
\label{eq:expre-E(Q)-R}
\mathbb{E}({\bf Q}(z)) - {\bf R}(z) =  \bs{\Delta}(z)  {\bf R}(z) + \sigma^{2} \frac{1}{N} \left[ \Tr \bs{\Delta}(z) \right]
\mathbb{E}({\bf Q}(z)) {\bf R}(z)
\end{equation}
and that
\begin{align}
\label{eq:forme-utile}
        \mathbb{E}\left[\frac{1}{N} \Tr\Q(z)\right] - \frac{1}{N} \Tr\R(z) =
        \frac{\sigma}{N}\Tr\left(\mathbb{E}\left[\Q(z)\right]\R(z)\right) \frac{\sigma}{N}\Tr\bs{\Delta}(z)
        + \frac{1}{N}\Tr\bs{\Delta}(z) \R(z)
\end{align}
The above expression of $ \mathbb{E}\left[\frac{1}{N} \Tr\Q(z)\right] - \frac{1}{N} \Tr\R(z)$ allows to prove
the following Proposition. 
\begin{proposition}
        \label{prop:Q-R}
        $\forall z \in \mathbb{C}_{+}$, we have
        \begin{align}
        \label{eq:majorationT-R}
                \left| \mathbb{E}\left[\frac{1}{N} \Tr\Q(z)\right] - \frac{1}{N} \Tr\R(z) \right| 
                \leq 
                \frac{1}{N^2} \mathrm{P}_1(|z)) \mathrm{P}_2( |\Im(z)|^{-1})
        \end{align}
\end{proposition}
\begin{IEEEproof}
We first prove the following preliminary result. 
\begin{lemma}
        \label{lemma_variances}
        Consider $M \times M$ matrices  $ \U_N$ and $M \times N$ matrices $\U_N^{'}$ 
satisfying  $\sup_{N} \| {\bf U}_N \| < \infty, \sup_{N} \| {\bf U}_N \| < \infty$. Then, we have
        $\forall z \in \mathbb{C}_{+}$
        \begin{align}
                \mathrm{Var}\left[\frac{1}{N} \Tr \Q(z) \U\right] &\leq C \|\U\|^2 \frac{1}{N^2} \mathrm{P}_1(|z|) \mathrm{P}_2(\frac{1}{|\mathrm{Im}(z)}|)\\
                \mathrm{Var}\left[\frac{1}{N} \Tr \bs{\bs{\Sigma}}^H \Q(z) \U^{'} \right] 
                &\leq C \frac{1}{N^2} \|\U^{'}\|^2  \mathrm{P}_1(|z|) \mathrm{P}_2(\frac{1}{|\mathrm{Im}(z)|})
        \end{align}     
        where the polynomials $\mathrm{P}_1$ and $\mathrm{P}_2$ and constant $C$ are independent of $M,N$ and $\U,\U^{'}$.
\end{lemma}
\begin{proof}
        As the proofs of the two statements are similar, we just prove the first statement of the Lemma. We first remark that
\begin{align}
\frac{\partial [\Q(z)]_{pq}}{\partial \W_{ij}} & = -  {\bf Q}_{pi} \left( \bs{\Sigma}^{H} {\bf Q} \right)_{jq} \\
\frac{\partial [\Q(z)]_{pq}}{\partial \W_{ij}^{*}} & = -  {\bf Q}_{iq}  \left( {\bf Q} \bs{\Sigma} \right)_{pj}
\end{align}
The Nash-Poincaré inequality gives
        \begin{align}
                \mathrm{Var}\left[\frac{1}{N} \Tr \Q(z) \U\right] 
                        &\leq \frac{\sigma^2}{N} \sum_{i,j}
                        \left[
                                \mathbb{E} \left|\frac{1}{N} \sum_{p,q} \frac{\partial [\Q(z)]_{pq}}{\partial \W_{ij}}\U_{qp}\right|^2
                                +
                                \mathbb{E} \left|\frac{1}{N} \sum_{p,q} \frac{\partial [\Q(z)]_{pq}}{\partial \W_{ij}^{*}}\U_{qp}\right|^2
                        \right]
                        \\
                        &\leq C \frac{1}{N^3} \sum_{i,j}
                        \left[
                                \mathbb{E} \left|\left[\bs{\bs{\Sigma}}^H \Q(z) \U\Q(z)\right]_{ji} \right|^2
                                +
                                \mathbb{E} \left|\left[\Q(z)\U\Q(z)\bs{\bs{\Sigma}}^H\right]_{ij}\right|^2
                        \right]
                        \\
                        &\leq C \frac{1}{N^3} \sum_{j}
                        \mathbb{E}
                        \left[ 
                                \left(\bs{\bs{\Sigma}}^H \Q(z) \U\Q(z) \Q(z)^H \U^H \Q(z)^H \bs{\bs{\Sigma}}\right)_{jj}  
                        \right]
                       + \\
                       & C \frac{1}{N^3} \sum_{j} \mathbb{E}
                        \left[ 
                                \left(\bs{\bs{\Sigma}}^H \Q(z)^H \U^H \Q(z)^H \Q(z) \U \Q(z) \bs{\bs{\Sigma}}\right)_{jj}
                        \right]
                        \\
                        &\leq C \frac{1}{N^3}
                               \mathbb{E}
                                \left[
                                \Tr 
                                \left(
                                        \Q(z) \U\Q(z) \Q(z)^H \U^H \Q(z)^H \bs{\bs{\Sigma}} \bs{\bs{\Sigma}}^H
                                \right)
                                \right]
                                +
                         \\
                          &    C \frac{1}{N^3}   \mathbb{E}
                                \left[ 
                                        \Tr 
                                        \left(
                                        \Q(z)^H \U^H \Q(z)^H \Q(z) \U \Q(z) \bs{\bs{\Sigma}} \bs{\bs{\Sigma}}^H
                                        \right)
                                \right]                    
         \end{align}
We use the resolvent identity 
\begin{equation}
\label{eq:resolvente}
{\bf Q}(z)  \bs{\bs{\Sigma}} \bs{\bs{\Sigma}}^H =  \bs{\bs{\Sigma}} \bs{\bs{\Sigma}}^H {\bf Q}(z) =  {\bf I} + z {\bf Q}(z)
\end{equation}
Therefore, 
         \begin{align}
          \mathrm{Var}\left[\frac{1}{N} \Tr \Q(z) \U\right]
                        &\leq C \frac{1}{N^3}
                                \mathbb{E}
                                \left|
                                \Tr \left(\Q(z) \U\Q(z) \Q(z)^H \U^H \left(\I + z^{*} \Q(z)^H \right)\right)
                                \right|
                                + \\
                           &    C \frac{1}{N^3}  \mathbb{E}
                                \left|
                                \Tr \left(\Q(z)^H \U^H \Q(z)^H \Q(z) \U \left(\I + z \Q(z) \right) \right)
                                \right|
                        \\
                        &\leq C \|\U\|^2\frac{1}{N^2} \left(\frac{|z|}{|\Im(z)|^4} + \frac{1}{|\Im(z)|^3} \right)
                        \\
                        & \leq C \|\U\|^2\frac{1}{N^2} \left(|z|+1 \right) \left(\frac{1}{|\mathrm{Im}(z)|^{4}} + \frac{1}{|\mathrm{Im}(z)|^{3}} \right)
        \end{align}
which establishes the first statement of Lemma \ref{lemma_variances}. 

\end{proof}
We now complete the proof of Proposition \ref{prop:Q-R}. For this, we 
use the inequalities $\| {\bf Q}(z) \| \leq \frac{1}{|\mathrm{Im}(z)|}$ and  $\| {\bf R}(z) \| \leq \frac{1}{|\mathrm{Im}(z)|}$
for $z \in \mathbb{C} - \mathbb{R}$. This leads to  
\begin{equation}
\label{eq:premier-terme}
        \left|\frac{\sigma}{N}\Tr\left(\mathbb{E}\left[\Q(z)\right]\R(z)\right) \frac{\sigma}{N}\Tr\bs{\Delta}(z)\right|
        \leq
        C \frac{1}{|\Im(z)|^2} 
        \left|
                \frac{1}{N}\Tr \bs{\Delta}_1(z) + \frac{1}{N}\Tr \bs{\Delta}_2(z) + \frac{1}{N}\Tr \bs{\Delta}_3(z)
        \right|
\end{equation}
We establish that 
\begin{equation}
\label{eq:inegalite-tracedelta}
\left| \frac{1}{N} \mathrm{Tr}({\bs \Delta}_i(z)) \right| \leq \frac{1}{N^{2}} \mathrm{P}_1(|z|) \mathrm{P}_2(|\mathrm{Im}(z)|^{-1})
\end{equation}
for $i=1,2,3$. 
In order to evaluate $\frac{1}{N} \mathrm{Tr}({\bs \Delta}_i(z))$ for $i=1,2,3$, we first remark that
\[
\frac{1}{|z(1 + \sigma \alpha(z))|} < \frac{1}{|\mathrm{Im}(z)|}
\] 
because $-\frac{1}{z(1 + \sigma \alpha(z))}$ is the Stieltjès transform of a probability measure. Therefore, we have
\begin{equation}
\label{eq:majoration-denominateur}
\frac{1}{|1 + \sigma \alpha(z)|} < \frac{|z|}{|\mathrm{Im}(z)|}
\end{equation}
The resolvent identity (\ref{eq:resolvente}) implies that 
\begin{align}
\frac{1}{N} \mathrm{Tr}({\bs \Delta}_1(z)) & =  -\frac{\sigma}{1+\sigma \alpha(z)} \mathbb{E} \left[ z \frac{1}{N} \mathrm{Tr}{\bf Q}(z) \, \frac{\sigma}{N}  \mathrm{Tr}\left( {\bf Q}(z) - \mathbb{E}{\bf Q}(z) \right) \right] \\
& =  -\frac{\sigma}{1+\sigma \alpha(z)} \mathbb{E} \left[ \frac{z}{N}  \left(\mathrm{Tr}\left({\bf Q}(z) - \mathbb{E}{\bf Q}(z)\right) \right) \, 
\frac{\sigma}{N}  \left( \mathrm{Tr}\left( {\bf Q}(z) - \mathbb{E}{\bf Q}(z) \right) \right) \right]
\end{align}
(\ref{eq:majoration-denominateur}) and the first statement of Lemma \ref{lemma_variances} give immediately (\ref{eq:inegalite-tracedelta}) for $i=1$. 
Similarly, $\frac{1}{N} \mathrm{Tr}({\bs \Delta}_2(z))$ can be written as 
$$
\frac{1}{N} \mathrm{Tr}({\bs \Delta}_2(z)) =  -\frac{\sigma^{2}}{1+\sigma \alpha(z)} \mathbb{E}\left[  \left(\frac{1}{N} \mathrm{Tr}{\bf Q}(z) - \mathbb{E}( \frac{1}{N} \mathrm{Tr}{\bf Q}(z) \right) \, 
\left(  \frac{\sigma}{N}  \mathrm{Tr} {\bs \Sigma}^{H} {\bf Q}(z) {\bf B} - \mathbb{E}(  \frac{\sigma}{N}  \mathrm{Tr} {\bs \Sigma}^{H} {\bf Q}(z) {\bf B}) \right) \right]
$$
Using again (\ref{eq:majoration-denominateur}), the Schwartz inequality, Lemma \ref{lemma_variances}, and the identity $(xy)^{1/2} \leq (\frac{x+y}{2})$ for $x \geq 0, y \geq 0$, 
we get  (\ref{eq:inegalite-tracedelta}) for $i=2$.  (\ref{eq:inegalite-tracedelta}) for $i=3$ is obtained similarly. This and (\ref{eq:premier-terme})
imply that 
$$
 \left|\frac{\sigma}{N}\Tr\left(\mathbb{E}\left[\Q(z)\right]\R(z)\right) \frac{\sigma}{N}\Tr\bs{\Delta}(z)\right|
        \leq  \frac{1}{N^{2}} \mathrm{P}_1(|z|) \mathrm{P}_2(|\mathrm{Im}(z)|^{-1})
$$
Using the same approach and the identity $\| {\bf R}(z) \| \leq (|\mathrm{Im}(z)|)^{-1}$, we obtain easily that
$$
        \left| \frac{1}{N}\Tr \bs{\Delta}(z) \R(z) \right|
        \leq
        \frac{1}{N^2} \,  \mathrm{P}_1(|z|) \mathrm{P}_2(|\mathrm{Im}(z)|^{-1})
$$     
(\ref{eq:forme-utile}) thus implies Proposition \ref{prop:Q-R}. 

\end{IEEEproof}

\begin{remark}
\label{rem:dualite}
It is also possible to establish that 
 $\forall z \in \mathbb{C}_{+}$, we have
        \begin{align}
        \label{eq:majorationtildeT-tildeR}
                \left| \mathbb{E}\left[\frac{1}{N} \Tr\tilde{\Q}(z)\right] - \frac{1}{N} \Tr\tilde{\R}(z) \right| 
                \leq 
                \frac{1}{N^2} \mathrm{P}_1(|z|) \mathrm{P}_2( |\Im(z)|^{-1})
        \end{align}
because it is shown in \cite{dumont2009capacity} that a relation similar to (\ref{eq:Delta_Q_1}) 
holds for $\mathbb{E}(\tilde{\bf Q}(z))$.  Following the derivation of (\ref{eq:expre-E(Q)-R}), we obtain an expression of  $\mathbb{E}\left[\frac{1}{N} \Tr\tilde{\Q}(z)\right] - \frac{1}{N} \Tr\tilde{\R}(z)$ similar to (\ref{eq:forme-utile}) which allows to establish (\ref{eq:majorationtildeT-tildeR}).
\end{remark}

\subsection{Study of $\mathbb{E}\left( \frac{1}{N} \mathrm{Tr} {\bf Q}(z) \right) - \frac{1}{N} \mathrm{Tr} {\bf T}(z)$}
In order to complete the proof of Proposition \ref{proposition:decomposition}, we show in this paragraph that 
\begin{equation}
\label{eq:majoration-biais}
\sigma \left|\mathbb{E}\left( \frac{1}{N} \mathrm{Tr} {\bf Q}(z) \right) - \frac{1}{N} \mathrm{Tr} {\bf T}(z) \right| =  |\alpha(z) - \delta(z)| \leq  \frac{1}{N^2} \,  \mathrm{P}_1(|z|) \mathrm{P}_2(|\mathrm{Im}(z)|^{-1})
\end{equation}
for each $z \in \mathbb{C}_{+}$. For this, we denote by $\epsilon(z)$ and 
$\tilde{\epsilon}(z)$ the terms defined by
\begin{align}
\epsilon(z) &= \alpha(z) - \sigma \frac{1}{N} \mathrm{Tr}({\bf R}(z)) = \sigma \left( \mathbb{E} \frac{1}{N} \mathrm{Tr}({\bf Q}(z)) -  \frac{1}{N} \mathrm{Tr}({\bf R}(z)) \right)\\
\tilde{\epsilon}(z) &= \tilde{\alpha}(z) - \sigma \frac{1}{N} \mathrm{Tr}(\tilde{{\bf R}}(z)) =  \sigma \left( \mathbb{E} \frac{1}{N} \mathrm{Tr}(\tilde{{\bf Q}}(z)) -  \frac{1}{N} \mathrm{Tr}(\tilde{{\bf R}}(z)) \right)
\end{align}
Proposition \ref{prop:Q-R} and Remark \ref{rem:dualite} imply that 
\begin{align}
\label{eq:majoration-epsilon}
|\epsilon(z)| & \leq \frac{1}{N^2} \,  \mathrm{P}_1(|z|) \mathrm{P}_2(|\mathrm{Im}(z)|^{-1})\\
\label{eq:majoration-tildeepsilon}
|\tilde{\epsilon}(z)| &  \leq \frac{1}{N^2} \,  \mathrm{P}_1(|z|) \mathrm{P}_2(|\mathrm{Im}(z)|^{-1})
\end{align}
for each $z \in \mathbb{C}_{+}$. In order to study $\alpha(z) - \delta(z)$, we express 
$\alpha(z)$ as $\alpha(z) = \sigma \frac{1}{N} \mathrm{Tr}({\bf R}(z)) + \epsilon(z)$. Therefore, 
$\alpha(z) - \delta(z) = \sigma \frac{1}{N} \mathrm{Tr}({\bf R}(z) - {\bf T}(z)) + \epsilon(z)$. 
We have similarly $\tilde{\alpha}(z) - \tilde{\delta}(z) = \sigma \frac{1}{N} \mathrm{Tr}(\tilde{{\bf R}}(z) - \tilde{{\bf T}}(z)) + \tilde{\epsilon}(z)$. We remark that ${\bf R}(z) - {\bf T}(z)$ can be written as ${\bf R}(z) \left( {\bf T}^{-1}(z) - {\bf R}^{-1}(z) \right) {\bf T}(z)$, and that $\tilde{{\bf R}}(z) - \tilde{{\bf T}}(z)$
is equal $\tilde{{\bf R}}(z) \left( \tilde{{\bf T}}^{-1}(z) - \tilde{{\bf R}}^{-1}(z) \right) \tilde{{\bf T}}(z)$.
Using the expression of ${\bf R}(z)^{-1}, {\bf T}(z)^{-1}, \tilde{{\bf R}}(z)^{-1}$ and $\tilde{{\bf T}}(z)^{-1}$, 
we obtain that 
\begin{equation}
\label{eq:expre-alpha-delta}
\left( \begin{array}{c} \alpha(z) - \delta(z) \\  \tilde{\alpha}(z) - \tilde{\delta}(z) \end{array} \right) = 
{\bf D}_0(z) \left( \begin{array}{c} \alpha(z) - \delta(z) \\  \tilde{\alpha}(z) - \tilde{\delta}(z) \end{array} \right) +  \left( \begin{array}{c} \epsilon(z) \\  \tilde{\epsilon}(z) \end{array} \right)
\end{equation}
where 
\begin{equation}
\label{eq:form-D0}
{\bf D}_0(z)  = \left( \begin{array}{cc} u_0(z) & z v_0(z) \\ z \tilde{v}_0(z) & \tilde{u}_0(z) \end{array} \right)
\end{equation}
with $u_0, \tilde{u}_0, v_0, \tilde{v}_0$ defined by 
\begin{align}
u_0(z) = & \frac{1}{N} \mathrm{Tr} \frac{ \sigma^{2} {\bf R}(z) {\bf B} {\bf B}^{H} {\bf T}(z)}{(1+\sigma \alpha(z))(1+\sigma \delta(z))} \\
\tilde{u}_0(z) = & \frac{1}{N} \mathrm{Tr} \frac{ \sigma^{2} \tilde{{\bf R}}(z) {\bf B}^{H} {\bf B} \tilde{{\bf T}}(z)}{(1+\sigma \tilde{\alpha}(z))(1+\sigma \tilde{\delta}(z))} \\
v_0(z) = & \frac{1}{N} \mathrm{Tr} \sigma^{2} {\bf R}(z) {\bf T}(z) \\
\tilde{v}_0(z) = & \frac{1}{N} \mathrm{Tr} \sigma^{2} \tilde{{\bf R}}(z) \tilde{{\bf T}}(z)
\end{align}
Using the matrix inversion lemma and the observation that matrices ${\bf R}, {\bf T}, {\bf B} {\bf B}^{H}$ commute, 
the reader can check easily than $u_0(z) = \tilde{u}_0(z)$. 

In order to establish (\ref{eq:majoration-biais}), we remark that (\ref{eq:expre-alpha-delta}) 
is equivalent to the linear system 
\begin{equation}
\label{eq:systeme-alpha-delta}
\left( {\bf I} - {\bf D}_0(z) \right) \left( \begin{array}{c} \alpha(z) - \delta(z) \\  \tilde{\alpha}(z) - \tilde{\delta}(z) \end{array} \right) =  \left( \begin{array}{c} \epsilon(z) \\  \tilde{\epsilon}(z) \end{array} \right)
\end{equation}
In the following, we show matrix $\left( {\bf I} - {\bf D}_0(z) \right)$ is invertible for $z \in \mathbb{C}_{+}$, and that 
the entries of its inverse can be bounded by terms such as  $\mathrm{P}_1(|z|) \mathrm{P}_2(|\mathrm{Im}(z)|^{-1})$.
Proposition  \ref{proposition:decomposition} will follow immediately from (\ref{eq:majoration-epsilon}) and (\ref{eq:majoration-tildeepsilon}). 

We first evaluate a lower bound of $\mathrm{det}\left( {\bf I} - {\bf D}_0(z) \right)$ for 
$z \in \mathbb{C}_{+}$. For this, we introduce matrix ${\bf D}(z)$ defined by
\begin{equation}
\label{eq:form-D}
{\bf D}(z)  = \left( \begin{array}{cc} u(z) &  v(z) \\ |z|^{2} \tilde{v}(z) & \tilde{u}(z) \end{array} \right)
\end{equation}
with $u, \tilde{u}, v, \tilde{v}$ defined by 
\begin{align}
u(z) = & \frac{1}{N} \mathrm{Tr} \frac{ \sigma^{2} {\bf T}(z) {\bf B} {\bf B}^{H} {\bf T}(z)^{H}}{|1+\sigma \delta(z)|^{2}} \\
\tilde{u}(z) = & \frac{1}{N} \mathrm{Tr} \frac{ \sigma^{2} \tilde{{\bf T}}(z) {\bf B}^{H} {\bf B} \tilde{{\bf T}}(z)^{H}}{|1+\sigma \tilde{\delta}(z)|^{2}} \\
v(z) = & \frac{1}{N} \mathrm{Tr} \sigma^{2} {\bf T}(z) {\bf T}(z)^{H} \\
\tilde{v}(z) = & \frac{1}{N} \mathrm{Tr} \sigma^{2} \tilde{{\bf T}}(z) \tilde{{\bf T}}(z)^{H}
\end{align}
and define matrix ${\bf D}^{'}(z)$ as the analogue of ${\bf D}(z)$  but in which ${\bf T}, \tilde{{\bf T}}, 
\delta, \tilde{\delta}$ are replaced by  ${\bf R}, \tilde{{\bf R}}, \alpha, \tilde{\alpha}$ respectively. 
The entries of ${\bf D}^{'}(z)$ are denoted by $u^{'},  v^{'}, |z|^{2} \tilde{v}^{'}, \tilde{u}^{'}$. 
We note that the entries of ${\bf D}(z)$ and ${\bf D}^{'}(z)$ are positive, and that, 
using the matrix inversion lemma, it is easily seen that $u=\tilde{u}$ and that $u^{'} = \tilde{u}^{'}$. 
These matrices are useful because we have the following proposition. 
\begin{proposition}
\label{prop:minoration-detI-D0}
There exists a strictly positive constant $\eta$ such that 
\begin{equation}
\label{eq:minoration-detI-D}
\mathrm{det}\left({\bf I} - {\bf D}(z) \right) \geq  \frac{1}{(16)^{2}} \frac{|\mathrm{Im}(z)|^{8}}{(\eta^{2} + |z|^{2})^{4}}
\end{equation}
for each $z \in \mathbb{C}_{+}$ and for each $N$. Moreover, there exist an integer $N_0$ and
2 polynomials  $\mathrm{Q}_1$ and $\mathrm{Q}_2$, independent of $N$, with positive coefficients, such that for each $N > N_0$, 
\begin{equation}
\label{eq:minoration-detI-D'}
\mathrm{det}\left({\bf I} - {\bf D}^{'}(z) \right) \geq \frac{1}{(64)^{2}} \frac{|\mathrm{Im}(z)|^{8}}{2(\eta^{2} + |z|^{2})^{4}}
\end{equation} 
for each element $z$ of the set $\mathbb{E}_N$ defined by
\begin{equation}
\label{eq:def-ensemble}
\mathbb{E}_N = \{ z \in \mathbb{C}_{+}, \, 1 - \frac{1}{N^{2}} \mathrm{Q}_1(|z|)  \mathrm{Q}_2(\mathrm{Im}(z)^{-1}) > 0 \}
\end{equation}
Finally, for each $N > N_0$, 
\begin{equation}
\label{eq:minoration-detI-D0}
\left| \mathrm{det}\left({\bf I} - {\bf D}_0(z) \right) \right| > \sqrt{\mathrm{det}({\bf I} - {\bf D}(z))} \sqrt{\mathrm{det}({\bf I} - {\bf D}^{'}(z))}  > \frac{1}{(32)^{2}} \frac{|\mathrm{Im}(z)|^{8}}{\sqrt{2} (\eta^{2} + |z|^{2})^{4}}
\end{equation}
if $z \in \mathbb{E}_N$.
\end{proposition}
\begin{IEEEproof}
We first establish (\ref{eq:minoration-detI-D}). For this, we express $\mathrm{Im}(\delta(z))$ and 
$\mathrm{Im}(z \tilde{\delta}(z))$ as 
\begin{align}
\mathrm{Im}(\delta(z)) & = \frac{1}{N} \mathrm{Tr}\left( \sigma \mathrm{Im}({\bf T}(z)) \right) \\   
\mathrm{Im}(z \tilde{\delta}(z)) & = \frac{1}{N} \mathrm{Tr}\left( \sigma \mathrm{Im}(z \tilde{{\bf T}}(z)) \right)
\end{align}
where for each matrix ${\bf U}$, we define $\mathrm{Im}({\bf U})$ by 
 $\mathrm{Im}({\bf U}) = \frac{{\bf U} - {\bf U}^{H}}{2i}$. Writing 
$\mathrm{Im}({\bf T}(z))$ as $\frac{1}{2i} {\bf T}(z)({\bf T}(z)^{-H} - {\bf T}(z)^{-1}){\bf T}(z)^{H}$
and $\mathrm{Im}(z \tilde{{\bf T}}(z))$ as $\frac{1}{2i} z{\bf T}(z)((z{\bf T}(z))^{-H} - (z{\bf T}(z))^{-1})(z{\bf T}(z))^{H}$, we get immediately that 
\begin{equation}
\label{eq:systeme-delta-tildedelta}
\left( \begin{array}{c} \mathrm{Im}(\delta(z)) \\ \mathrm{Im}(z \tilde{\delta}(z)) \end{array} \right) = 
{\bf D}(z) \left( \begin{array}{c} \mathrm{Im}(\delta(z)) \\ \mathrm{Im}(z \tilde{\delta}(z)) \end{array} \right)
+ \left( \begin{array}{c} w(z) \\ \tilde{w}(z) \end{array} \right) \mathrm{Im}(z)
\end{equation}
where $w(z)$ and $\tilde{w}(z)$ are defined by
\begin{equation}
\label{eq:def-w-tildew}
\begin{array}{cc}
w(z) = \frac{1}{N} \mathrm{Tr} \left( \sigma^{2} {\bf T}(z) {\bf T}(z)^{H} \right) & 
\tilde{w}(z) = \frac{1}{N} \mathrm{Tr} \left( \frac{\sigma \tilde{{\bf T}}(z) {\bf B}^{H} {\bf B} \tilde{{\bf T}}(z)^{H}}{|1+\sigma \tilde{\delta}|^{2}} \right)
\end{array}
\end{equation}  
This is equivalent to 
\begin{align}
\label{eq:expre-Imdelta}
(1 - u) \, \mathrm{Im} \delta & = v \,  \mathrm{Im} (z \tilde{\delta}) +  w \,  \mathrm{Im} z \\
\label{eq:expre-Imtildedelta}
(1 - \tilde{u}) \,  \mathrm{Im} (z \tilde{\delta}) & = |z|^{2} \, \tilde{v} \, \mathrm{Im} \delta + \tilde{w} \, \mathrm{Im} z
\end{align}
As $\delta$ and $\tilde{\delta}$ are proportional to the Stieltjès transform of probability measures carried by $\mathbb{R}_{+}$, 
$\mathrm{Im}(\delta) > 0, \mathrm{Im} (z \tilde{\delta}) > 0$ for $z \in \mathbb{C}_{+}$ (see Property \ref{enu:Imzpsi}
of Lemma \ref{property:stieltjes}). Therefore, (\ref{eq:expre-Imdelta}, \ref{eq:expre-Imtildedelta}) imply that 
$1 - u = 1 - \tilde{u}$ is strictly positive. After some algebra, we also obtain that $\mathrm{det}\left( {\bf I} - {\bf D} \right) = (1-u)(1-\tilde{u}) - |z|^{2} v \tilde{v}$ coincides with 
\begin{equation}
\label{eq:expre-detI-D}
\mathrm{det}\left( {\bf I} - {\bf D} \right)=  \left( v \tilde{w} + (1-\tilde{u}) w \right) \, \frac{\mathrm{Im} z}{\mathrm{Im} \delta}
\end{equation}
Therefore, 
$$\mathrm{det}\left( {\bf I} - {\bf D} \right) \geq (1-\tilde{u}) w \, \frac{\mathrm{Im} z}{\mathrm{Im} \delta}$$
As $\delta(z) = \sigma c m(z)$, Property \ref{enu:majoration} of Lemma \ref{property:stieltjes} implies that 
$\mathrm{Im}(\delta(z)) \leq \frac{\sigma c}{\mathrm{Im}(z)}$ or equivalently that $\frac{\mathrm{Im} z}{\mathrm{Im} \delta} \geq (\mathrm{Im}(z))^{2} / \sigma c$. Hence, 
$$ \mathrm{det}\left( {\bf I} - {\bf D} \right) \geq \frac{(1-\tilde{u}) w (\mathrm{Im}(z))^{2}}{\sigma c}$$
(\ref{eq:expre-Imdelta}) implies that 
$$1 - u = 1 - \tilde{u} > w \, \frac{\mathrm{Im} z}{\mathrm{Im} \delta}  \geq \frac{w (\mathrm{Im}(z))^{2}}{\sigma c}$$
We finally get that 
\begin{equation}
\label{eq:minoration-1-detI-D}
\mathrm{det}\left( {\bf I} - {\bf D} \right) \geq \frac{w^{2} (\mathrm{Im}(z))^{4}}{(\sigma c)^{2}}
\end{equation}
In order to obtain a lower bound of $w = \frac{1}{N} \mathrm{Tr} \sigma {\bf T} {\bf T}^{H}$, we 
first remark that $ \frac{1}{M} \mathrm{Tr}  {\bf T} {\bf T}^{H} \geq \left| \frac{1}{M} \mathrm{Tr} {\bf T} \right|^{2} = |m|^{2}$
by the Jensen inequality. Therefore, $w \geq \sigma c |m|^{2} \geq \sigma c |\mathrm{Im}(m)|^{2}$. 
$\mathrm{Im}(m(z))$ can be written as 
\[
\mathrm{Im}(m(z)) = \mathrm{Im}(z) \, \int_{\mathbb{R}_{+}} \frac{d \mu_N(\lambda)}{|\lambda - z|^{2}}
\]
We recall that it is shown in \cite{hachem2007deterministic} that the sequence $(\mu_N)_{N \geq 0}$ 
is tight. This implies that it exists $\eta > 0$ for which $\mu_N(]\eta, +\infty[) \leq 1/2$ for each 
$N \in \mathbb{N}$, or equivalently
for which
\begin{equation}
\label{eq:prop-eta}
\mu_N([0, \eta]) > 1/2
\end{equation}
for each integer $N$. It is clear that
$$\int_{\mathbb{R}_{+}} \frac{d \mu_N(\lambda)}{|\lambda - z|^{2}} > \int_{0}^{\eta} \frac{d \mu_N(\lambda)}{|\lambda - z|^{2}} > \frac{1}{2(\eta^{2} + |z|^{2})} \, \mu_N([0, \eta]) > \frac{1}{4(\eta^{2} + |z|^{2})}$$
Therefore, $w > \frac{\sigma c (\mathrm{Im}(z))^{2}}{16(\eta^{2} + |z|^{2})^{2}}$ and Eq. (\ref{eq:minoration-1-detI-D})
gives (\ref{eq:minoration-detI-D}). 

We now establish (\ref{eq:minoration-detI-D'}). For this, we express that
$\mathrm{Im}(\alpha(z))$ and 
$\mathrm{Im}(z \tilde{\alpha}(z))$ as 
\begin{align}
\mathrm{Im}(\alpha(z)) & = \frac{1}{N} \mathrm{Tr}\left( \sigma \mathrm{Im}({\bf R}(z)) \right) + \mathrm{Im}(\epsilon(z)) \\   
\mathrm{Im}(z \tilde{\alpha}(z)) & = \frac{1}{N} \mathrm{Tr}\left( \sigma \mathrm{Im}(z \tilde{{\bf R}}(z)) \right) +
\mathrm{Im}(z \tilde{\epsilon}(z))
\end{align}
After some algebra, we obtain that 
\begin{equation}
\label{eq:systeme-alpha-tildealpha}
\left( \begin{array}{c} \mathrm{Im}(\alpha(z)) \\ \mathrm{Im}(z \tilde{\alpha}(z)) \end{array} \right) = 
{\bf D}^{'}(z) \left( \begin{array}{c} \mathrm{Im}(\alpha(z)) \\ \mathrm{Im}(z \tilde{\alpha}(z)) \end{array} \right)
+ \left( \begin{array}{c} w^{'}(z) \\ \tilde{w}^{'} (z) \end{array} \right) \mathrm{Im}(z) + 
\left( \begin{array}{c} \mathrm{Im}(\epsilon(z)  \\  \mathrm{Im}(z \tilde{\epsilon}(z)) \end{array} \right) 
\end{equation}
where $w^{'}(z)$ and $\tilde{w}^{'}(z)$ are defined as $w(z)$ and $\tilde{w}(z)$ by replacing 
${\bf T}(z), \tilde{{\bf T}}(z), \delta(z), \tilde{\delta}(z)$ by
${\bf R}(z), \tilde{{\bf R}}(z), \alpha(z), \tilde{\alpha}(z)$ respectively. This is equivalent to 
\begin{align}
\label{eq:expre-Imalpha}
(1 - u^{'}) \, \mathrm{Im} \alpha & = v \,  \mathrm{Im} (z \tilde{\alpha}) +  w^{'} \,  \mathrm{Im} z  + \mathrm{Im}(\epsilon(z)) \\
\label{eq:expre-Imtildealpha}
(1 - \tilde{u}^{'}) \,  \mathrm{Im} (z \tilde{\alpha}) & = |z|^{2} \, \tilde{v}^{'} \, \mathrm{Im} \alpha + \tilde{w}^{'} \, \mathrm{Im} z + \mathrm{Im}(z \tilde{\epsilon}(z))
\end{align}
These equations are of course similar to (\ref{eq:expre-Imdelta}, \ref{eq:expre-Imtildedelta})
except that the righthandsides of (\ref{eq:expre-Imalpha}, \ref{eq:expre-Imtildealpha}) are 
corrupted by the two error terms $ \mathrm{Im}(\epsilon(z))$ and  $\mathrm{Im}(z \tilde{\epsilon}(z))$. 
In order to prove (\ref{eq:minoration-detI-D'}), we follow the proof of (\ref{eq:minoration-detI-D})
but take into account the presence of the error terms in (\ref{eq:expre-Imalpha}, \ref{eq:expre-Imtildealpha}).
As $\alpha$ and $\tilde{\alpha}$ are proportional to the Stieltjès transform of probability measures carried by $\mathbb{R}_{+}$, 
$\mathrm{Im}(\alpha) > 0, \mathrm{Im} (z \tilde{\alpha}) > 0$ for $z \in \mathbb{C}_{+}$. 
Therefore, (\ref{eq:expre-Imalpha}) implies that 
\begin{equation}
\label{eq:inequation-Imalpha}
(1 - u^{'}) \, \mathrm{Im} \alpha > w^{'} \, \mathrm{Im}(z) - |\epsilon(z)|
\end{equation}
In order to determine a subset of $\mathbb{C}_{+}$ on which $1 - u^{'} = 1 - \tilde{u}^{'}$ is strictly positive,
we evaluate a lower bound of $w^{'}(z) = \frac{1}{N} \mathrm{Tr}(\sigma {\bf R}(z) {\bf R}(z)^{H})$. 
For this, we follow what preceds. We express $w^{'}$ as $w^{'} = \sigma c   \frac{1}{M} \mathrm{Tr}( {\bf R}(z) {\bf R}(z)^{H})$ and note that $w^{'} \geq  \sigma c \left|  \frac{1}{M} \mathrm{Tr} {\bf R} \right|^{2}$. As ${\bf R}(z)$ is the Stieltjès transform of
a matrix valued measure whose mass is the matrix ${\bf I}_M$, 
$\frac{1}{M} \mathrm{Tr} {\bf R}(z))$ is the Stieltjès transform of a probability measure $\xi_N$. 
It is shown in \cite{dumont2009capacity} that $\frac{1}{M} \mathrm{Tr} {\bf R}(z) - m_N(z) \rightarrow 0$ for 
each $z \in \mathbb{C} - \mathbb{R}_{+}$. Therefore, the sequence $(\xi_N - \mu_N)_{N \geq 0}$ converges weakly 
torwards 0. $\eta>0$ being defined by (\ref{eq:prop-eta}), it thus exists an integer $N_1$ for which 
\begin{equation}
\label{eq:propbisprime-eta}
\xi_N([0,\eta]) > \frac{1}{4}
\end{equation} 
for each $N > N_1$. Using the same calculations as above, we obtain that $w^{'} >   \frac{\sigma c (\mathrm{Im}(z))^{2}}{64(\eta^{2} + |z|^{2})^{2}}$. Hence, using (\ref{eq:inequation-Imalpha}) and (\ref{eq:majoration-epsilon}), we get 
\begin{equation}
\label{eq:inequationbis-Imalpha}
(1 - u^{'}) \mathrm{Im}(\alpha) \, > \, \frac{\sigma c (\mathrm{Im}(z))^{3}}{64(\eta^{2} + |z|^{2})^{2}}   - \frac{1}{N^{2}} \mathrm{P}_1(|z|) \mathrm{P}_1((\mathrm{Im}(z))^{-1})
\end{equation}
If we denote by $\mathbb{E}_{1,N}$ the subset of $\mathbb{C}_{+}$ defined by 
\begin{equation}
\label{eq:def-E1N}
\frac{\sigma c (\mathrm{Im}(z))^{3}}{64(\eta^{2} + |z|^{2})^{2}}   - \frac{1}{N^{2}} \mathrm{P}_1(|z|) \mathrm{P}_1((\mathrm{Im}(z))^{-1}) > 0
\end{equation}
it is clear that  $1 - u^{'} = 1 - \tilde{u}^{'} > 0$ for each $N > N_1$ and each $z \in \mathbb{E}_{1,N}$. We note that $\mathbb{E}_{1,N}$ can be written as
\begin{equation}
\label{eq:def-bis-E1N}
\left \{z \in \mathbb{C}_{+}, 1 - \frac{1}{N^{2}} \mathrm{S}_1(|z|) \mathrm{S}_2((\mathrm{Im}(z))^{-1}) > 0 \right \}
\end{equation}
for some polynomials with positive coefficients.

Using some algebra as well as the identity 
$u^{'} = \tilde{u}^{'}$, we get that 
\begin{equation}
\label{eq:expre-detI-D'}
\mathrm{det}\left( {\bf I} - {\bf D}^{'} \right)=  \left( v^{'} \tilde{w}^{'} + (1-u^{'}) w^{'} \right) \, \frac{\mathrm{Im} z}{\mathrm{Im} \alpha} + v^{'} \mathrm{Im}(z \tilde{\epsilon}) + (1 - u^{'}) \mathrm{Im}(\epsilon)
\end{equation}  
Therefore,  for each $N > N_1$ and each $z \in \mathbb{E}_{1,N}$, we have
\[
\mathrm{det}\left( {\bf I} - {\bf D}^{'} \right) > (1-u^{'}) w^{'}  \, \frac{\mathrm{Im} z}{\mathrm{Im} \alpha} - v^{'} |z \tilde{\epsilon}| - |\epsilon|
\]

Moreover, as $\frac{\mathrm{Im}(\alpha)}{\mathrm{Im}(z)} \leq \frac{\sigma c}{(\mathrm{Im}(z))^{2}}$, using (\ref{eq:inequation-Imalpha}), we get
\[
(1-u^{'})  > \frac{w^{'} (\mathrm{Im}(z))^{2}}{\sigma c} - \frac{|\epsilon|}{\mathrm{Im}(\alpha)}
\]
It is shown in  \cite{dumont2009capacity} that $\frac{1}{M} \mathrm{Tr} (\mathbb{E}({\bf Q}(z))) - m_N(z) \rightarrow 0$ for 
each $z \in \mathbb{C} - \mathbb{R}_{+}$. Therefore, the sequence $(\omega_N - \mu_N)_{N \geq 0}$ converges weakly 
torwards 0 where measure $\omega_N$ is defined by (\ref{eq:def-omega}). $\eta>0$ being defined by (\ref{eq:prop-eta}), it thus exists an integer $N_0 \geq N_1$ for which 
\begin{equation}
\label{eq:propter-eta}
\omega_N([0,\eta]) > \frac{1}{4}
\end{equation} 
for each $N > N_0$. This allows to show that $\mathrm{Im}(\alpha) > \frac{\sigma c \mathrm{Im}(z)}{8(\eta^{2} + |z|^{2})}$ for $N > N_0$, and that 
\[
(1-u^{'})  > \frac{w^{'} (\mathrm{Im}(z))^{2}}{\sigma c} -  \frac{8(\eta^{2} + |z|^{2})}{\sigma c \mathrm{Im}(z)} \, |\epsilon(z)|
\]
As $\| {\bf R}(z) \| \leq (\mathrm{Im}(z))^{-1}$, $v^{'} = \frac{1}{N} \mathrm{Tr} \sigma^{2} {\bf R} {\bf R}^{H}$ verifies $v^{'} \leq \sigma^{2} c  (\mathrm{Im}(z))^{-2}$
while $w^{'} =  \frac{1}{N} \mathrm{Tr} \sigma {\bf R} {\bf R}^{H}$ is less than $\sigma c  (\mathrm{Im}(z))^{-2}$. 
Putting all the pieces together, we obtain that
\begin{equation}
\label{eq:minoration-intermediaire}
 (1-u^{'}) w^{'}  \, \frac{\mathrm{Im} z}{\mathrm{Im} \alpha} \, > \frac{\mathrm{Im}(z)^{8}}{(64)^{2}(\eta^{2} + |z|^{2})^{4}} - \frac{64(\eta^{2} + |z|^{2})^{2}}{\sigma c (\mathrm{Im}(z))^{4}} \, |\epsilon(z)|
\end{equation}
and 
\begin{equation}
\label{eq:minoration-finale-detI-D'}
\mathrm{det}\left( {\bf I} - {\bf D}^{'} \right) \, > \, \frac{\mathrm{Im}(z)^{8}}{(64)^{2}(\eta^{2} + |z|^{2})^{4}} - \left(1 + \frac{64(\eta^{2} + |z|^{2})^{2}}{\sigma c (\mathrm{Im}(z))^{4}} \right) \, |\epsilon(z)| 
- \frac{\sigma^{2} c}{(\mathrm{Im}(z))^{2}} |z| |\tilde{\epsilon}(z)|
\end{equation}
for $N > N_0$ and for $z \in \mathbb{E}_{1,N}$. (\ref{eq:minoration-finale-detI-D'}) can also be written as
\[
\mathrm{det}\left( {\bf I} - {\bf D}^{'} \right) \, > \, \frac{\mathrm{Im}(z)^{8}}{(64)^{2}(\eta^{2} + |z|^{2})^{4}}  \left( 1 - \frac{1}{N^{2}} \mathrm{S}^{'}_1(|z|) \mathrm{S}^{'}_2((\mathrm{Im}(z))^{-1}) \right)
\]
for $N > N_0$ and for $z \in \mathbb{E}_{1,N}$ for some polynomials with positive coefficients independent of $N$ $\mathrm{S}^{'}_1$ and  $\mathrm{S}^{'}_2$. We denote by $\mathbb{E}_{2,N}$ the set 
\[
\mathbb{E}_{2,N} = \left\{ z \in \mathbb{C}_{+}, \left( 1 - \frac{1}{N^{2}} \mathrm{S}^{'}_1(|z|) \mathrm{S}^{'}_2((\mathrm{Im}(z))^{-1}) \right) > \frac{1}{2} \right \}
\]
We remark that 
$$
\left \{ z \in \mathbb{C}_{+}, 1 -  \frac{1}{N^{2}} \mathrm{S}_1(|z|) \mathrm{S}_2((\mathrm{Im}(z))^{-1}) - \frac{2}{N^{2}} \mathrm{S}^{'}_1(|z|) \mathrm{S}^{'}_2((\mathrm{Im}(z))^{-1}) > 0 \right \} \subset \mathbb{E}_{1,N} \cap \mathbb{E}_{2,N}
$$
We consider polynomials $\mathrm{Q}_1$ and $\mathrm{Q}_2$ defined by $\mathrm{Q}_i = \mathrm{S}_i + \sqrt{2} \mathrm{S}^{'}_i$
for $i=1,2$ and define the set $\mathbb{E}_N$ by 
$$
\mathbb{E}_N = \left \{ z \in \mathbb{C}_+,  1 - \frac{1}{N^{2}}  \mathrm{Q}_1(|z|) \mathrm{Q}_2((\mathrm{Im}(z))^{-1})
> 0 \right \}
$$
which is included into  $\mathbb{E}_{1,N} \cap \mathbb{E}_{2,N}$. It is clear that (\ref{eq:minoration-detI-D'})
holds.  \\

In order to verify  (\ref{eq:minoration-detI-D0}), we first remark that the following inequalities hold:
\begin{align}
\left| \mathrm{det} \, ({\bf I} - {\bf D}_0(z)) \right| & = \left|(1-u_0)(1-\tilde{u}_0) - z^{2} v_0 \tilde{v}_0 \right| \\
  & \geq |1 - u_0| |1 - \tilde{u}_0| - |z|^{2} |v_0| |\tilde{v}_0| \\
  & \geq (1 - |u_0|) (1 - |\tilde{u}_0|) - |z|^{2} |v_0| |\tilde{v}_0|
 \end{align}
Using the Schwartz inequality, we get that $|u_0| = |\tilde{u}_0| \leq |u|^{1/2} |u^{'}|^{1/2} =   |\tilde{u}|^{1/2} |\tilde{u}^{'}|^{1/2}$, 
$|v_0| \leq |v|^{1/2} |v^{'}|^{1/2}$, and $|\tilde{v}_0| \leq |\tilde{v}|^{1/2} |\tilde{v}^{'}|^{1/2}$. For $N > N_0$ and for $z \in \mathbb{E}_N$, 
$u=\tilde{u} < 1$ and $u^{'} = \tilde{u}^{'} < 1$ hold. Therefore, we obtain that
\begin{equation}
\label{eq:inegalite-utile}
\left| \mathrm{det} \, ({\bf I} - {\bf D}_0(z)) \right| \geq (1-|u|^{1/2} |u^{'}|^{1/2})(1- |\tilde{u}|^{1/2} |\tilde{u}^{'}|^{1/2}) - |z|^{2}  |v|^{1/2} |v^{'}|^{1/2} |\tilde{v}|^{1/2} |\tilde{v}^{'}|^{1/2}
\end{equation}
As $\mathrm{det}({\bf I} - {\bf D}(z)) = (1-u)(1-\tilde{u}) - |z|^{2} v \tilde{v}$ and  $\mathrm{det}({\bf I} - {\bf D}^{'}(z)) = (1-u^{'})(1-\tilde{u}^{'}) - |z|^{2} v^{'} \tilde{v}^{'}$
are positive for $N > N_0$ and for $z \in \mathbb{E}_N$, it is easy to check that the righthandside of (\ref{eq:inegalite-utile}) is greater than 
$ \left(\mathrm{det} \, ({\bf I} - {\bf D}(z)) \mathrm{det} \, ({\bf I} - {\bf D}^{'}(z)) \right)^{1/2}$ for $N > N_2$ and for $z \in \mathbb{E}_N$. This shows (\ref{eq:minoration-detI-D0}). 
\end{IEEEproof}

In order to complete the proof of (\ref{eq:majoration-biais}), we express $\alpha(z) - \delta(z)$ 
as 
$$
\alpha(z) - \delta(z) = \frac{1}{\mathrm{det}({\bf I} - {\bf D}_0(z))} \left[ (1 - \tilde{u}_0(z)) \epsilon(z)
+ z v_0(z) \tilde{\epsilon}(z) \right]
$$
If $N > N_2$, and if $z \in \mathbb{E}_N$, 
(\ref{eq:minoration-detI-D0}), (\ref{eq:majoration-epsilon}, \ref{eq:majoration-tildeepsilon}), $|v_0(z)| \leq \frac{\sigma^{2} c}{(\mathrm{Im}(z))^{2}}$ 
and $|u_0(z)| \leq \frac{\sigma^{2} b_{max}^{2} |z|^{2}}{(\mathrm{Im}(z))^{2}}$ (recall that $b_{max}$ is defined by (\ref{eq:Bbornee}))
give immediately
\begin{equation}
\label{eq:majoration-biais-2}   
|\alpha(z) - \delta(z)| \leq \frac{1}{N^{2}}  \mathrm{P}_1(|z|) \mathrm{P}_2((\mathrm{Im}(z))^{-1})
\end{equation}
for some polynomials $\mathrm{P}_i$, $i=1,2$ with positive coefficients. If $z \in \mathbb{C}_{+} \, \backslash \, \mathbb{E}_N$, 
we follow the trick of \cite{haagerup2005new} and \cite{capitaine2009largest}, and remark that 
$$
|\alpha(z) - \delta(z)|  \leq |\alpha(z)| + |\delta(z)| \leq \frac{2 \sigma c}{\mathrm{Im}(z)}
$$
If $z \in \mathbb{C}_{+} \, \backslash \, \mathbb{E}_N$, $2 \leq \frac{2}{N^{2}}  \mathrm{Q}_1(|z|) \mathrm{Q}_2((\mathrm{Im}(z))^{-1})$
so that
$$
|\alpha(z) - \delta(z)|  \leq  \frac{ 2 \sigma c}{\mathrm{Im}(z)} \frac{1}{N^{2}}  \mathrm{Q}_1(|z|) \mathrm{Q}_2((\mathrm{Im}(z))^{-1})
$$
Therefore, for $N > N_0$, and for each $z \in \mathbb{C}_{+}$, 
$$
|\alpha(z) - \delta(z)|  \leq \frac{1}{N^{2}} \left( \mathrm{P}_1(|z|) \mathrm{P}_2((\mathrm{Im}(z))^{-1}) + \frac{ 2 \sigma c}{\mathrm{Im}(z)}
 \mathrm{Q}_1(|z|) \mathrm{Q}_2((\mathrm{Im}(z))^{-1}) \right) \leq \frac{1}{N^{2}} (|z| + C)^{k} \mathrm{Q}((\mathrm{Im}(z))^{-1})
$$
where $k$ is an integer, $C$ is a positive constant and $\mathrm{Q}$ is a positive coefficients polynomial. 
Proposition \ref{proposition:decomposition} follows directly from the identity 
$\alpha(z) - \delta(z) = \sigma c \left(  \mathbb{E}(\frac{1}{M} \mathrm{Tr} {\bf Q}(z)) - \frac{1}{M} \mathrm{Tr} {\bf T}(z) \right)$.


\section{Proof of (\ref{eq:convergence-Qij}).}
\label{sec:individual-entries}
We first show that for each $z \in \mathbb{C}_{+}$, ${\bf u}_N^{H} ({\bf Q}_N(z) - {\bf T}_N(z)) {\bf v}_N$ 
converges towards 0 on a set of probability 1 which, in principle, depends on $z$. In order to obtain the 
almost sure convergence towards 0 for each $z \in \mathbb{C} - \mathbb{R}_{+}$, we use a standard argument
based on Montel's theorem. 

We first write 
\begin{equation}
\label{eq:erreur-Qij}
{\bf u}_N^{H} \left( {\bf Q}_N(z) - {\bf T}_N(z) \right) {\bf v}_N = {\bf u}_N^{H} \left( {\bf Q}_N(z) - \mathbb{E}({\bf Q}_N(z)) \right) {\bf v}_N +
{\bf u}_N^{H} \left( \mathbb{E}({\bf Q}_N(z)) - {\bf T}_N(z) \right) {\bf v}_N
\end{equation}
We study the second term of the righthandside of (\ref{eq:erreur-Qij}) and write
$$
{\bf u}_N^{H} \left( \mathbb{E}({\bf Q}_N(z)) - {\bf T}_N(z) \right) {\bf v}_N = {\bf u}_N^{H} \left( \mathbb{E}({\bf Q}_N(z)) - {\bf T}_N(z) \right) {\bf v}_N +  {\bf u}_N^{H} \left( {\bf R}_N(z)) - {\bf T}_N(z) \right) {\bf v}_N
$$
where we recall that matrix ${\bf R}_N(z)$ is defined by (\ref{def:R}). (\ref{eq:majoration-biais}) implies that $\alpha_N(z) - \delta_N(z)$ and $\tilde{\alpha}_N(z) - \tilde{\delta}_N(z)$ converge towards $0$ ($\alpha_N, \delta_N, \tilde{\alpha}_N, \tilde{\delta}_N$ are defined 
by (\ref{def:alpha}, \ref{def:delta}, \ref{def:tildealpha}, \ref{def:tildedelta}) respectively) . Using the identity  ${\bf R}_N(z) - {\bf T}_N(z) = {\bf R}_N(z) \left( {\bf T}_N^{-1}(z) - {\bf R}_N^{-1}(z) \right) {\bf T}_N(z)$ allows to express 
${\bf u}_N^{H} \left( {\bf R}_N(z)) - {\bf T}_N(z) \right) {\bf v}_N$ as a linear combination of $\alpha_N(z) - \delta_N(z)$ and $\tilde{\alpha}_N(z) - \tilde{\delta}_N(z)$. As $\| {\bf R}_N(z)) \| \leq |\mathrm{Im}(z)|^{-1},  \| {\bf T}_N(z)) \| \leq |\mathrm{Im}(z)|^{-1}$, the coefficients 
of this linear combination remain bounded when $N \rightarrow +\infty$. This shows that  
${\bf u}_N^{H} \left( {\bf R}_N(z)) - {\bf T}_N(z) \right) {\bf v}_N$ converges towards 0. 

In order to study ${\bf u}_N^{H} \left( \mathbb{E}({\bf Q}_N(z)) - {\bf R}_N(z) \right) {\bf v}_N$, we use relation (\ref{eq:expre-E(Q)-R}). 
Using the Nash-Poincaré inequality, it is easy to check that $ {\bf u}_N^{H} {\bf R}_N(z) \bs{\Delta}_N(z) {\bf v}_N \rightarrow 0$. 
(\ref{eq:inegalite-tracedelta}) implies moreover that $\frac{1}{N} \Tr \bs{\Delta}_N(z) \rightarrow 0$.  (\ref{eq:expre-E(Q)-R})
thus shows that  ${\bf u}_N^{H} \left( \mathbb{E}({\bf Q}_N(z)) - {\bf R}_N(z) \right) {\bf v}_N \rightarrow 0$. 

It remains to prove that $x_N(z) = {\bf u}_N^{H} \left( {\bf Q}_N(z) - \mathbb{E}({\bf Q}_N(z)) \right) {\bf v}_N$ converges towards 0 almost surely. 
For this, it is sufficient to show that 
\begin{equation}
\label{eq:moment-ordre-4}
\mathbb{E}|x_N(z)|^{4} \leq \frac{C(z)}{N^{2}}
\end{equation}
where $C(z)$ does not depend on $N$. We express $\mathbb{E}|x_N(z)|^{4}$ as
$$
\mathbb{E}|x_N(z)|^{4} = \left| \mathbb{E}(x_N(z)^{2}) \right|^{2} + \mathrm{Var} \left( x_N(z) \right)^{2}
$$
We remark that  $\left| \mathbb{E}(x_N(z)^{2}) \right|^{2} \leq \left( \mathbb{E}|x_N(z)|^{2} \right)^{2}$. Moreover,  
$\mathbb{E}(x_N(z)) = 0$ implies that $\mathbb{E}|x_N(z)|^{2} = \mathrm{Var}(x_N(z))$. Therefore, 
$$
\mathbb{E}|x_N(z)|^{4} \leq \left( \mathrm{Var}(x_N(z)) \right)^{2} +  \mathrm{Var} \left[\left( x_N(z) \right)^{2} \right]
$$
Using the Nash-Poincaré inequality, it is easy to show that $\mathrm{Var}(x_N(z)) \leq  \frac{C(z)}{N}$ and
that $\mathrm{Var} \left( x_N(z)^{2} \right ) \leq \frac{C(z)}{N^{2}}$. This establishes (\ref{eq:moment-ordre-4})
and that ${\bf u}_N^{H} \left( {\bf Q}_N(z) - {\bf T}_N(z) \right) {\bf v}_N$ converges towards 0 on a set of probability 1 
depending on $z$. 

In order to prove the almost sure convergence for each $z \in \mathbb{C} - \mathbb{R}_{+}$, we use the following standard 
argument. We consider a countable subset ${\cal Z}_c \subset \mathbb{C}_{+}$ having an accumulation point. On a set $\Omega$ of probability 1 , 
${\bf u}_N^{H} \left( {\bf Q}_N(z) - {\bf T}_N(z) \right) {\bf v}_N \rightarrow 0$ for each $z \in {\cal Z}_c$. We fix a realization 
of the set $\Omega$.
We denote by $y_N(z)$ the function $y_N(z) = {\bf u}_N^{H} \left( {\bf Q}_N(z) - {\bf T}_N(z) \right) {\bf v}_N$. 
Functions $z \rightarrow {\bf u}_N^{H} {\bf Q}_N(z) {\bf v}_N$ and $z \rightarrow {\bf u}_N^{H} {\bf T}_N(z) {\bf v}_N$ are 
Stieltjès transforms of bounded measures carried by $\mathbb{R}_{+}$. Therefore, 
function $y_N$ is analytic on $\mathbb{C} - \mathbb{R}_{+}$, 
and for each compact subset ${\cal K}$ of $\mathbb{C} - \mathbb{R}_{+}$, it holds that 
$$
|y_N(z)| \leq \frac{C}{\mathrm{dist}({\cal K}, \mathbb{R}_{+})}
$$
for some constant $C$ (this is a trivial generalization of (\ref{eq:inegalite-amelioree}) to the Stieltjès transform of a non necessarily positive
bounded measure carried by $\mathbb{R}_{+}$). Montel's theorem (\cite{conway}) thus implies that it exists a subsequence $y_{\psi(N)}$ extracted from $y_N$ which converges 
uniformly on each compact subset of $\mathbb{C} - \mathbb{R}_{+}$ towards a certain function $y_*$ which is analytic 
on $\mathbb{C} - \mathbb{R}_{+}$. However, $y_*(z) = 0$ for each $z \in {\cal Z}_c$, thus showing that $y_*$ is identically
0 on  $\mathbb{C} - \mathbb{R}_{+}$. The limit of each converging subsequence extracted from  $y_N$ is thus identically 0. We thus 
obtain that the whole sequence $y_N$ converges uniformly towards 0 on each  compact subset of  $\mathbb{C} - \mathbb{R}_{+}$. 
Therefore, for each realization of the probability 1 set $\Omega$, we have shown that 
$$
{\bf u}_N^{H} \left( {\bf Q}_N(z) - {\bf T}_N(z) \right) {\bf v}_N \rightarrow 0
$$
for each $z \in \mathbb{C} - \mathbb{R}_{+}$. This completes the proof of  (\ref{eq:convergence-Qij}).


\section{Proof of Lemma \ref{lemma:poles_localization}}

\label{section:proof_lemma_poles_localization} An elementary study of function
$x\rightarrow\hat{m}_{N}(x)$ shows that $\hat{\omega}_{k}\in\left]
\hat{\lambda}_{k}^{(N)},\hat{\lambda}_{k+1}^{(N)}\right[  $, $\forall
k=1,\ldots,M-1$ and that $\hat{\omega}_{M}^{(N)}>\hat{\lambda}_{M}^{(N)}$.
Therefore, by Theorem \ref{theo:exact-separation}, we only need to prove that
$\hat{\omega}_{M-K}^{(N)}<t_{1}^{+}$ almost surely for all sufficiently large
$N$. \newline

Consider the contour $\mathcal{C}$ defined in Proposition \ref{prop:contour}.
Noting that $\mathcal{C}$ encloses $\left\{  0\right\}  $ on the complex plane
and that $\mathrm{Ind}_{\mathcal{C}}(0)=1$, we can write%
\begin{align}
1  &  =\frac{1}{2\pi\mathrm{i}}\oint_{\mathcal{C}^{+}}\lambda^{-1}%
\mathrm{d}\lambda\\
&  =\frac{1}{2\pi\mathrm{i}}\int_{t_{1}^{-}}^{t_{1}^{+}}\left(  \frac
{w_{N}^{\prime}(x)}{w_{N}(x)}\right)  ^{\ast}\mathrm{d}x-\frac{1}%
{2\pi\mathrm{i}}\int_{t_{1}^{-}}^{t_{1}^{+}}\frac{w_{N}^{\prime}(x)}{w_{N}%
(x)}\mathrm{d}x
\end{align}
where the notation $\mathcal{C}^{+}$ means that the contour $\mathcal{C}$ is
counterclockwise oriented. Since functions $h\mapsto w_{N}(x+\mathrm{i}h)$ and
$h\mapsto w_{N}^{\prime}(x+\mathrm{i}h)$ are continuous at $h=0$ for all
$x\in]t_{1}^{-},t_{1}^{+}[$ (except for the points $x\in\left\{  x_{1}%
^{(N)-},x_{1}^{(N)+}\right\}  $), Lemma \ref{lemma:domination} together with
the Dominated Convergence Theorem imply that
\begin{align}
1  &  =\lim_{y\downarrow0}\left[  \frac{1}{2\pi\mathrm{i}}\int_{t_{1}^{-}%
}^{t_{1}^{+}}\left(  \frac{w_{N}^{\prime}(x+\mathrm{i}y)}{w_{N}(x+\mathrm{i}%
y)}\right)  ^{\ast}\mathrm{d}x-\frac{1}{2\pi\mathrm{i}}\int_{t_{1}^{-}}%
^{t_{1}^{+}}\frac{w_{N}^{\prime}(x+\mathrm{i}y)}{w_{N}(x+\mathrm{i}%
y)}\mathrm{d}x\right] \\
&  =\lim_{y\downarrow0}\left[  \frac{1}{2\pi\mathrm{i}}\oint_{\partial
\mathcal{R}_{y}^{+}}\frac{w_{N}^{\prime}(z)}{w_{N}(z)}\mathrm{d}\lambda
+\frac{1}{2\pi}\int_{-y}^{y}\frac{w_{N}^{\prime}(t_{1}^{-}-\mathrm{i}h)}%
{w_{N}(t_{1}^{-}-\mathrm{i}h)}\mathrm{d}h-\frac{1}{2\pi}\int_{-y}^{y}%
\frac{w_{N}^{\prime}(t_{1}^{+}+\mathrm{i}h)}{w_{N}(t_{1}^{+}+\mathrm{i}%
h)}\mathrm{d}h\right]
\end{align}
where $\partial\mathcal{R}_{y}^{+}$ denotes the contour of the rectangle
defined in \eqref{definition:rectangle} counterclockwise oriented. The
function $h\mapsto\frac{w^{\prime}(x+\mathrm{i}h)}{w(x+\mathrm{i}h)}$ is a
continuous function on the compact set $[-y,y]$ for $x=t_{1}^{-}$ or
$t_{1}^{+}$, and therefore the two last integrals vanish as $y\downarrow0$, so
that we can write
\[
1=\lim_{y\downarrow0}\frac{1}{2\pi\mathrm{i}}\oint_{\partial\mathcal{R}%
_{y}^{+}}\frac{w_{N}^{\prime}(z)}{w_{N}(z)}\mathrm{d}z.
\]
Since the function $\frac{w_{N}^{\prime}(\lambda)}{w_{N}(\lambda)}$ is
holomorphic on $\mathbb{C}\backslash [x_1^{(N)-}, x_1^{(N)+}]$, the last integral does not
depend on the value of $y>0$, and thus we can drop the limit, i.e.
\begin{equation}
\label{eq:argument}
1=\frac{1}{2\pi\mathrm{i}}\oint_{\partial\mathcal{R}_{y}^{+}}\frac
{w_{N}^{\prime}(z)}{w_{N}(z)}\mathrm{d}z.
\end{equation}
This identity will be key in order to prove that $\hat{\omega}_{M-K}<t_{1}%
^{+}$ almost surely for all sufficiently large $N$. \newline

Before going further into the proof of this result, let us first examine the
function $\hat{w}_{N}\left(  z\right)  $ defined by (\ref{eq:def-what}) when
$z\in\mathbb{R}$. The following result follows from elementary
analysis:\begin{figure}[h]
\centering
\par
\includegraphics[width=15cm]{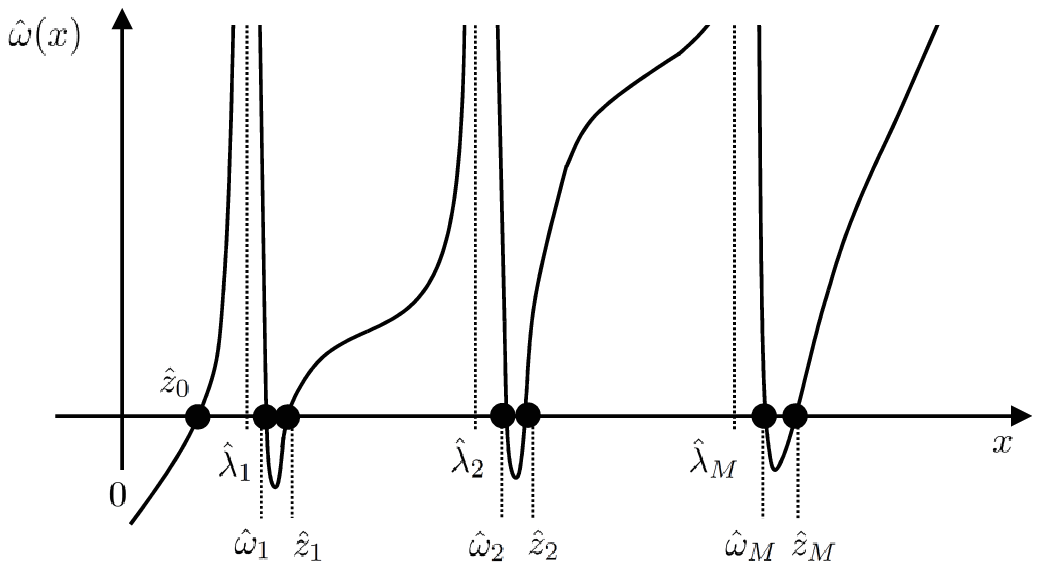}\caption{Typical representation
of $\hat{w}_{N}\left(  x\right)  $ as a function of $x$ for $M=3$ (we drop the
dependence with $N$ from all quantities for clarity). }%
\label{figure:what}%
\end{figure}

\begin{lemma}
\label{property:hatw}The function $\hat{w}_{N}$ defined on $\mathbb{R}$ by
\[
\hat{w}_{N}(x)=x\left(  1+\sigma^{2}c_{N}\hat{m}_{N}(x)\right)  ^{2}%
-\sigma^{2}(1-c_{N})\left(  1+\sigma^{2}c_{N}\hat{m}_{N}(x)\right)
\]
satisfies (see further Figure \ref{figure:what})
\begin{align}
\lim_{x\downarrow\hat{\lambda}_{k}}\hat{w}_{N}(x)  &  =+\infty,\quad
\lim_{x\uparrow\hat{\lambda}_{k}}\hat{w}_{N}(x)=+\infty\\
\lim_{x\rightarrow+\infty}\hat{w}_{N}(x)  &  =+\infty,\quad\lim_{x\rightarrow
-\infty}\hat{w}_{N}(x)=-\infty.
\end{align}
Moreover, $\hat{w}_{N}(x)=0$ is a polynomial equation with degree $2M+1$ with
the following zeros:

\begin{itemize}
\item One zero in $\left]  0,\hat{\lambda}_{1}^{(N)}\right[  $, denoted as
$\hat{z}_{0}^{(N)}$.

\item Two zeros in each interval $\left]  \hat{\lambda}_{k}^{(N)},\hat
{\lambda}_{k+1}^{(N)}\right[  $, denoted as $\hat{\omega}_{k}^{(N)}$, $\hat
{z}_{k}^{(N)}$, $k=1\ldots M-1.$

\item Two zeros in $\left]  \hat{\lambda}_{M}^{(N)},+\infty\right[  $, denoted
as $\hat{\omega}_{M}^{(N)}$, $\hat{z}_{M}^{(N)}$.
\end{itemize}

Furthermore, we have
\begin{multline*}
0<\hat{z}_{0}^{(N)}<\hat{\lambda}_{1}^{(N)}<\hat{\omega}_{1}^{(N)}<\hat{z}%
_{1}^{(N)}<\hat{\lambda}_{2}^{(N)}<\ldots\\
\ldots<\hat{\lambda}_{k}^{(N)}<\hat{\omega}_{k}^{(N)}<\hat{z}_{k}^{(N)}%
<\hat{\lambda}_{k+1}^{(N)}<\ldots<\hat{\lambda}_{M}^{(N)}<\hat{\omega}%
_{M}^{(N)}<\hat{z}_{M}^{(N)}.
\end{multline*}

\end{lemma}

Now, the function $z\rightarrow\hat{w}_{N}(z)$, defined on $\mathbb{C}$, is
holomorphic everywhere except at poles (of order $2$) $\hat{\lambda}_{1}%
^{(N)}$, \ldots, $\hat{\lambda}_{M}^{(N)}$. Moreover,
function $z\rightarrow\frac{\hat{w}_{N}^{\prime}(z)}%
{\hat{w}_{N}(z)}$ is holomorphic everywhere except at the zeros of $\hat
{w}_{N}$ and at the sample eigenvalues $\hat{\lambda}_{1}^{(N)}$, \ldots,
$\hat{\lambda}_{M}^{(N)}$.

\begin{figure}[h]
\centering
\par
\includegraphics[width=15cm]{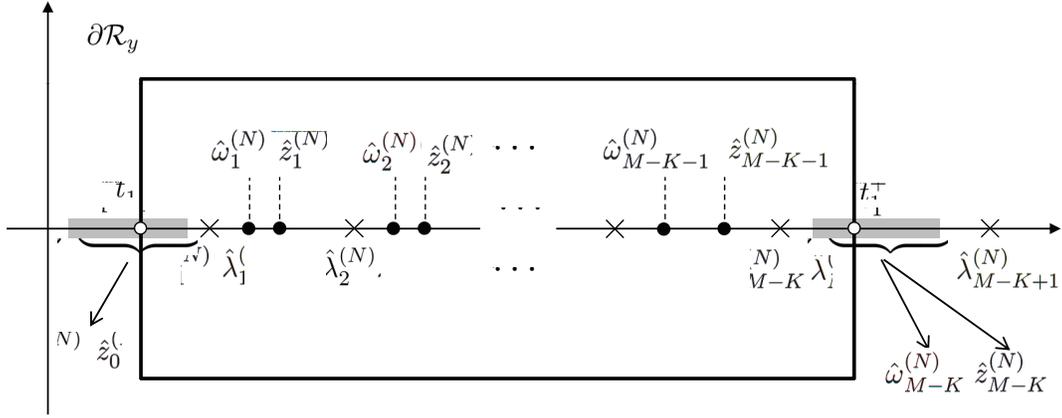}\caption{Schematic
representation of the position of the zeros (circles) and poles (crosses) of
the function $\hat{w}_{N}(z)$ on the region enclosed by $\partial
\mathcal{R}_{y}$. }%
\label{figure:zerospoles}%
\end{figure}

Figure \ref{figure:zerospoles} gives an schematic representation of the
positions of the zeros and poles of $\hat{w}_{N}(x)$ in terms of the contour
$\partial\mathcal{R}_{y}$. Observe that, for sufficiently high $N$, Theorem
\ref{theo:exact-separation} ensures that $\left\{  \hat{\lambda}_{1}%
^{(N)},\ldots,\hat{\lambda}_{M-K}^{(N)}\right\}  $ will be inside
$\partial\mathcal{R}_{y}$, whereas the rest of the sample eigenvalues will be
outside. Given the position of the zeros $\hat{\omega}_{k}^{(N)}$, $\hat
{z}_{k}^{(N)}$ established in Lemma \ref{property:hatw}, we see that the
position of the sample eigenvalues determines that the zeros $\left\{
\hat{\omega}_{k}^{(N)},\hat{z}_{k}^{(N)},k=1\ldots M-K-1\right\}  $ will also
be inside $\partial\mathcal{R}_{y}$ for all $N$ sufficiently high.
Furthermore, the remaining zeros will be outside $\partial\mathcal{R}_{y}$,
except for the zeros $\hat{z}_{0}^{(N)}$, $\hat{\omega}_{M-K}^{(N)}$ and
$\hat{z}_{M-K}^{(N)}$, for which we can not state anything. In what follows,
we will see that these three zeros are in fact located inside $\partial
\mathcal{R}_{y}$ with probability one for all large $N$, which will conclude
the proof of Lemma \ref{lemma:poles_localization}. As a first step, we
introduce an intermediate result that establishes that none of these zeros can
converge to a the boundary point of $\partial\mathcal{R}_{y}$ when
$N\rightarrow+\infty$.

\begin{lemma}
\label{lemma:nozerosonthecontour}For all $N$ large enough, $\hat{z}_{0}%
^{(N)}\neq t_{1}^{-}$, $\hat{\omega}_{M-K}^{(N)}\neq t_{1}^{+}$ and $\hat
{z}_{M-K}^{(N)}\neq t_{1}^{+}$.
\end{lemma}

%

\begin{IEEEproof}%
We will just establish that $\hat{\omega}_{M-K}^{(N)}\neq t_{1}^{+}$ and
$\hat{z}_{M-K}^{(N)}\neq t_{1}^{+}$, since the proof that $\hat{z}_{0}%
^{(N)}\neq t_{1}^{-}$ is quite similar. For this, we prove the following:
\begin{eqnarray}
\label{eq:infwN>0}
\inf_N \inf_{x \in [t_1^{+}, t_2^{-}]} |w_N(x)| > 0 \\
\label{eq:cvuniforme}
\lim_{N \rightarrow +\infty} \sup_{x \in [t_1^{+}, t_2^{-}]} |w_N(x) - \hat{w}_N(x)| = 0 \; a.s
\end{eqnarray}
If (\ref{eq:infwN>0}, \ref{eq:cvuniforme}) hold true, it is clear that 
almost surely, it exists $N_1 \in \mathbb{N}$ for which 
\begin{equation}
\label{eq:infwNhat}
\inf_{N > N_1}  \inf_{x \in [t_1^{+}, t_2^{-}]} |\hat{w}_N(x)| > 0 \; a.s.
\end{equation}
a property which implies that  $\hat{\omega}_{M-K}^{(N)}\neq t_{1}^{+}$ and $\hat
{z}_{M-K}^{(N)}\neq t_{1}^{+}$ for $N > N_1$. 

In order to prove  (\ref{eq:infwN>0}),  we note that Assumptions \ref{assumption:exact_sep_1}
and \ref{assumption:exact_sep_2} imply the existence of $\epsilon > 0$ such that $w_{N}(x)>0$ if 
$x\in\left[t_{1}^{+} - \epsilon, t_{2}^{-} + \epsilon \right]$ and $N > N_0$. Now, we write $w_N(z)$ as
\begin{equation}
\label{eq:expre-bis-w_N}
w_N(z) = z (1 + \sigma \delta_N(z)) (1 + \sigma \tilde{\delta}_N(z)) = z(1+\sigma^{2} c_N m_N(z))(1 + \sigma^{2} c_N m_N(z) - \frac{\sigma^{2} (1 - c_N)}{z})
\end{equation}
where we recall that $\delta_N$ and $\tilde{\delta}_N$ are defined by 
(\ref{def:delta}) and (\ref{def:tildedelta}) respectively. It has been mentioned
in Appendix \ref{appendix:proof_polynomial_bounding} that 
function $z \rightarrow -\frac{1}{z(1+\sigma \delta_N(z))} = -\frac{1}{z(1+\sigma^{2} c_N m_N(z))}$ coincides with the 
Stieltjès transform of a probability measure carried by $\mathbb{R}_{+}$. We denote by $\gamma_N$ this measure. As $w_N(x) > 0$ if $x\in\left[t_{1}^{+} - \epsilon, t_{2}^{-} + \epsilon \right]$,
function $z \rightarrow -\frac{1}{z(1+\sigma \delta_N(z))}$ is analytic on 
$\mathbb{C}_{+} \cup \mathbb{C}_{-} \cup ]t_{1}^{+} - \epsilon, t_{2}^{-} + \epsilon[$ and is real-valued on $[t_{1}^{+} - \epsilon, t_{2}^{-} + \epsilon]$.
The support of measure $\gamma_N$ is thus included into $\mathbb{R}_{+} - ]t_{1}^{+} - \epsilon, t_{2}^{-} + \epsilon[$. Therefore, 
Property \ref{eq:inegalite-amelioree} of Lemma \ref{property:stieltjes} implies that
\begin{equation}
\label{eq:bornesup1}
\left| x(1+\sigma^{2} c_N m_N(x)) \right|^{-1} \leq \frac{1}{\epsilon}
\end{equation}
for each $x \in [t_1^{+}, t_2^{-}]$. It can also be shown that $z \rightarrow -\frac{1}{z(1+\sigma \tilde{\delta}_N(z))} 
= -\frac{1}{z(1+\sigma^{2} c_N m_N(z))- \sigma^{2}(1 - c_N)}$ coincides with the Stieltjès transform of a probability measure carried by $\mathbb{R}_{+}$.
Using the same approach as above, we obtain that 
\begin{equation}
\label{eq:bornesup2}
\left| x(1+\sigma^{2} c_N m_N(x)) - \sigma^{2} (1 - c_N)\right|^{-1} \leq \frac{1}{\epsilon}
\end{equation}
for  each $x \in [t_1^{+}, t_2^{-}]$. This, in turn, implies (\ref{eq:infwN>0}). 

In order to establish (\ref{eq:cvuniforme}), we note that it is sufficient to establish that 
\begin{equation}
\label{eq:cvuniforme-m}
\lim_{N \rightarrow +\infty} \sup_{x \in [t_1^{+}, t_2^{-}]} |m_N(x) - \hat{m}_N(x)| = 0 \; a.s
\end{equation}
Theorem \ref{theo:exact-separation} implies the existence of
$\epsilon > 0$ for which, almost surely, function $z \rightarrow \hat{m}_N(z)$ is analytic 
on $\mathbb{C}_{+} \cup \mathbb{C}_{-} \cup ]t_1^{+} - \epsilon, t_2^{-} + \epsilon[$ for $N > N_1$ where $N_1 > N_0$ is a certain integer.  
Eq. (\ref{eq:inegalite-amelioree}) implies that for each compact 
subset ${\cal K}$ of  $\mathbb{C}_{+} \cup \mathbb{C}_{-} \cup ]t_1^{+} - \epsilon, t_2^{-} + \epsilon[$, there exists a constant 
$C({\cal K})$ for which almost surely $\sup_{N > N_1} \sup_{z \in {\cal K}} |\hat{m}_N(z)| \leq  C({\cal K}) $. For the same reasons, 
it holds that $\sup_{N > N_1} \sup_{z \in {\cal K}}  |m_N(z)|\leq  C({\cal K})$. Montel's Theorem (\cite{conway}) 
thus implies that it exists a subsequence $\hat{m}_{\psi(N)} - m_{\psi(N)}$ extracted from $(\hat{m}_{N} - m_{N})_{N > N_1}$
which converges uniformly on each compact subset of $\mathbb{C}_{+} \cup \mathbb{C}_{-} \cup ]t_1^{+} - \epsilon, t_2^{-} + \epsilon[$  torwards a function $p_{*}(z)$, analytic 
on  $\mathbb{C}_{+} \cup \mathbb{C}_{-}  \cup ]t_1^{+} - \epsilon, t_2^{-} + \epsilon[$. Proposition \ref{prop:m} implies 
that almost surely, $\hat{m}_N(z) - m_N(z) \rightarrow 0$ for each $z \in \mathbb{C} \backslash \mathbb{R}_{+}$. This implies that 
$p_*(z)$ is identically zero. As the limit of each convergent subsequence extracted from  $\hat{m}_{N} - m_{N}$ is 0, 
the whole sequence  $(\hat{m}_{N} - m_{N})_{N > N_1}$ converges uniformly torwards 0 on each compact subset of $\mathbb{C}_{+} \cup \mathbb{C}_{-}  \cup ]t_1^{+} - \epsilon, t_2^{-} + \epsilon[$. 
This, of course, implies (\ref{eq:cvuniforme-m}).  This completes the proof of Lemma \ref{lemma:nozerosonthecontour}.

\end{IEEEproof}

Using the same arguments as above, it is easy to show that  there exists $N_2 \in \mathbb{N}$ 
such that $\inf_{N >N_2 } \inf_{z \in \partial\mathcal{R}_{y}} |w_N(z)| > 0$ 
and such that, almost surely,
$\inf_{N > N_2} \inf_{z \in \partial\mathcal{R}_{y}} |\hat{w}_N(z)| > 0$. 
It also holds that $\sup_{N > N_2} \sup_{z \in \partial\mathcal{R}_{y}} |w^{'}_N(z)| < +\infty$
and  $\sup_{N > N_2} \sup_{z \in \partial\mathcal{R}_{y}} |\hat{w}^{'}_N(z)| < +\infty$ almost surely. 
Since almost surely the function $\frac{\hat{w}_{N}^{\prime}(z)}{\hat{w}_{N}(z)}-\frac
{w_{N}^{\prime}(z)}{w_{N}(z)}$ converges to $0$ for each $z \in  \partial\mathcal{R}_{y}$, 
the Dominated Convergence Theorem ensures that, with probability one,
\[
\left\vert \frac{1}{2\pi\mathrm{i}}\oint_{\partial\mathcal{R}_{y}^{+}}\left[
\frac{w_{N}^{\prime}(z)}{w_{N}(z)}-\frac{\hat{w}_{N}^{\prime}(z)}{\hat{w}%
_{N}(z)}\right]  \mathrm{d}z\right\vert \xrightarrow[N\to +\infty]{}0
\]
Now, according to Lemma \ref{lemma:nozerosonthecontour}, $\hat{z}_{0}%
^{(N)}\neq t_{1}^{-},\hat{\omega}_{M-K}^{(N)}\neq t_{1}^{+},\hat{z}%
_{M-K}^{(N)}\neq t_{1}^{+}$ with probability one for all large $N$. Hence, it
is possible to use the argument principle to function $\frac{\hat{w}^{\prime
}(z)}{\hat{w}(z)}$ on contour $\partial\mathcal{R}_{y}$. More precisely,
\[
\frac{1}{2\pi\mathrm{i}}\oint_{\partial\mathcal{R}_{y}^{+}}\frac{\hat{w}%
_{N}^{\prime}(z)}{\hat{w}_{N}(z)}\mathrm{d}z=\mathrm{card}\left\{  z\in
:\hat{w}_{N}(z)=0\right\}  -2(M-K)
\]
and since the previous integral is an integer, using (\ref{eq:argument}), we finally have with
probability one for $N$ large enough
\[
2(M-K)+1=\mathrm{card}\left\{  z\in\mathcal{R}_{y}:\hat{w}_{N}(z)=0\right\}
.
\]
We already know that $\hat{z}_{1}^{(N)}$,\ldots,$\hat{z}_{M-K-1}^{(N)}$ and
$\hat{\omega}_{1}^{(N)}$,\ldots,$\hat{\omega}_{M-K-1}^{(N)}$, which are zeros
of $\hat{w}_{N}(z)$, belong to $\mathcal{R}_{y}$. Since the total number of
zeros is $2M+1$, $3$ other zeros of $\hat{w}_{N}(z)$ belong to $\mathcal{R}%
_{y}$ with probability one for $N$ large enough. However, all the zeros of
$\hat{w}_{N}(z)$ are real-valued, which implies that the $3$ additional zeros
necessarily include $\hat{\omega}_{M-K}^{(N)}$. This concludes the proof Lemma
\ref{lemma:poles_localization}.

\section{Proof of (\ref{eq:borne-g}) and (\ref{eq:borne-hatg}).}
\label{sec:borne-g-hatg}
We first establish (\ref{eq:borne-g}). For this, we 
recall that  ${\bf T}_N(z)$ is the 
Stieltjès transform of a positive matrix valued measure ${\bs \Gamma}_N$ with mass 
${\bf I}_N$. Therefore,
function $z \rightarrow {\bf b}_N^{H} {\bf T}_N(z) {\bf b}_N$
coincides with the Stieltjès transform of the positive measure 
${\bf b}_N^{H} {\bs \Gamma}_N {\bf b}_N$. This measure is clearly 
absolutely continuous w.r.t. measure $\mathrm{Tr}(\bs{\Gamma}_N)$, 
or equivalently w.r.t. measure $\mu_N = \frac{1}{M} \mathrm{Tr}(\bs{\Gamma}_N)$. 
The support of ${\bf b}_N^{H} {\bs \Gamma}_N {\bf b}_N$ is thus contained into 
${\cal S}_N$. Therefore, it holds that 
\[
|{\bf b}_N^{H} {\bf T}_N(z) {\bf b}_N| \leq \frac{\| {\bf b}_N \|^{2}}{\mathrm{dist}(z, {\cal S}_N)}
\]
(see (\ref{eq:inegalite-amelioree}). We have already mentioned in Appendix \ref{appendix:proof_polynomial_bounding} 
and in Appendix \ref{section:proof_lemma_poles_localization} that function $z \rightarrow \left( -z(1+\sigma^{2} c_N m_N(z)) \right)^{-1}$
is the Stieltjès transform of a probability measure carried by $\mathbb{R}_{+}$. 
This function is moreover analytic in $\mathbb{C} - {\cal S}_N$ because $1 + \sigma^{2} c_N m_N(z) \neq 0$ 
on $\mathbb{C} - {\cal S}_N$ (see Property \ref{enu:Re_m} of Proposition \ref{prop:m}), a property which 
implies that the support of its associated measure is included into ${\cal S}_N$. Therefore, 
we have 
\[
\left| -z(1+\sigma^{2} c_N m_N(z)) \right|^{-1} \leq \frac{1}{\mathrm{dist}(z, {\cal S}_N)}
\]
or equivalently
\[
\left| 1+\sigma^{2} c_N m_N(z)) \right|^{-1} \leq \frac{|z|}{\mathrm{dist}(z, {\cal S}_N)}
\]
Assumptions (\ref{assumption:exact_sep_1}) and  (\ref{assumption:exact_sep_2}) imply that $\inf_{N>N_0} \mathrm{dist}( \partial {\cal R}_y, {\cal S}_N) > 0$. We thus obtain that 
\[
\sup_{N>N_0} \sup_{z \in \partial {\cal R}_y} \frac{| {\bf b}_N^{H} {\bf T}_N(z) {\bf b}_N|}{| 1+\sigma^{2} c_N m_N(z))|} < +\infty
\] 
Using again that  $\inf_{N>N_0} \mathrm{dist}( \partial {\cal R}_y, {\cal S}_N) > 0$, it can be checked that 
$\sup_{N>N_0} \sup_{z \in \partial {\cal R}_y} |w_N^{'}(z)| < +\infty$. This in turn establishes (\ref{eq:borne-g}). 

In order to prove (\ref{eq:borne-hatg}), we recall that $\hat{m}_N(z)$ 
is the Stieltjès transform of the probability measure $\hat{\mu}_N = \frac{1}{M} \sum_{k=1}^{M} \delta(\lambda - \hat{\lambda}_k^{(N)})$. 
Assumptions (\ref{assumption:exact_sep_1}) and  (\ref{assumption:exact_sep_2}) imply it exists $N_0 \in \mathbb{N}$ such that 
the distance between  $\partial {\cal R}_y$
and the support of $\hat{\mu}_N$ is lower bounded by a strictly positive term independent of $N \geq N_0$. It is easily seen that 
$z \rightarrow {\bf b}_N^{H} {\bf Q}_N(z) {\bf b}_N$ is the Stieltjès transform of measure 
$\frac{1}{M} \sum_{k=1}^{M}  |{\bf b}_N^{H} \hat{{\bf e}}_k^{(N)}|^{2} 
\delta(\lambda - \hat{\lambda}_k^{(N)})$. The support of this measure is included into $\{ \hat{\lambda}_1^{(N)}, \ldots, \hat{\lambda}_M^{(N)} \}$.
Using (\ref{eq:inegalite-amelioree}) as above, we deduce from this that 
$$
\sup_{N \geq N_0} \sup_{z \in \partial {\cal R}_y} {\bf b}_N^{H} {\bf Q}_N(z) {\bf b}_N < +\infty
$$ 
The same arguments can be used to show that $\sup_{N \geq N_0} \sup_{z \in \partial {\cal R}_y} |\hat{w}_N^{'}(z)| < +\infty$.

Finally, using Property \ref{enu:converse-stieljes} of Lemma \ref{property:stieltjes}, it is easily seen that function $z \rightarrow \left( -z(1+\sigma^{2} c_N \hat{m}_N(z)) \right)^{-1}$ 
is the Stieltjès transform of a probability measure. Its support is included into the set 
$\{  \hat{\lambda}_1^{(N)}, \ldots, \hat{\lambda}_M^{(N)}, \hat{\omega}_1^{(N)}, \ldots, \hat{\omega}_M^{(N)} \}$.  
Moreover, in the statement of Lemma \ref{lemma:poles_localization}, $t_1^{-}$ and $t_1^{+}$ can be replaced by 
$t_1^{-} + \epsilon_1$ and $t_1^{+}- \epsilon_1$ where $\epsilon_1$ is chosen in such a way that 
$t_1^{-} + \epsilon_1 < \inf_{N>N_0} x_1^{(N)-} < \sup_{N>N_0} x_1^{(N)+} < t_1^{+}- \epsilon_1$. Therefore, 
the distance between $\partial {\cal R}_y$ and 
$\{  \hat{\lambda}_1^{(N)}, \ldots, \hat{\lambda}_M^{(N)}, \hat{\omega}_1^{(N)}, \ldots, \hat{\omega}_M^{(N)} \}$
is lower bounded by  a strictly positive term independent of $N \geq N_0$. This implies that 
$$
\sup_{N \geq N_0} \sup_{z \in \partial {\cal R}_y} \left| 1+\sigma^{2} c_N \hat{m}_N(z) \right|^{-1} < +\infty
$$
This completes the proof of (\ref{eq:borne-hatg}).

\section{Proof of Lemma \ref{lemma:xi_transformation}}

\label{section:proof_lemma_xi_transformation}

We first write the equation in $\omega$, $1+\sigma^{2}c\hat{m}_{N}(\omega)=0$
as
\begin{equation}
\frac{\sigma^{2}c_{N}}{M}\sum_{j=1}^{M}\frac{1}{\hat{\lambda}_{j}-\omega
}+1=0\label{equation:mu}%
\end{equation}
and by multiplying the left hand side by $\prod_{i=1}^{M}\left(  \hat{\lambda
}_{j}-\omega\right)  $, we define a new polynomial $Q(\omega)$, by
\[
Q(\omega)=\frac{\sigma^{2}c_{N}}{M}\sum_{j=1}^{M}\prod_{\substack{l=1\\l\neq
j}}^{M}\left(  \hat{\lambda}_{l}-\omega\right)  +\prod_{l=1}^{M}\left(
\hat{\lambda}_{l}-\omega\right)  .
\]
As the monic polynomial function $Q$ has $M$ roots at $\hat{\omega}_{1}%
,\ldots,\hat{\omega}_{M}$, we can write
\[
Q(\omega)=\prod_{l=1}^{M}\left(  \hat{\omega}_{l}-\omega\right)
\]
Therefore,
\begin{equation}
Q(\hat{\lambda}_{k})=\prod_{l=1}^{M}\left(  \hat{\omega}_{l}-\hat{\lambda}%
_{k}\right)  =\frac{\sigma^{2}c_{N}}{M}\prod_{\substack{l=1\\l\neq k}}^{M}\left(  \hat{\lambda}_{l}-\hat{\lambda}_{k}\right)
\label{equation:Q_lambda_k}%
\end{equation}
which will be useful later on. Let us now consider the derivative of $Q$ given
by
\begin{equation}
Q^{\prime}(\omega)=-\sum_{j=1}^{M}\prod_{\substack{l=1\\l\neq j}}^{M}\left(
\hat{\omega}_{l}-\omega\right)  =-\sum_{j=1}^{M}\prod_{\substack{l\neq
j\\l=1}}^{M}\left(  \hat{\lambda}_{l}-\omega\right)  -\frac{\sigma^{2}c_{N}%
}{M}\sum_{m=1}^{M}\sum_{\substack{l=1\\l\neq m}}^{M}\prod
_{\substack{j=1\\j\neq m,l}}^{M}\left(  \hat{\lambda}_{j}-\omega\right)
\end{equation}
Evaluating again this function at point $\hat{\lambda}_{k}$, we obtain
\begin{equation}
Q^{\prime}(\hat{\lambda}_{k})=-\sum_{j=1}^{M}\prod_{\substack{l=1\\l\neq
j}}^{M}\left(  \hat{\omega}_{l}-\hat{\lambda}_{k}\right)  =-\prod
_{\substack{l=1\\l\neq k}}^{M}\left(  \hat{\lambda}_{l}-\hat{\lambda}%
_{k}\right)  -\frac{2\sigma^{2}c_{N}}{M}\sum_{\substack{l=1\\l\neq k}%
}^{M}\prod_{\substack{j=1\\j\neq k,l}}^{M}\left(  \hat{\lambda}_{j}%
-\hat{\lambda}_{k}\right)
\end{equation}
or, dividing both sides by the first term on the right hand side of the
equation,%
\[
\frac{\sum_{j=1}^{M}\prod_{\substack{l=1\\l\neq j}}^{M}\left(  \hat{\omega
}_{l}-\hat{\lambda}_{k}\right)  }{\prod_{\substack{l=1\\l\neq k}}^{M}\left(
\hat{\lambda}_{l}-\hat{\lambda}_{k}\right)  }=1+\frac{2\sigma^{2}c_{N}}{M}%
\sum_{\substack{l=1\\l\neq k}}^{M}\frac{1}{\hat{\lambda}_{l}-\hat{\lambda}%
_{k}}%
\]
Going back to equation \eqref{equation:Q_lambda_k}, one can also write
\begin{equation}
\frac{\sum_{j=1}^{M}\prod_{\substack{l=1\\l\neq j}}^{M}\left(  \hat{\omega
}_{l}-\hat{\lambda}_{k}\right)  }{\prod_{\substack{l=1\\l\neq k}}^{M}\left(
\hat{\lambda}_{l}-\hat{\lambda}_{k}\right)  }=\frac{\sigma^{2}c_{N}}{M}%
\frac{\sum_{j=1}^{M}\prod_{\substack{l=1\\l\neq j}}^{M}\left(  \hat{\omega
}_{l}-\hat{\lambda}_{k}\right)  }{\prod_{l=1}^{M}\left(  \hat{\omega}_{l}%
-\hat{\lambda}_{k}\right)  }=\frac{\sigma^{2}c_{N}}{M}\sum_{l=1}^{M}\frac
{1}{\hat{\omega}_{l}-\hat{\lambda}_{k}}.
\end{equation}
Consequently, we\ see that we can write%
\[
1+\frac{2\sigma^{2}c_{N}}{M}\sum_{\substack{l=1\\l\neq k}}^{M}\frac{1}%
{\hat{\lambda}_{l}-\hat{\lambda}_{k}}=\frac{\sigma^{2}c_{N}}{M}\frac{1}%
{\hat{\omega}_{k}-\hat{\lambda}_{k}}+\frac{\sigma^{2}c_{N}}{M}\sum
_{\substack{l=1\\l\neq k}}^{M}\frac{1}{\hat{\omega}_{l}-\hat{\lambda}_{k}}%
\]
or, reorganizing the terms of this expression in a convenient way,%
\begin{equation}
1+\frac{\sigma^{2}c_{N}}{M}\frac{1}{\hat{\lambda}_{k}-\hat{\omega}_{k}}%
+\frac{\sigma^{2}c_{N}}{M}\sum_{\substack{l=1\\l\neq k}}^{M}\frac{1}%
{\hat{\lambda}_{l}-\hat{\lambda}_{k}}=\frac{\sigma^{2}c_{N}}{M}\sum
_{\substack{l=1\\l\neq k}}^{M}\frac{1}{\hat{\omega}_{l}-\hat{\lambda}_{k}%
}-\frac{\sigma^{2}c_{N}}{M}\sum_{\substack{l=1\\l\neq k}}^{M}\frac{1}%
{\hat{\lambda}_{l}-\hat{\lambda}_{k}}.\label{equation:tmp1_lemme_simp}%
\end{equation}
But from the equation in $\omega$ \eqref{equation:mu}, we obtain
\[
1+\frac{\sigma^{2}c_{N}}{M}\frac{1}{\hat{\lambda}_{k}-\hat{\omega}_{k}}%
+\frac{\sigma^{2}c_{N}}{M}\sum_{\substack{l=1\\l\neq k}}^{M}\frac{1}%
{\hat{\lambda}_{l}-\hat{\omega}_{k}}=0
\]
and by inserting this expression into \eqref{equation:tmp1_lemme_simp}, we
finally get the expression in the lemma.

\bibliographystyle{IEEEtran}
\bibliography{biblio}

\end{document}